\documentclass{aa}  

\usepackage{graphicx}
\usepackage{txfonts}
\usepackage{booktabs,threeparttable}
\usepackage{sistyle}
\usepackage{amsmath}
\usepackage{array,multirow,makecell}
\usepackage{subfig}
\usepackage{color}
\usepackage{gensymb}
\usepackage[utf8]{inputenc}
\usepackage[debug]{hyperref}
\hypersetup{
    colorlinks=true,
    linkcolor=blue,
    filecolor=magenta,      
    citecolor=blue,
}
\usepackage[capitalise]{cleveref} 

\newcommand{\Msol}{\ensuremath{M_{\odot}}}

\begin{document}

   \title{Mass segregation and sequential star formation in NGC~2264 revealed by \emph{Herschel}\thanks{\emph{Herschel} is an ESA space observatory with science instruments provided by European-led Principal Investigator consortia and with important participation from NASA.} }


  \author{T. Nony \inst{1}  
           \and J.-F. Robitaille \inst{2}        
           \and F. Motte\inst{2} 
           \and M. Gonzalez\inst{2}
           \and I. Joncour\inst{2}
           \and E. Moraux\inst{2}
           \and A. Men'shchikov\inst{3} 
           \and P. Didelon\inst{3} 
           \and F. Louvet\inst{3} 
           \and A. S. M. Buckner\inst{4} 
           \and N. Schneider\inst{5} 
           \and S. L. Lumsden\inst{6}  
           \and S. Bontemps\inst{7}   
           \and Y. Pouteau\inst{2}
           \and N. Cunningham\inst{2}
           \and E. Fiorellino\inst{8}
           \and R. Oudmaijer\inst{6}
           \and P. André\inst{3}
           \and B. Thomasson\inst{2}}

    \institute{Instituto de Radioastronomía y Astrofísica, Universidad Nacional Autónoma de México, Apdo. Postal 3-72, 58089 Morelia, Michoacán, México
    \email{t.nony@irya.unam.mx}
    \and Univ. Grenoble Alpes, CNRS, IPAG, 38000 Grenoble, France
    \and AIM, CEA, CNRS, Université Paris-Saclay, Université Paris Diderot, Sorbonne Paris Cité, 91191 Gif-sur-Yvette, France
     \and School of Physics and Astronomy, University of Exeter, Stocker Road, Exeter, EX4 4QL, UK
     \and Physikalisches Institut, Universit{\"a}t zu K{\"o}ln, Z{\"u}lpicher Str. 77, 50937 K{\"o}ln, Germany
     \and School of Physics and Astronomy, University of Leeds, Leeds LS2 9JT, U.K.
    \and Laboratoire d'astrophysique de Bordeaux, Univ. Bordeaux, CNRS, B18N, allée Geoffroy Saint-Hilaire, 33615 Pessac, France
    \and INAF - Istituto di Astrofisica e Planetologia Spaziali (IAPS), via Fosso del Cavaliere 100, 00133 Roma, Italy
             }

   \date{}

  \abstract
  {The mass segregation of stellar clusters could be primordial rather than dynamical. Despite the abundance of studies of mass segregation for stellar clusters, those for stellar progenitors are still scarce, so the question on the origin and evolution of mass segregation is still open.}
  %
   {Our goal is to characterize the structure of the NGC~2264 molecular cloud and compare the populations of clumps and young stellar objects (YSOs) in this region whose rich YSO population has shown evidence of sequential star formation.}
  %
   {We separated the \emph{Herschel} column density map of NGC 2264 in three subregions and compared their cloud power spectra using a multiscale segmentation technique. 
   We extracted compact cloud fragments from the column density image,
   measured their basic properties, and studied their spatial and mass distributions.
   }
   %
   {We identified in the whole NGC~2264 cloud a population of 256 clumps with typical sizes of $\sim$0.1 pc and masses ranging from 0.08~$\Msol$ to 53~$\Msol$. 
   Although clumps have been detected all over the cloud, the central subregion of NGC 2264 concentrates most of the massive, bound clumps.
   The local surface density and the mass segregation ratio indeed indicate a strong degree of mass segregation for the 15 most massive clumps, with a median $\Sigma_{6}$ three time that of the whole clumps population and $\Lambda_{\rm MSR} \simeq 8$. We showed that this cluster of massive clumps is forming within a high-density cloud ridge, itself formed and probably still fed by the high concentration of gas observed on larger scales in the central subregion. 
   The time sequence obtained from the combined study of the clump and YSO populations in NGC~2264 suggests that the star formation started in the northern subregion, that it is now actively developing at the center and will soon start in the southern subregion.
   }
   {
   Taken together, the cloud structure and the clump and YSO populations in NGC~2264 argue for a dynamical scenario of star formation. Cloud could first undergo global collapse, driving most clumps to centrally concentrated ridges/hubs. After their main accretion phase, some, and probably the most massive, YSOs would stay clustered while others would be ejected from their birth sites. We propose that the mass segregation observed in some star clusters could be inherited from that of clumps, originating from the mass assembly phase of molecular clouds.}
 

   \maketitle
%

\section{Introduction}
\label{sec:intro}

Among the major open questions in the field of stellar cluster formation today is to define whether stellar properties, and especially their clustering characteristics and dynamical state, could be inherited from the properties of the clouds themselves. 
Mass segregation has been studied extensively in stellar clusters for decades, to investigate if the energy equipartition associated with two body relaxation \citep{Spitzer69} 
causes high-mass stars to `fall' toward the center of mass of clusters. As a matter of fact, mass segregation has been observed, as expected, in dynamically old globular and open clusters, but also in very young regions, suggesting it was not due to dynamical evolution \citep[see, e.g the review by][and references therein]{Meylan00}. Primordial mass segregation prompts the question of whether there are preferential sites for high-mass star formation \citep{Murray96} or whether the observed segregation is related to energy equipartition at all \citep[e.g.,][]{Parker16}.
Simulations indicate that primordial structure, if present, can be rapidly erased through dynamical interactions of the stars and expulsion of the gas \citep{Parker12,Fujii16}. Observational studies, however, show a weak, if any, correlation between the substructure and the age of the stellar clusters  \citep{Sanchez09,Dib18}. \cite{Hetem19} indicate that stellar clusters did not significantly change in terms of structure within their first 10~Myrs, and corroborate the absence of correlation between mass segregation and the age of clusters, as initially reported by \cite{Dib18}. 
Specifically, measuring the spatial distribution and mass segregation of molecular cloud fragments and comparing it to that of very young stellar objects is essential to study this link. 
A handful of studies have started investigating the mass segregation of molecular cores \citep{Plunkett18,Dib19, Roman19}, mostly focusing, first, on low-mass star-forming regions. 

The \emph{Herschel} observatory imaged, with an unprecedented sensitivity and angular resolution, the low- to high-mass star-forming clouds located at 100--500~pc and 0.7--3~kpc from the Sun \citep{Andre10, Motte10}. These surveys provide the opportunity to build large, if not complete, catalogs of cloud fragments with 0.01-0.1~pc (qualified of cores) to 0.1-1~pc sizes (called clumps).
 At these scales, clumps will likely fragment into cores, which could themselves be the direct progenitors of 
stars or small stellar systems.
HOBYS\footnote{see http://hobys-herschel.cea.fr}, the \emph{Herschel} imaging survey of OB young stellar objects  \citep[][]{Motte10}, is the first systematic survey of a complete sample of nearby molecular cloud complexes forming high-mass stars. The wide-field photometric imaging, performed with both the SPIRE and PACS cameras of \emph{Herschel}, aims at completing the census of high-mass star progenitors at 0.1~pc scales in essentially all the molecular cloud complexes at less than 3~kpc. The HOBYS sample contains seven massive, $3\times 10^5$ to $3\times 10^6$~\Msol, molecular complexes at 1.4-3~kpc from the Sun, including Cygnus~X and NGC~6334, and two intermediate-mass, a few $10^5$~\Msol, cloud complexes at 0.7-0.8~kpc, Vela~C and Monoceros (Mon~R1, Mon~R2, NGC~2264).

HOBYS notably revealed a tight link between the density, dynamics, and clump population of the so-called ridges \citep{Motte18a}. The latter are high-density filaments ($n>10^5$~cm$^{-3}$ 
over $\sim$5~pc$^3$) forming clusters of high-mass stars \citep[e.g.,][]{Schneider10, Hill11, Nguyen11, Hennemann12, Nguyen13, Tige17}, whereas hubs are more spherical smaller structures forming at most a couple of high-mass stars \citep[e.g.,][]{Schneider12,Peretto13, Rivera13, Didelon15, Rayner17}. 
The existence of ridges and hubs is predicted by dynamical models of cloud formation such as colliding flow or gravitationally-driven gas inflows simulations \citep[e.g.,][]{Heitsch08,Smith09,Hartmann12} and some analytical theories of filament collapse/conveyor belt \citep{Myers09,Krumholz20}.

The Monoceros OB1 (Mon~OB1) cloud complex is located at a Gaia DR2 determined distance of 723$^{+56}_{-49}$ pc \citep[][]{Cantat18}. 
In the mid-infrared, it displays a 2.5$\degree$ ($\sim$30~pc) diameter half loop \citep{Schwartz87},
of which three 10~pc clouds have been imaged by \emph{Herschel} as part of the HOBYS and Galactic Cold Cores key programs \citep{Motte10,Juvela12}. 
NGC~2264 and G202.3+2.5 lie at the eastern extremity of Mon~OB1, Monoceros R1 is in the west.
NGC~2264 is the best-studied cloud of the Mon~OB1 complex \citep[see the review by][]{Dahm08}. Over the past decade, a population of more than 1000 Young Stellar Objects (YSOs) at various evolutionary stages
has been revealed \citep{Teixeira12,Povitch13,Rapson14,Venuti18}. 
The NGC~2264 cloud is elongated in the NW-SE direction, with a gradient of star formation observed for YSOs
from north to south \citep{Sung10,Venuti18}.
NGC~2264 hosts the massive O7-type binary star S~Monocerotis (S~Mon), in the north, and the famous Cone Nebula, a triangular projection of molecular gas lying $\sim$8~pc south of S Mon. 
The YSO population of NGC 2264 is mainly distributed in three sub-clusters: a cluster of pre-main sequence stars in the north, in the vicinity of S Mon \citep{Sung09} and two clusters in the center. 
The central sub-clusters are often labeled according to their dominant objects, NGC~2264-IRS1, a B2-type star also known as Allen's source and -IRS2, a Class I protostar. IRS1 and IRS2 areas are also referred as NGC~2264-C and -D, or the Spokes cluster for NGC~2264-IRS2 region \citep{Teixeira06}. 
These regions are still active sites of star formation, with many embedded Class~0/I protostars observed in millimeter continuum and lines \citep{Williams02,Peretto06,Young06,Cunningham16}. 
Gas motions have also been reported from large scales, like the global collapse of NGC~2264-C
\citep{Williams02, Peretto06}, to small scales like protostellar outflows \citep{Maury09, Cunningham16}.

The paper is organized as follows. We introduce in \cref{sec:data} the \emph{Herschel} column density map and clump catalog of the NGC~2264 cloud. In \cref{sec:cloud}, we analyze the cloud structure using a multi-scale decomposition tool and divide the NGC~2264 cloud into three subregions. We then compare in \cref{sec:clump_anal} the 
physical properties of clumps in these subregions and discuss their spatial distribution and mass segregation. In \cref{sec:YSO-clumps}, we compare the YSO and clump populations, in \cref{sec:discu} we discuss mass segregation and propose a revised history for star formation in the NGC~2264 cloud.
Finally, we summarize our conclusions in \cref{sec:conclu}.


\section{The NGC~2264 data set}
\label{sec:data}

\subsection{\emph{Herschel} images, derived column density and temperature maps}
\label{sub:maps}

\begin{figure*}
    \centering
    \includegraphics[scale=0.5]{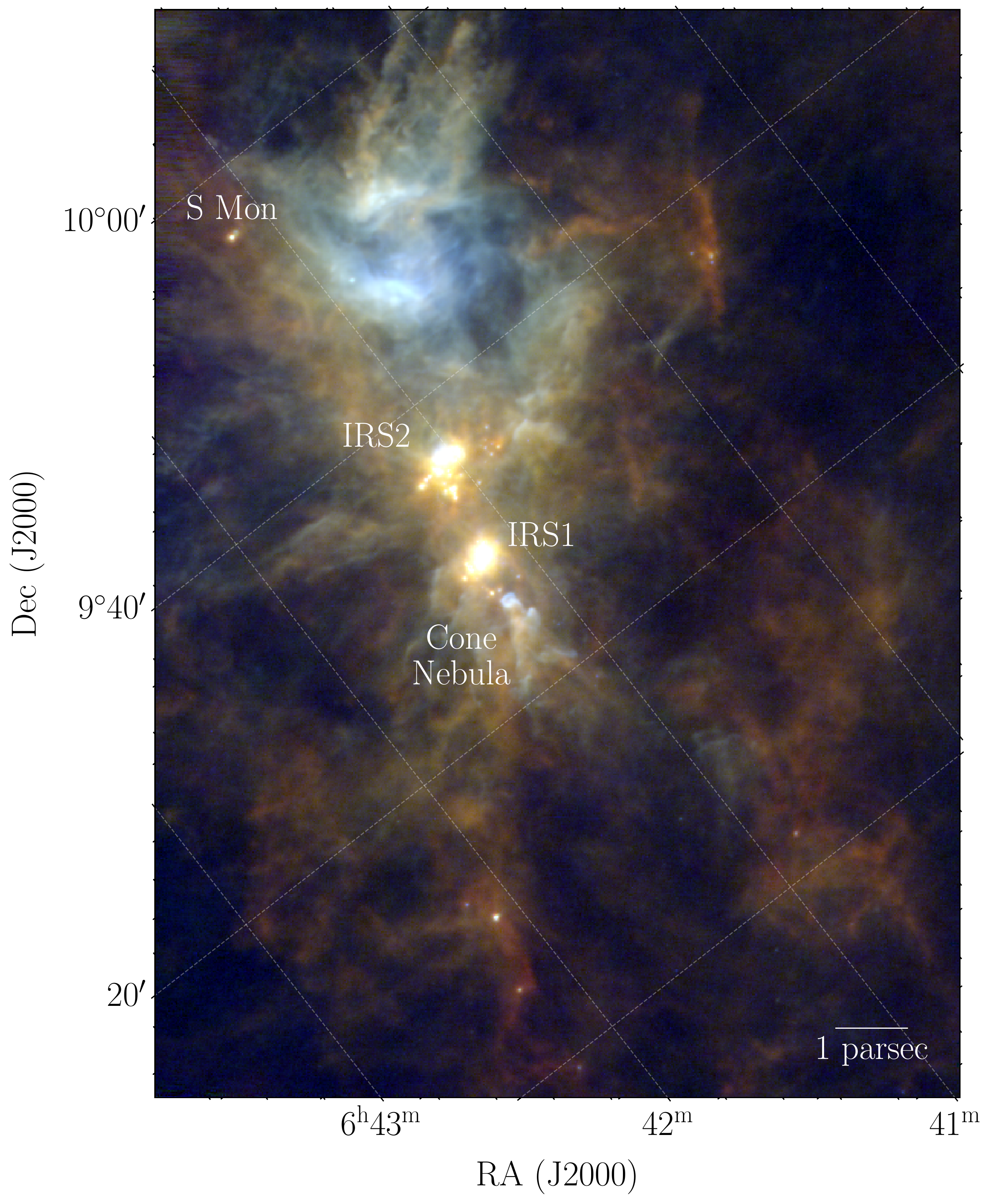}
    \caption{Composite three-color \emph{Herschel} image of NGC~2264: PACS 70~$\mu$m and 160~$\mu$m in blue and green, SPIRE 250~$\mu$m in red. The massive star S~Mon, the central protoclusters NGC2264~IRS1 and IRS2 and the Cone Nebula are labeled. The blue component traces heated regions while earlier stage star-forming sites, such as clumps and filaments, are traced by the red component. The image has been rotated from the RA-Dec grid. 
    }
    \label{fig:3col}
\end{figure*}

\begin{figure*}[h!]
\includegraphics[width=0.94\hsize]{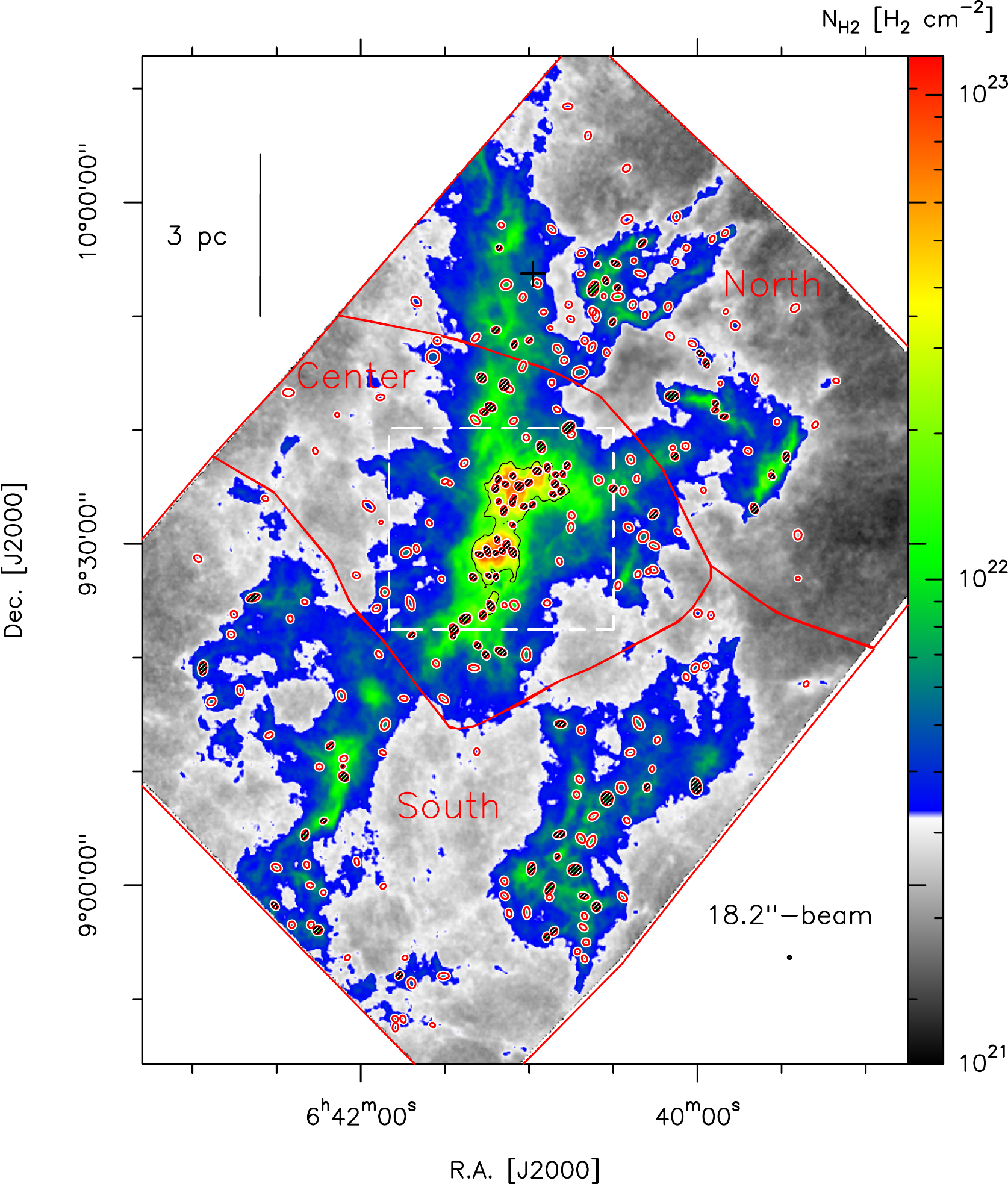}
\caption{\label{fig:map-MDC} 
The NGC~2264 cloud traced by its \emph{Herschel} column density.  
The clumps extracted by \textsl{getsf} are indicated by ellipses, hatched for gravitationally bound clumps, empty otherwise. A contour at $2 \times 10^{22}$~cm$^{-2}$ is drawn to highlight the brightest parts.
The resolution of the map, 18.2\arcsec, is shown in the lower-right corner and a scale bar is given in the upper-left corner.
The outline and labels of the three subregions  defined in \cref{sub:3reg} and used in the analysis are shown in red. The location of the zoom of \cref{fig:map-MDC-z} is shown with a white box.
}
\end{figure*}

\begin{figure*}[]
\centering
\includegraphics[width=0.95\hsize]{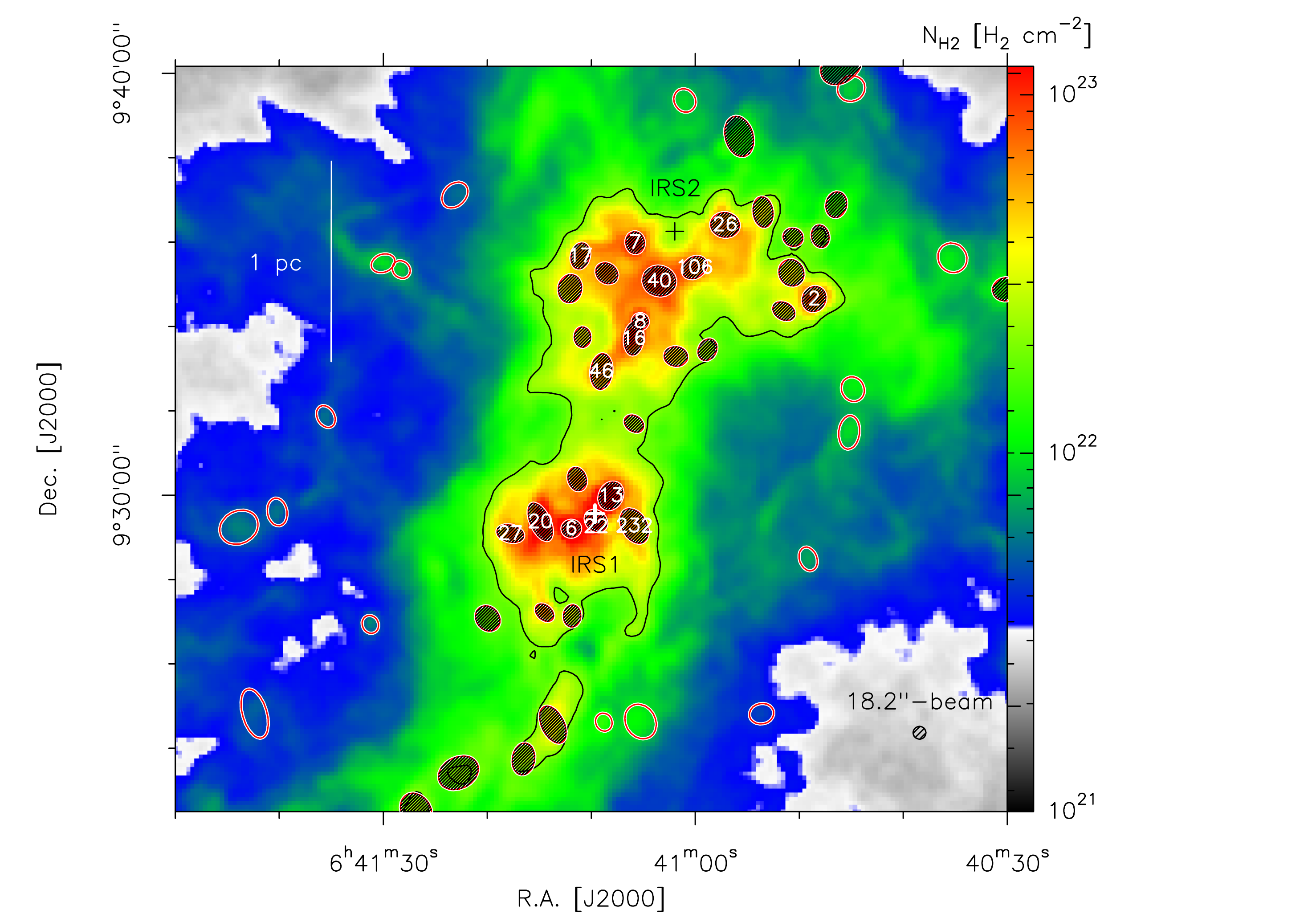}
\caption{\label{fig:map-MDC-z} Zoom toward the central part of NGC~2264, imaged through its \emph{Herschel} column density. Same caption as \cref{fig:map-MDC}. The white and black crosses show the location of the NGC2264~IRS1 and IRS2 sources, respectively. The associated IRS1 and IRS2 protoclusters contain most of the gravitationally bound clumps of this NGC~2264 subregion. The 15 most massive clumps ($M > 9.3\,\Msol$) are pinpointed with their number, 
the 16th is outside this zoom. 
}
\end{figure*}

NGC~2264 was observed by the \emph{Herschel} space observatory with the PACS \citep{Poglitsch10} and SPIRE \citep{Griffin10} instruments as part of the HOBYS \citep{Motte10} Key Program (OBSIDs: 1342205056 and 1342205057). 
Data were taken in five bands: 70 and 160 $\mu$m for PACS and 250, 350, and 500 $\mu$m for SPIRE with FWHM resolutions of 5.9$\arcsec$, 11.7$\arcsec$, 18.2$\arcsec$, 24.9$\arcsec$, and  36.3$\arcsec$, respectively. Observations were performed in parallel mode, using both instruments simultaneously, with a scanning speed of 20$\arcsec$s$^{-1}$. This results in a common PACS and SPIRE area of $1.0\degree \times 1.2\degree$, which corresponds to 10~pc $\times$ 13~pc at a distance of 723~pc.

Data were reduced using the \emph{Herschel} Interactive Processing Environment \citep[HIPE,][]{Ott10} software, version 10.0.2751. 
SPIRE  nominal  and orthogonal  maps  were  separately  processed  and  subsequently combined and reduced for de-stripping, relative gains, and color correction with HIPE. PACS maps were reduced with HIPE up to Level 1 and, from there up to their final version (Level 3), using Scanamorphos v21.0 \citep{Roussel13}.
The composite three-color (RGB = 250~$\mu$m/160~ $\mu$m/70~$\mu$m) \emph{Herschel} image of the NGC~2264 cloud is presented in \cref{fig:3col}.

High-resolution column density and temperature images, at the 18.2$\arcsec$ resolution of the 250~$\mu$m image, were built using the methods presented in \cite{Palmeirin13} and Men’shchikov (in prep.). 
In short, we fitted pixel-by-pixel spectral energy distributions of the \emph{Herschel} images with modified blackbody models to four, three, and two wavelengths (160/250/350/500~$\mu$m, 160/250/350~$\mu$m, and 160/250~$\mu$m) that were convolved to the lowest resolution of the images available in the three sets of wavebands. The higher-resolution information contained in the resulting density images, obtained with the smaller number of wavelengths, was transferred differentially to the 36$\arcsec$ resolution density image derived by fitting four wavebands, thereby increasing its resolution by a factor of two.
We adopted a power law approximation to the dust opacity law per unit of mass at submillimeter wavelengths, $\kappa_\nu = 0.1 \times (\lambda/ \rm 300~\mu m)^{-\beta}$~cm$^2\,$g$^{-1}$, assuming a dust emissivity index $\beta=2$ and a gas-to-dust ratio of 100.
The resulting column density and dust temperature images are presented in Figs.~\ref{fig:map-MDC} and \ref{fig:map-temp}.
The column density map traces cloud structures from $\sim2\times10^{21}$~cm$^{-2}$ to $2.7\times 10^{23}$~cm$^{-2}$. The main cloud structure above $\sim$4$\times 10^{21}$~cm$^{-2}$ displays a "Y" shape and consists of a hub of three filaments
connecting toward a NW-SE ridge.
The highest column densities, $N_{\rm H_2}= 2-30 \times 10^{22}$~cm$^{-2}$, are reached at the center of NGC~2264, toward the two protostellar clusters NGC~2264-IRS1 and IRS2.
Most of the low to medium (column) density gas has a dust temperature of $\sim$14.5~K, decreasing down to 11-12~K within some high-density filaments and increasing up to 20-24~K toward IRS1 and IRS2. In the north, a large bubble of low density, heated gas, with 18-27~K over $\sim$4~pc, is associated with a rosette-shaped nebula observed in infrared and a couple of OB stars including the Herbig Ae/Be star called W90 \citep{Dahm08}.

Since Mon~OB1 lies above the Galactic plane, the contamination of \emph{Herschel} column density and temperature images by other clouds along the same line of sight is minimal (see Schneider et al. in prep. for a complete discussion of contamination effects). 
Absolute uncertainties on $N_{\rm H_2}$, a factor of 2, and $T_{\rm dust}$, a few degrees, depend mostly on the assumption taken for the dust mass opacity.


\subsection{Clump catalog}
\label{sub:catalog}

\begin{table*}[]
\small
     \caption{Sample catalog of clumps identified in NGC~2264 using the \emph{Herschel} column density map. The full table is shown in \cref{tab:clumps-long} and available in electronic form through CDS.}
     \makebox[\textwidth][c]{
    \begin{tabular}{|c|c|c|c|c|c|c|c|c|c|c|c|c|} 
\hline    
\#    &     R.A.  &   Dec.  &    $A \times B$     &     PA        &      Sig    &   $N^{\rm peak}_{\rm H_2}$  &   $N^{\rm int}_{\rm H_2}$  &    FWHM$_{\rm dec}$\tablefootmark{a}       &   $T_{\rm dust}$  &    $ M$   &   $\alpha_{\rm BE}$  &   $N^{\rm back}_{\rm H_2}$\\
          &  [J2000]         &  [J2000]            &       [$\arcsec \times \arcsec$]     &  [$\degree$]  &       &   [$\times10^{21}\,\rm 
          cm^{-2}$] & [$\times10^{21}\,\rm cm^{-2}$ ] & [pc] &   [K]    &  [$\Msol$]&      &  [$\times10^{21}\,\rm cm^{-2}$]\\
\hline    
1  &      100.5263   &   9.1742   &     $18.2 \times 18.2$   &    18.3   &     80.5   &    $18 \pm 2$    &    $18 \pm 2$   &     0.03   &    $20 \pm 2$    &    $1.8 \pm 0.2$    &   0.3   &  $11.5 \pm 0.2$  \\
2  &      100.2023   &   9.5775   &     $36.6 \times 32.4$   &    152.2   &     174.4   &    $80 \pm 6$    &    $286 \pm 10$   &     0.1   &    $11.8 \pm 0.3$    &    $30 \pm 1$    &   0.03   &  $17.4 \pm 0.6$  \\
3  &      100.2944   &   9.9336   &     $27.4 \times 26.2$   &    24.5   &     104.8   &    $19 \pm 2$    &    $41 \pm 2$   &     0.07   &    $15.7 \pm 0.7$    &    $4.3 \pm 0.2$    &   0.21   &  $8.8 \pm 0.2$  \\
4  &      99.8894   &   9.5995   &     $33.6 \times 23.2$   &    44.3   &     94.8   &    $30 \pm 3$    &    $68 \pm 4$   &     0.07   &    $11.5 \pm 0.4$    &    $7.0 \pm 0.4$    &   0.1   &  $13.0 \pm 0.3$  \\
5  &      100.2081   &   9.0423   &     $30.1 \times 23.4$   &    168.7   &     76.2   &    $11.2 \pm 0.8$    &    $21.0 \pm 0.9$   &     0.07   &    $12.3 \pm 0.2$    &    $2.2 \pm 0.1$    &   0.31   &  $6.5 \pm 0.1$  \\
6  &      100.2997   &   9.4868   &     $26.5 \times 23.8$   &    93.5   &     64.4   &    $209 \pm 12$    &    $400 \pm 15$   &     0.06   &    $17.3 \pm 0.7$    &    $41 \pm 2$    &   0.02   &  $61 \pm 1$  \\
7  &      100.2741   &   9.5998   &     $31.3 \times 26.3$   &    7.2   &     56.5   &    $82 \pm 11$    &    $195 \pm 14$   &     0.08   &    $15.0 \pm 0.4$    &    $20 \pm 1$    &   0.05   &  $46 \pm 3$  \\
8  &      100.272   &   9.5686   &     $22.8 \times 22.8$   &    22.6   &     47.7   &    $77 \pm 11$    &    $90 \pm 13$   &     0.05   &    $15.4 \pm 0.7$    &    $9 \pm 1$    &   0.07   &  $52 \pm 2$  \\
9  &      100.3104   &   9.4536   &     $29.5 \times 20.2$   &    48.4   &     50.1   &    $44 \pm 8$    &    $77 \pm 9$   &     0.06   &    $13.3 \pm 0.4$    &    $8.0 \pm 0.9$    &   0.08   &  $22.6 \pm 0.6$  \\
10  &      100.5549   &   9.0945   &     $31.8 \times 22.5$   &    115.5   &     42.3   &    $11 \pm 2$    &    $22 \pm 3$   &     0.07   &    $12.6 \pm 0.4$    &    $2.3 \pm 0.3$    &   0.31   &  $9.7 \pm 0.5$  \\
11  &      99.9603   &   9.6866   &     $40.8 \times 24.9$   &    93.2   &     45.8   &    $13 \pm 1$    &    $35 \pm 1$   &     0.09   &    $12.2 \pm 0.2$    &    $3.6 \pm 0.1$    &   0.26   &  $6.4 \pm 0.2$  \\
12  &      99.9953   &   9.779   &     $38.2 \times 25.4$   &    56.7   &     46.9   &    $5.9 \pm 0.2$    &    $17.0 \pm 0.3$   &     0.09   &    $13.5 \pm 0.3$    &    $1.7 \pm 0.0$    &   0.58   &  $2.9 \pm 0.0$  \\
13  &      100.2838   &   9.4998   &     $41.0 \times 32.3$   &    153.3   &     48.6   &    $138 \pm 13$    &    $514 \pm 21$   &     0.11   &    $16 \pm 2$    &    $53 \pm 2$    &   0.03   &  $41 \pm 3$  \\
14  &      100.25   &   9.7985   &     $31.9 \times 27.8$   &    129.0   &     40.5   &    $5.9 \pm 0.9$    &    $15 \pm 1$   &     0.08   &    $15.1 \pm 0.3$    &    $1.6 \pm 0.1$    &   0.65   &  $7.1 \pm 0.2$  \\
15  &      100.224   &   8.9245   &     $38.6 \times 28.0$   &    174.2   &     44.6   &    $5.3 \pm 0.4$    &    $16.0 \pm 0.6$   &     0.1   &    $13.0 \pm 0.2$    &    $1.7 \pm 0.1$    &   0.61   &  $3.9 \pm 0.1$  \\
16  &      100.2746   &   9.5619   &     $46.7 \times 26.6$   &    174.2   &     32.2   &    $67 \pm 11$    &    $204 \pm 15$   &     0.11   &    $13.7 \pm 0.7$    &    $21 \pm 2$    &   0.06   &  $45 \pm 3$  \\
17  &      100.2961   &   9.5946   &     $37.4 \times 25.9$   &    166.2   &     31.2   &    $46 \pm 13$    &    $100 \pm 13$   &     0.09   &    $14.0 \pm 0.9$    &    $10 \pm 1$    &   0.1   &  $35 \pm 3$  \\
18  &      100.1488   &   9.9097   &     $23.8 \times 22.8$   &    113.5   &     31.8   &    $5 \pm 1$    &    $9 \pm 2$   &     0.05   &    $15.8 \pm 0.4$    &    $0.9 \pm 0.2$    &   0.74   &  $5.2 \pm 0.3$  \\
19  &      100.1259   &   9.8254   &     $37.1 \times 24.9$   &    155.3   &     39.4   &    $6 \pm 1$    &    $16 \pm 2$   &     0.09   &    $16.5 \pm 0.3$    &    $1.7 \pm 0.2$    &   0.69   &  $5.9 \pm 0.2$  \\
20  &      100.3121   &   9.4895   &     $57.4 \times 27.1$   &    25.2   &     36.5   &    $105 \pm 13$    &    $382 \pm 19$   &     0.12   &    $14.3 \pm 0.4$    &    $40 \pm 2$    &   0.04   &  $55 \pm 3$  \\
    \hline
    \end{tabular} 
    }\\
    \tablefootmark{a}{source deconvolved size at FWHM, $FWHM_{\rm dec}= (A\times B - 18.2^2)^{1/2}$, minimized at half the beam size} \\
    \label{tab:MDCs}
\end{table*}

Compact sources were extracted from the column density map at 18.2$\arcsec$ resolution (see \cref{sub:maps}) using \textsl{getsf} (v200124, Men'shchikov in prep.), 
a new multi-scale, multi-wavelength source and filament extraction algorithm based on the separation of structural components. The method is a major improvement over its predecessors \textsl{getsources}, \textsl{getfilaments}, and \textsl{getimages} \citep{Mensh12, Mensh13, Mensh17}. Original images are spatially decomposed into a set of $99$ single-scale images, from two pixels to a maximum scale of 4 times the maximum size of sources of interest, 5.6$\arcsec$ to 232$\arcsec$ (or 0.8~pc) for the \object{NGC~2264} HOBYS field. 
The decomposed images are processed and analyzed by \textsl{getsf} to separate three structural components -- sources, filaments, and background. The method flattens the sources and filament components to equalize their noise and background fluctuations, in order to reliably detect sources and filaments using intensity (here, column density) thresholds. 

Detection of sources and filaments is done in the single-scale flattened images, whereas their measurements are done in the background-subtracted original image. 
The resulting catalog contains, for each extracted source, its peak and integrated column densities, $N_{\rm H_2}^{\rm peak}$ and $N_{\rm H_2}^{\rm int}$, with uncertainties, its major and minor axis sizes at FWHM, $A$ and $B$, and the position angle of its elliptical footprint (PA).
Also given is the source detection significance, $S$, which is a single-scale analog of the signal-to-noise ratio and the column density of the subtracted background, $N_{\rm H_2}^{\rm back}$. To measure the latter we averaged the background map provided by \emph{getsf} over the source {\it FWHM}.
 A subset of this catalog is given in \cref{tab:MDCs} and the complete table can be found in \cref{tab:clumps-long}.

 We also list in \cref{tab:MDCs} the source FWHM sizes deconvolved from the beam, $FWHM_{\rm dec}$, which provide an estimation of the physical sizes and range from 0.03~pc to 0.2~pc with a median value of 0.11~pc.
The source average dust temperatures, $T_{\rm dust}$, are listed with their associated uncertainties. They are defined as the mean and dispersion, respectively,  measured over the source FWHM areas on the temperature map of Fig.~\ref{fig:map-temp}.
The measured temperature are averaged along the line of sight of each clump and generally overestimate by a few degrees the mean dust temperature of cold, shielded clumps \citep{Hill12}.
$T_{\rm dust}$ ranges from 11.5~K to 22.7~K with a median of 13.9~K. The temperature distribution in \cref{fig:temp-dist} show little variations,  as 85\% of the clumps have a temperature between 12 and 16 K. 

A total of 256 sources has been extracted in NGC~2264 using \textsl{getsf}, they are plotted over the column density map in \cref{fig:map-MDC,fig:map-MDC-z}. 
The mass of the clumps is calculated from the integrated column density given by \emph{getsf}, $N_{\rm H_2}^{\rm int}$ (see \cref{tab:MDCs}), following the equation:
\begin{eqnarray}
   M & = & \mu_{\rm H_2}\, m_{\rm H}\, d^2 \,\Omega_{\rm b}\, N_{\rm H_2}^{\rm int}\\
  & = & 0.10~\Msol\, \left(\frac{\Theta_{\rm HPBW}}{18.2\arcsec}\right)^2\,\left(\frac{N_{\rm H_2}^{\rm int}}{10^{21}\,\rm H_2\,cm^{-2}}\right),\nonumber
\end{eqnarray}
with $\mu_{\rm H_2}=2.8$ the molecular weight per hydrogen molecule, $m_{\rm H}$ the hydrogen mass, $d=723$~pc the distance to NGC~2264 and $\Omega_{\rm b}$ the beam solid angle. Masses range from 0.08~$\Msol$ to 53~$\Msol$ with a median of 0.9~$\Msol$. 
The relative uncertainty of clump masses is estimated to be less than 50\% but the absolute uncertainty could be of a factor of 2.

\subsection{YSO catalog}
\label{sub:YSOcat}

YSOs are selected from the 2MASS/\emph{Spitzer} catalog established by \cite{Rapson14} on the eastern part of the Mon~OB1 molecular complex. The \emph{Spitzer} map used by \cite{Rapson14} covers most of the \emph{Herschel} field, but small areas in the south, east and west, within which 10 clumps have been detected.
Other YSO catalogs built over smaller areas \citep[like one of, e.g.,][]{Kuhn14} are not used here to keep the homogeneity of the YSO catalog.
\cite{Rapson14} use \emph{Spitzer} IRAC (3.6-8$\micro$m) and MIPS 24$\micro$m data to select YSOs from their IR excess. They are thus sensitive to class 0/I sources while \cite{Cody14} are not (since they use only IRAC data). Moreover, as opposed to \cite{Venuti18} that present a spectroscopic study based on Gaia-ESO-Survey data, their spatial coverage is complete which is important for our study.

Among the 6381 YSOs within the field of view of the \emph{Herschel} column density map, there are 87 Class~0/I protostars and 398 Class~II pre-main sequence stars. Class~III sources have been excluded from the studied sample because they are evolved YSOs, less relevant for comparison with the gas distribution. Besides, the Class~III population in the catalog of \cite{Rapson14} is significantly contaminated with field stars. 
The classification of YSOs was carried out by \cite{Rapson14} using a color-based method developed by \cite{Gutermuth09}. 
Our sample also does not include the $\sim$20 intermediate- and high-mass sources that started forming a few $10^6$~years ago, i.e. at the same time as the Class~IIs of \citet{Rapson14}.

\section{Analysis of the cloud structure in NGC~2264}
\label{sec:cloud}

Based on the YSO spatial distribution, the NGC~2264 star-forming region separates into two main subregions \citep{Sung08,Sung09}: the S~Mon area 
and the region which encloses the two clusters associated with IRS1 and IRS2 (see \cref{sec:intro}).
In line with these studies, the three-color, temperature and column-density \emph{Herschel} images of Figs.~\ref{fig:3col}-\ref{fig:map-MDC} and \ref{fig:map-temp} argue to separate the intermediate-column density, hotter northern subregion from the high-column density central subregion hosting the IRS1 and IRS2 protoclusters.
The remaining part of the NGC~2264 cloud, which has intermediate column density and colder temperatures, is then labeled the southern subregion.

\subsection{Splitting the NGC~2264 cloud into 3 subregions}
\label{sub:3reg}

\begin{table*}

\small
\caption{Global characteristics and cloud structure of NGC~2264 and its three subregions and fit values for the total wavelet power spectrum of NGC 2264 and for the Gaussian and coherent power spectra of the northern, southern and central regions. 
}
\label{tab:pow_spec_fit} 
\begin{tabular}{c|clcl|lcclc}
\hline \hline
Region & Area & $\overline{N_{\rm H2}}$\tablefootmark{\,a} & $N_{\rm H2}^{\rm back}$\tablefootmark{\,b} & $\overline{T_{\rm dust}}$\tablefootmark{\,a} & MnGSeg & $\gamma$ & $A$ & $P_0$ & Noise \\
name & [\degree$^2$], [pc$^2$] & \multicolumn{2}{c}{[$\times 10^{21}$~cm$^{-2}$]} & [K] & components & powerlaw & $[$ & $\times 10^{42}~$(H$_2$ cm$^{-2}$)$^2$ & $]$\\
\hline
NGC~2264 & 1.16, $\sim$185 & 3.9 & 4.7 & 14.7 & all & $-2.3\pm0.1$ & $35\pm7$ & $ - $ & $(5.0\pm0.5) \times 10^{-3}$ \\
Northern & 0.38, $\sim$61 & 2.9 & 4.0 & 15.4 & Gaussian &  $-4.8\pm0.1$ & $0.06\pm0.01$ & $0.10\pm0.02$ & $(3.3\pm0.5) \times 10^{-3}$\\
& & & & & Coherent & $-2.31\pm0.09$ & $9.0\pm0.5$ & $-$ & $(3.3\pm0.5) \times 10^{-3}$\\
Central & 0.26, $\sim$41 & 6.6 & 7.3 & 14.9 & Gaussian & $-3.6\pm0.1$& $0.08\pm0.02$ & $0.09\pm0.02$ & $(3.3\pm0.5) \times 10^{-3}$\\
& & & & & Coherent & $-2.3\pm0.1$ & $299.0\pm0.6$ & $-$ &$(3.3\pm0.5) \times 10^{-3}$\\
Southern & 0.52, $\sim$83 & 3.3 & 4.3 & 14.1 & Gaussian & $-4.02\pm0.07$ & $0.09\pm0.01$ & $0.13\pm0.01$ & $(3.3\pm0.5) \times 10^{-3}$\\
& & & & & Coherent & $-2.43\pm0.09$ & $3.4\pm0.5$ & $-$ &$(3.3\pm0.5) \times 10^{-3}$\\
\hline
\end{tabular}
\\
\tablefootmark{a}{Typical column density and temperature, calculated as the average of the map in this area} \\
\tablefootmark{b}{Typical background column density, calculated as the median background of clumps located in this area 
}
\end{table*}

In order to refine the outlines of the three NGC~2264 subregions, we performed a multiscale analysis of the NGC~2264 cloud.
We used complex wavelet transforms described by \citet{Robitaille14, Robitaille19} to first calculate the power spectrum of the complete NGC~2264 region. While corresponding very well to the classic Fourier power spectrum, this method
goes further by making it possible to visualize the spatial distribution of density fluctuations for any scale of the power spectrum. The decomposition is done with the directional and complex Morlet wavelet defined in the Fourier space as
\begin{equation}
\begin{split}
\hat{\psi}(\vec{k}) & =\textrm{e}^{-|\vec{k}-\vec{k}_0|^2/2} \\
                                     & =\textrm{e}^{-[(u-|\vec{k}_0|\cos \theta)^2+(v-|\vec{k}_0|\sin \theta)^2]/2},
\end{split}
\label{eq:Morlet_Fourier}
\end{equation}
\noindent where $\vec{k} = [u,v]$ is the wavenumber, $\theta$ is the wavelet azimuthal direction and the constant $|\vec{k}_0|$ is set to $\pi\sqrt{2/\ln2} \approx 5.336$ to ensure that the admissibility condition is almost met \citep{2005CG.....31..846K}. From this decomposition, the amount of power for the density fluctuations as a function of spatial scales is calculated following the relation
\begin{equation}
P^W(\ell)=\left\langle \left\langle |\tilde{f}(\ell,\vec{x},\theta)|^2\right\rangle_{\theta} \right\rangle_{\vec{x}},
\label{eq:wavelet_powspec}
\end{equation}
\noindent where $\langle\rangle$ represents the averaging operation. The spatial scale $\ell$ is then converted to the Fourier wavenumber $k$ following the relation $k=|\vec{k}_0|/\ell$.

\begin{figure}
    \centering
    \includegraphics[width=.48\textwidth]{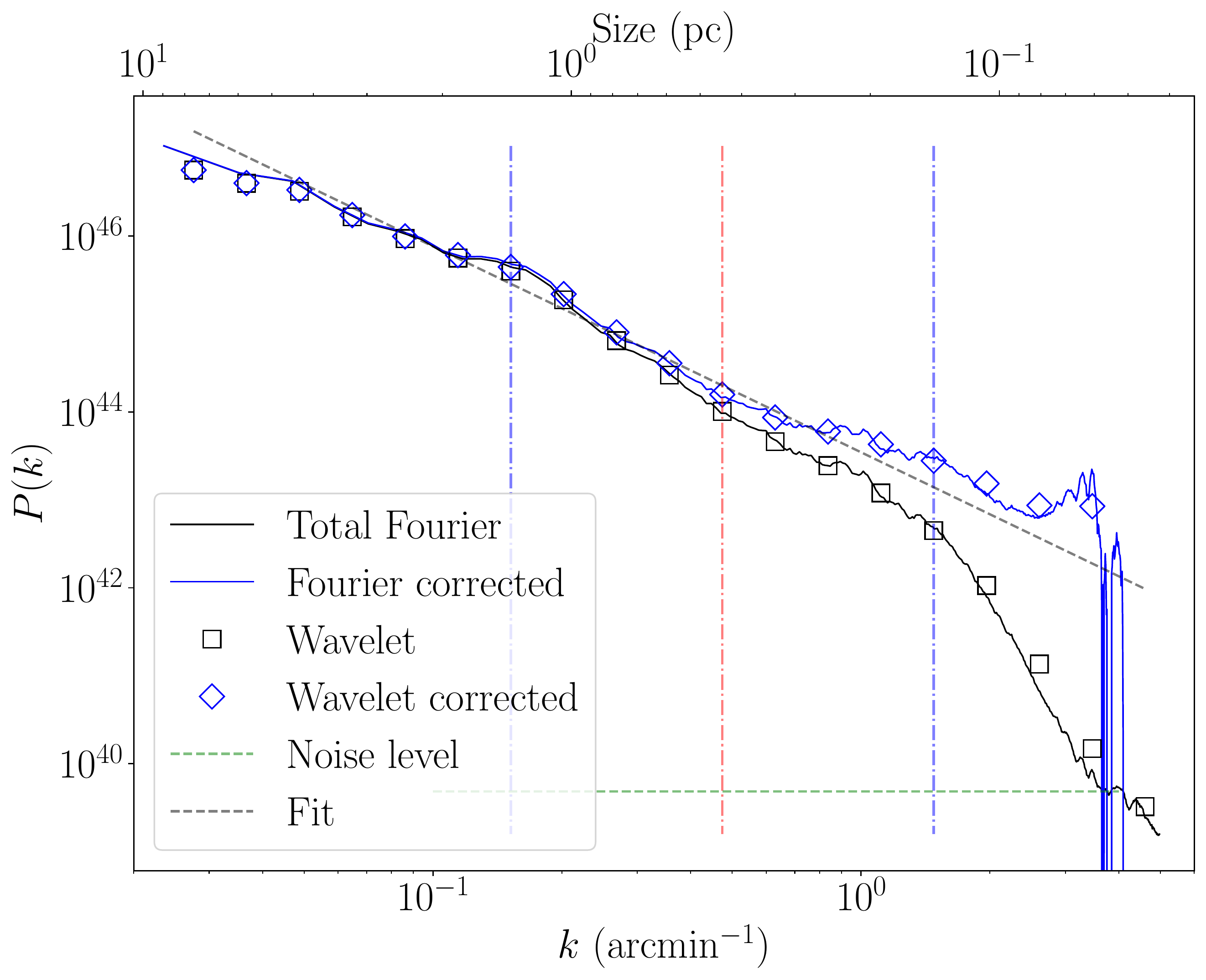}
    \caption{The Fourier (solid lines) and wavelet (symbols) power spectra of the NGC~2264 region.
    Before fitting a $P(k) = A\,k^{\gamma}$ relation (dashed curve), the corrected power spectra are subtracted by the noise level (plateau at the end of the original power spectrum, green dotted horizontal line) and divided by the empirical SPIRE 250~$\mu$m beam. The power spectra present two bumps, located at 1.4~pc and 0.1 pc (vertical blue dot-dashed lines). The intermediate scale chosen to separate the three NGC~2264 subregions is indicated by a red dot-dashed vertical line.
    }
    \label{fig:powspec_tot}
\end{figure}

\begin{figure*}
    \centering
    \includegraphics[width=1.0\textwidth]{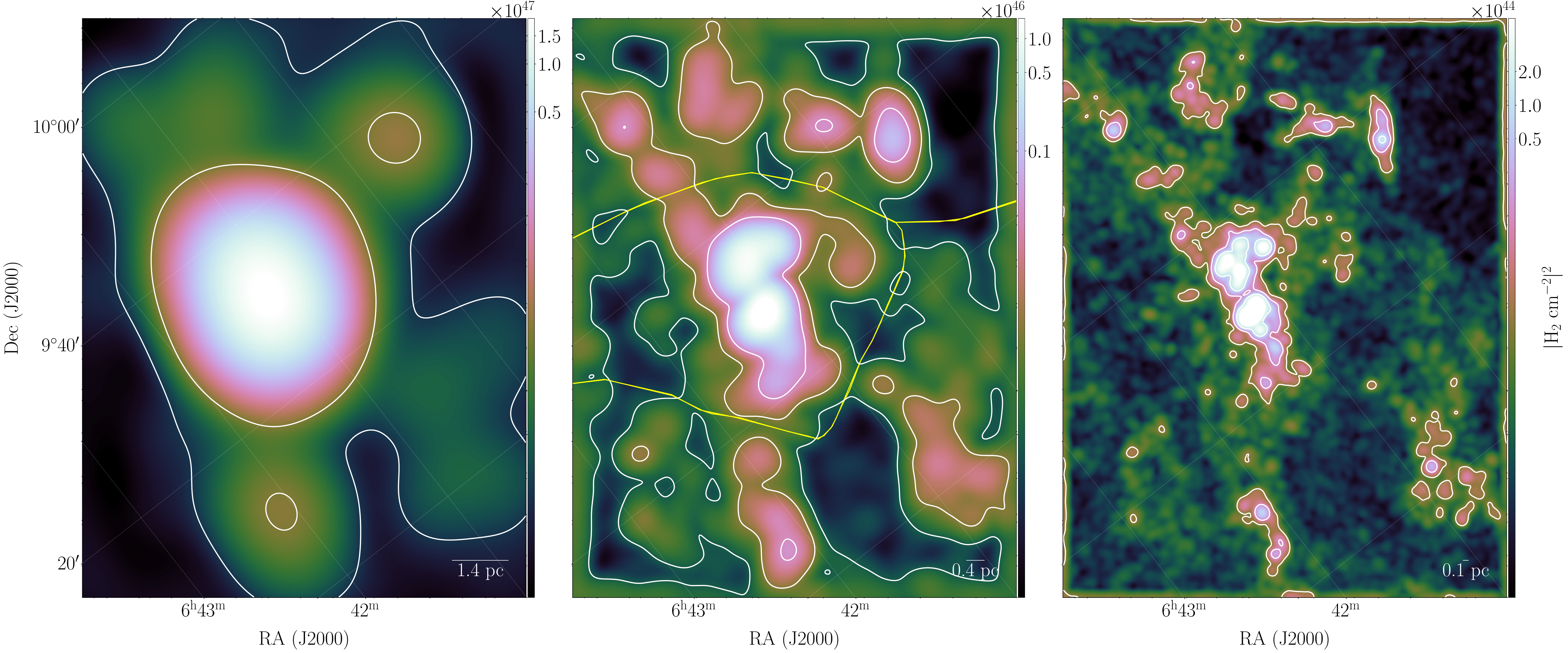}
    \caption{
    Power density fluctuations of NGC~2264 measured by MnGSeg and averaged over $\theta$
    for three scales: 0.15~arcmin$^{-1}$ (or 1.4~pc, left panel), 0.47~arcmin$^{-1}$ (or 0.4~pc, middle panel) and 1.5~arcmin$^{-1}$ (or 0.1~pc, right panel). 
    Images have been rotated from the RA-Dec grid (gray lines) to facilitate the wavelet decomposition.
    The yellow lines in the middle panel separate the three subregions of NGC~2264.
    }
    \label{fig:spatial_scales}
\end{figure*}

\Cref{fig:powspec_tot} shows the wavelet power spectrum of NGC~2264, corrected for the noise level and the $18.2\arcsec$ beam.  It can be fitted by a power law relation $P(k) = A\,k^{\gamma}$ (see fit parameters in \cref{tab:pow_spec_fit}), except for excess of power located at $\sim0.15$~arcmin$^{-1}$ and $\sim1.5$~arcmin$^{-1}$corresponding respectively to scales of $\sim1.4$~pc and $\sim0.1$~pc. \Cref{fig:spatial_scales} shows the spatial distribution of the power density fluctuations averaged over $\theta$, $|\tilde{f}(\ell,\vec{x})|^2$, for these two scales as well as the $0.47$ arcmin$^{-1}$ (or 0.4~pc) scale located between the two power excess. The first excess at 1.4~pc corresponds to the large-scale mass reservoir associated with the central part of the NGC~2264 cloud. It dominates in terms of density fluctuation at this spatial scale (see \cref{fig:spatial_scales}, Left panel) like do massive hubs or ridges in, e.g., the Cygnus~X, Vela~C, and NGC~6334 cloud complexes \citep{Schneider10, Hill11, Tige17}. This excess is also measured on NGC~2264 using the $\Delta$-variance method (Schneider et al. in prep.). The 0.1~pc-scale image (see \cref{fig:spatial_scales}, Right panel) displays smaller scale structures, associated with clumps.

The intermediate scale of 0.4~pc, which separates the scales associated with the mass reservoir of the central ridge and the mass reservoir of small-scale clumps, was chosen to refine the division of the NGC~2264 cloud into three subregions.
The outline of the northern, central, and southern subregions are set to go through saddle points (see \cref{fig:spatial_scales}, middle panel). 
The global characteristics of these subregions observed on the \emph{Herschel} column density and temperature images are listed in \cref{tab:pow_spec_fit}.


\subsection{Segmentation of the coherent and Gaussian components of the cloud structure}
\label{sub:MnGSeg}

We applied the Multiscale non-Gaussian Segmentation (MnGSeg) technique in order to investigate the hierarchical properties of the cloud structure in the three subregions of NGC~2264. This technique allows to separate the random spatial density fluctuations of a cloud from the dense, coherent structures, usually associated with star formation activities. 
While we expect the Gaussian component of a cloud to only consist of evanescent structures of the interstellar medium, its coherent component contains cores, clumps, hubs, ridges and other filamentary structures which correspond to the gas mass reservoir of present star formation. 
The non-Gaussian segmentation is done on wavelet coefficients as a function of orientations $\theta$ and scales $\ell$ (see \citealt{Robitaille19} for more details on the technique). The amount of power for the density fluctuations as a function of spatial scales, $|\tilde{f}(\ell,\vec{x})|^2$, can be calculated according to the spatial coverage of areas of interest following the relation
\begin{equation}
P^W(l)=\frac{1}{N_{\mathbb{L}}}\sum_{\vec{x}} \left\langle \mathbb{L}( |\tilde{f}(\ell,\vec{x},\theta) |)|\tilde{f}(\ell,\vec{x},\theta)|^2\right\rangle_{\theta},
\label{eq:conditinal_wavelet_powspec}
\end{equation}
where
\begin{equation}
\mathbb{L}(|\tilde{f}_{l,\theta}(\vec{x})|)=\left\{
        									          \begin{array}{ll}
  									          1 & \textrm{if $\vec{x}$ is inside the area of interest} \\
 									          0 & \textrm{if $\vec{x}$ is outside the area of interest}, \\
 									         \end{array}
  									         \right.
\end{equation}
and
\begin{equation}
    N_{\mathbb{L}} = \sum_{\vec{x}} \mathbb{L}(|\tilde{f}_{l,\theta}(\vec{x})|).
\end{equation}

\begin{figure*}
    \centering
    \includegraphics[width=0.45\hsize]{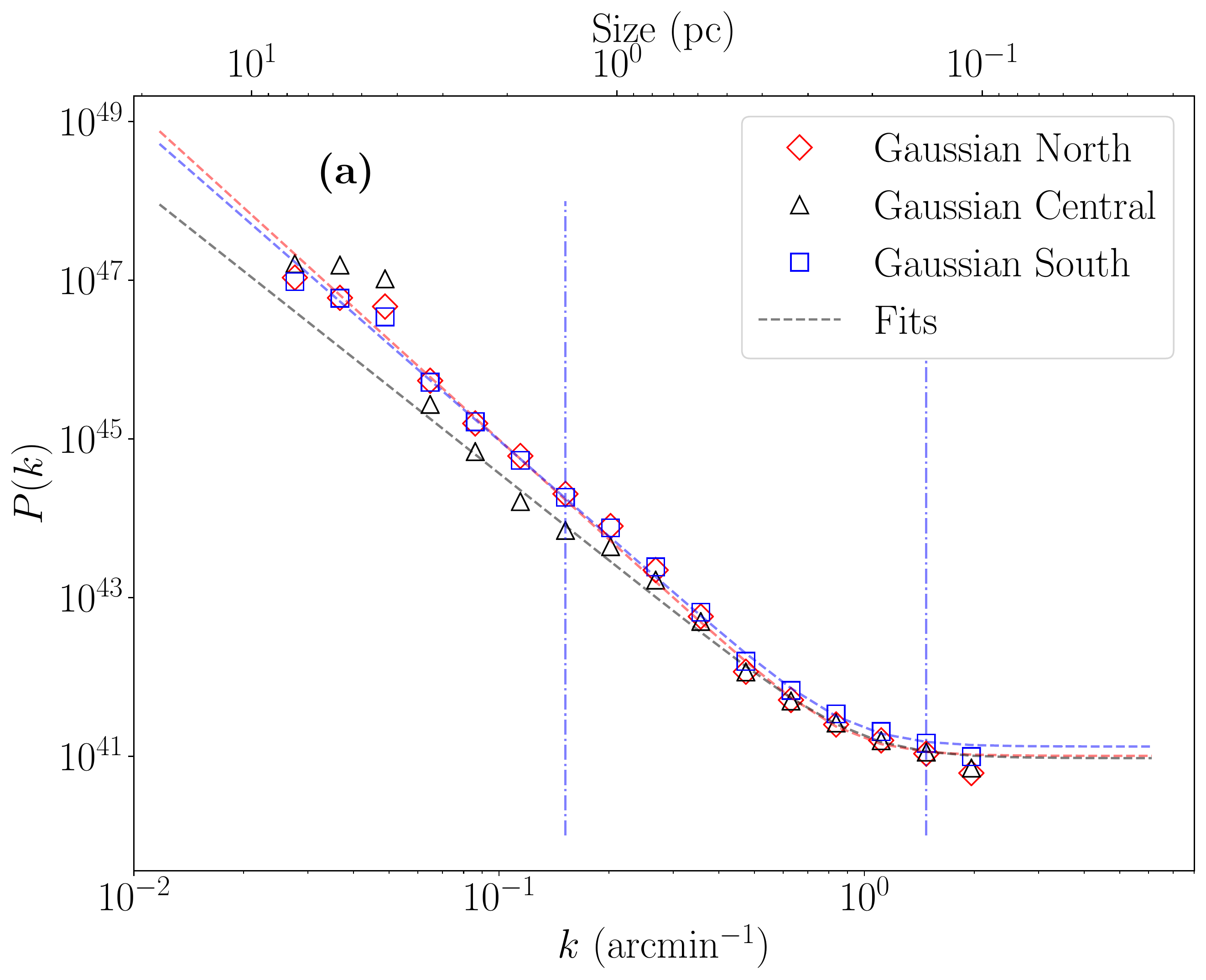} \hskip 0.5cm
    \includegraphics[width=0.45\hsize]{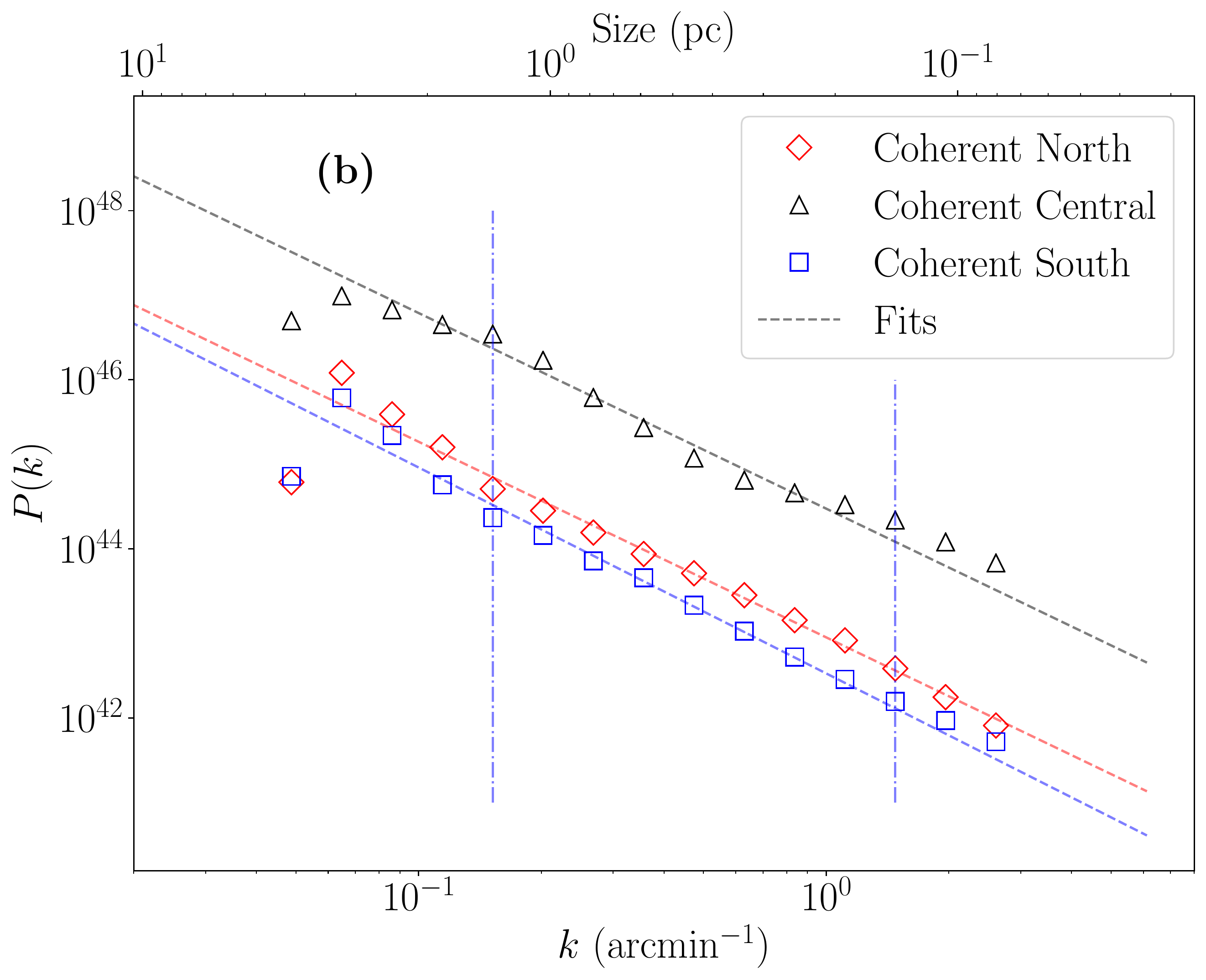}
    \caption{The random Gaussian (in \textbf{a}) and coherent (in \textbf{b}) wavelet power spectra calculated by MnGseg for the three NGC~2264 subregions. The black and red dotted curve represents the fitted power law relations to the northern and central subregions. The coherent power spectrum of the central subregion stands out with $\sim$100 times more power and bumps located at 1.4~pc and 0.1~pc (vertical blue dashed lines).}
    \label{fig:gc-powspec}
\end{figure*}

\Cref{fig:gc-powspec} presents the Gaussian (random) and coherent wavelet power spectra for the three subregions of NGC~2264 and \cref{tab:pow_spec_fit} lists the parameters of their fitted relation $P(k) = A\,k^{\gamma} + P_{0}$ for the Gaussian component and $P(k) = A\,k^{\gamma}$ for the coherent part.
For all subregions, the coherent component dominates in terms of power over the Gaussian components. Moreover, the central subregion presents a Gaussian component which resembles, in terms of both power level and power law index, those fitted for the northern and southern subregions. 
In contrast, the coherent component of the central subregion possesses a power level several hundred times higher than that of the northern and southern subregions. 
The coherent spectrum of the central subregion also displays more irregularities compared to the coherent spectra of the northern and southern parts (see  \cref{fig:gc-powspec}b). We 
notably recover the two power excess observed in \cref{fig:powspec_tot}. As shown by \citet{Robitaille19}, the small-scale spectra flattening modelled by the variable $P_0$ for the Gaussian components only is associated to the cosmic infrared background signal.

 Recently \cite{Robitaille20} discussed and proved, using statistical models, that the non-Gaussian segmentation performed by MnGSeg can also be interpreted as the separation of a component of multifractal nature, the coherent part, from a monofractal component, the random Gaussian part. The multifractal nature of the coherent component means that its hierarchical structures are defined by a collection of power law indexes rather than by a single one, as it is the case for the monofractal geometry of the Gaussian component.
 These different fractal properties explain why the power spectra of the coherent components in \cref{fig:gc-powspec}b are more diverse and complex than those of the Gaussian components shown in \cref{fig:gc-powspec}a. 
 It also suggests that the coherent component of a cloud (see \cref{fig:gc-powspec}b and \cref{tab:pow_spec_fit})
is the one that could reveal the complexity level of cloud structures, which increases 
 with the creation of large gravity potentials such as hubs and ridges.
 A more detailed analysis of the multifractal properties of the NGC~2264 cloud, its subregions, and clumps' local environment will be done by Robitaille et al. (in prep.).

\section{Analysis of the population of clumps in NGC~2264}
\label{sec:clump_anal}

After estimating the completeness level of the catalogs of clumps identified in the three subregions of NGC~2264 (see \cref{sub:compl}), we characterize the main physical properties of clumps (see \cref{sub:clumps_prop}) and evaluate their boundedness (see \cref{sub:clump_vir}). We then investigate the spatial distribution of clumps (see \cref{sub:clump_spacial}) and their mass segregation (see \cref{sub:mass_seg}).
\Cref{tab:statMDC} lists the main statistical parameters of the NGC~2264 cloud and its three subregions.

\subsection{Completeness of the clump catalogs}
\label{sub:compl}

To accurately compare the clump's properties between subregions, we estimated the completeness level and how it varies across the map. For this purpose, we injected a synthetic population of 1170 sources over the background image of NGC~2264, which is produced by getsf during the clump extraction process (see \cref{sub:catalog}).
Synthetic sources were split into 9 bins logarithmically spaced between 0.25~$\Msol$ and 4.0~$\Msol$, with a constant number of 130 sources per bin. Their chosen density profile is that of Bonnor-Ebert spheres with a central density increasing with the clump mass, in agreement with the profile observed for clumps. These synthetic clumps, once convolved by our 18.2$\arcsec$ angular resolution, display FWHM sizes of about 0.12~pc (or 35$\arcsec$) corresponding to the median size of extracted sources. 
We ran the extraction algorithm \textsl{getsf} on the synthetic image with the same parameters as for the observations. In total, 695 out of 1170 (59\%) sources have been recovered.

Figures~\ref{fig:comp}a-b show the detection rate of synthetic sources injected on the NGC~2264 background image.
From the detection rate versus mass curve of \cref{fig:comp}a, we estimated a global 90\% completeness level of $\sim$1.7~$\Msol$. 74 (29\%) clumps of \cref{tab:MDCs} lie above this completeness level.

The completeness level of any source extraction procedure does however depend on the intensity of the source background. We therefore computed, for each identified sources, its contrast, defined as $C=N_{\rm H_2}^{\rm peak}/N_{\rm H_2}^{\rm back}$ and plot the detection rate against source contrast in \cref{fig:comp}b. A minimum contrast of 0.4 is required to detect a source, while all sources with $C > 1.7$ are detected. 
The 90\% completeness level is reached for a source contrast of $C=1.1$, that is for clumps with $N_{\rm H_2}^{\rm peak}=1.1 \times N_{\rm H_2}^{\rm back}$. 
As shown in \cref{fig:map-MDC} and \cref{tab:MDCs,tab:pow_spec_fit}, the background level varies strongly from the northern or southern subregions and central part of NGC~2264. 
We therefore used the median value of the clumps' background to characterize each subregion: $\overline{N_{\rm H_2}^{\rm back}}=4\times 10^{21}$, $4.3\times 10^{21}$ and $7.3\times 10^{21}$~cm$^{-2}$ in the northern, southern and central subregions, respectively. 
Using the relation of mass versus peak column density found for both synthetic and observed clumps ($M\propto (N_{\rm H2}^{\rm peak})^{0.95}$, see \cref{fig:Mpic}a), we computed the associated mass threshold for clumps. 
We obtained a 90\% completeness level of $\sim$1.5~$\Msol$ for both the northern and southern subregions and
a completeness level about two times larger, 2.7~$\Msol$, for the central subregion of NGC~2264.
As a consequence, some low mass clumps in the center may be overlooked by the extraction algorithm. The number of clumps with masses above these 90\% completeness levels are 19, 22, and 33 in the northern, southern, and central subregions, respectively. 

\subsection{Main physical properties of clumps}
\label{sub:clumps_prop}

\begin{table*}[]
 \caption{
    Distribution of clumps and YSOs in NGC~2264 and its subregions.
    }
        \label{tab:statMDC}
\begin{tabular}{c|c c c c c c| c c c}
    \hline
    \hline
      Region & \multicolumn{6}{c}{Clumps} &  \multicolumn{3}{c}{YSOs}    \\ 
         name & Total\tablefootmark{a,b} & Bound\tablefootmark{b,c} & Unbound\tablefootmark{b,c} & 
         $\overline{\Sigma_{\rm clump}}$ & $Q$\tablefootmark{d}  & $\Lambda_{\rm MSR}^{\rm med}$\tablefootmark{d} & Total\tablefootmark{e} & Class~0/I & Class~II \\
       \hline
        NGC~2264 & 256/100\% & 92/36\% & 164 & $\sim$1.4 pc$^{-2}$ & $\simeq0.7$ & 7.8  & 485/100\% & 87 & 398 \\ 
        Northern  & 74/29\% & 21/28\% & 53 & $\sim$1.2 pc$^{-2}$ & $\simeq0.8$ & $\simeq$1 & 156/32\% & 10 & 146 \\
        Central & 97(96)/38\% & 48/50\% & 49(48) & $\sim$2.4 pc$^{-2}$ & $\simeq0.8$ & 3.7 & 302/62\% & 69 & 233 \\
        Southern  & 85(76)/33\% & 23(22)/27\% & 62(54) & $\sim$0.9 pc$^{-2}$ & $\simeq0.6$ & $\simeq$1 & 27/6\% & 8 & 19 \\

    \hline
    \end{tabular} \\
     \tablefootmark{a}{Number of clumps in each subregions and their corresponding fraction of the total population.}\\
    %
    \tablefootmark{b}{The number of clumps in the area where YSOs have been detected is given in parentheses.}\\
    \tablefootmark{c}{Gravitationally bound clumps have a $\alpha_{\rm BE} < 2$ parameter in \cref{tab:MDCs}. The fraction of bound clumps in the subregion population is indicated.}\\
    \tablefootmark{d}{The clustering $Q$ parameter is calculated from \cref{eq:Q-param}. The median mass segregation ratio, $\Lambda_{\rm MSR}^{\rm med}$, measured for the $N_{\rm MST}$ = 4 to 15 most massive clumps, are estimated from \cref{fig:lambda}.}\\
    \tablefootmark{e}{Total number of Class~0/I and Class~II YSOs in each subregions and their corresponding fraction of the total population.}
\end{table*}

\begin{figure*}
    \centering
    \includegraphics[width=0.47\hsize]{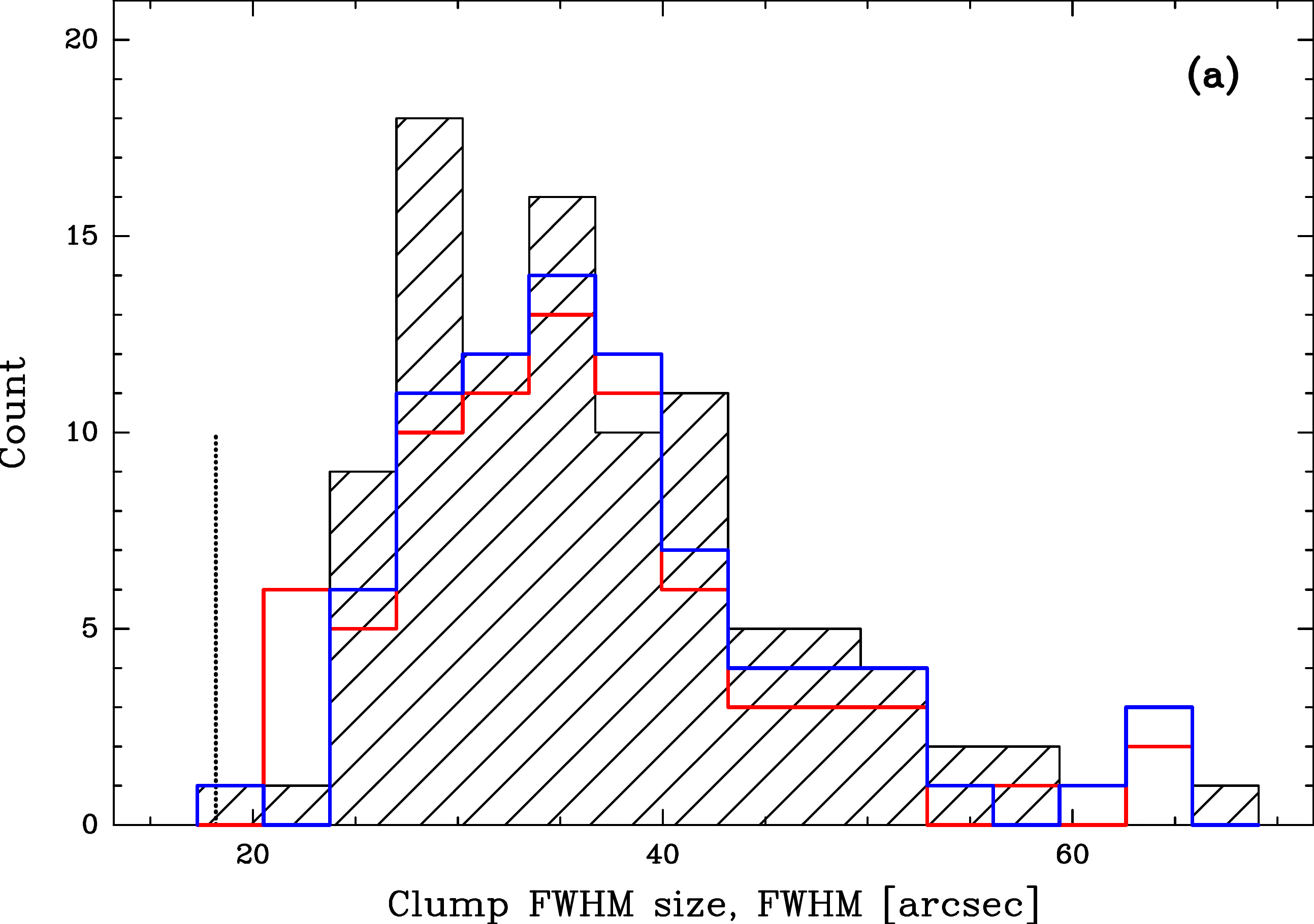}\hskip 0.7cm
    \includegraphics[width=0.47\hsize]{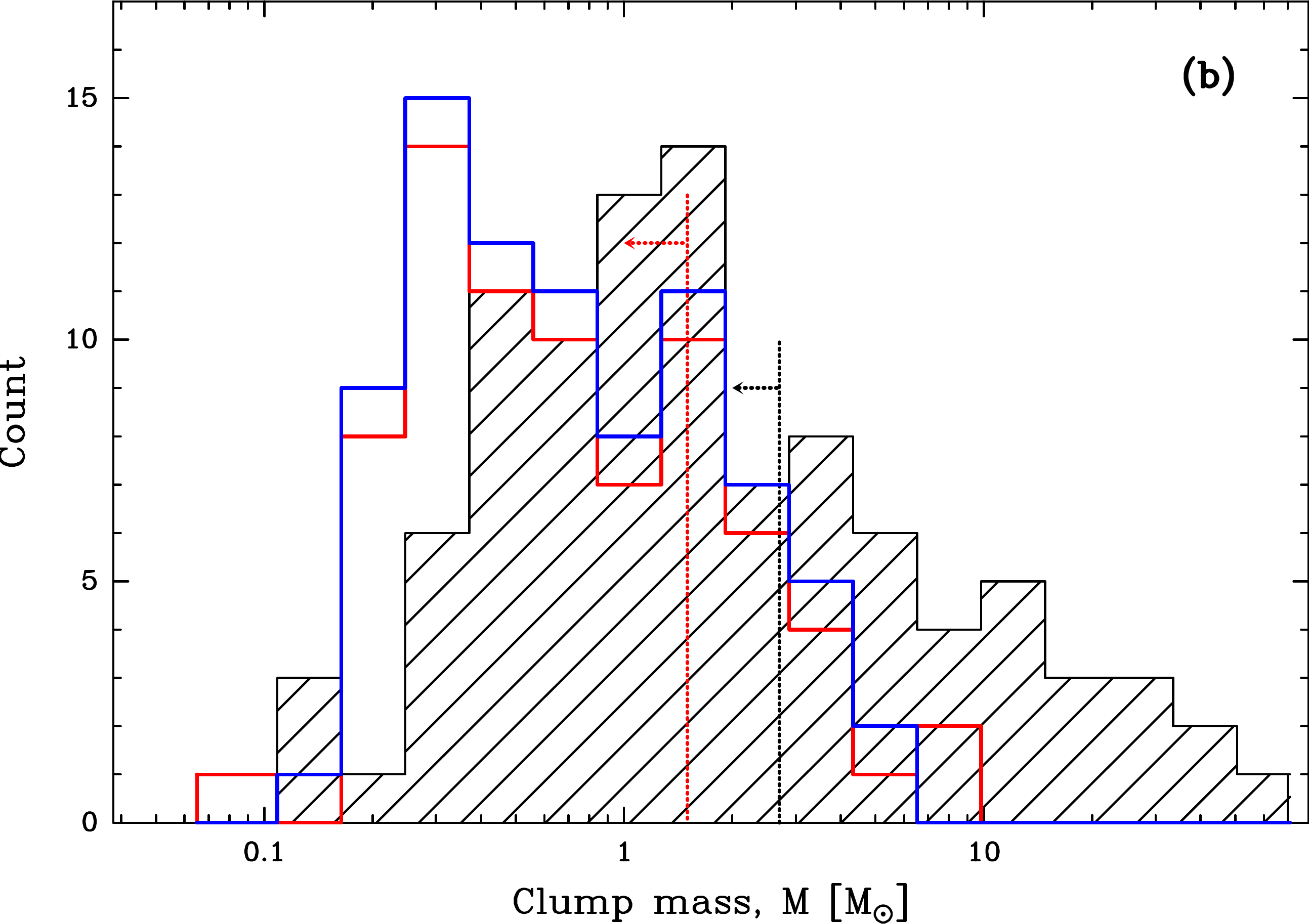}
    \caption{Distribution of clump size (in \textbf{a}) and clump mass (in \textbf{b}) for the three subregions of NGC~2264: central (black hatched histograms), northern (red histograms), and southern (blue histograms) subregions.
    The central subregion has an excess of high-mass clumps compared the northern and southern subregions. 
    The angular resolution of our observations, 18.2$\arcsec$, is given in \textbf{a}. The completeness levels, $\sim$2.7$\Msol$ in the central subregion and $\sim$1.5$\Msol$ in the northern and southern subregions, are indicated in \textbf{b} with dotted lines. 
    }
    \label{fig:mass-reg}
\end{figure*}

The size and mass distributions of clumps in the northern, central and southern subregions are presented in Figs.~\ref{fig:mass-reg}a-b.
As for the clumps sizes no significant variations have been observed between subregions (see \cref{fig:mass-reg}a). The size distributions are all centered around a median value of $\sim$35\arcsec, corresponding to a deconvolved size of {\it FWHM}$_{\rm dec}\sim$\,0.1~pc at 723~pc. The three subregions also host a couple of clumps with {\it FWHM} sizes close to the beam, 18.2$\arcsec$, and 9 clumps with sizes larger than 57$\arcsec$ (or 0.2~pc). 
At a scale of $\sim$ pc, clumps will likely fragment into smaller and denser cores, which could be the main mass reservoirs of individual protostars. 
Such cores have been observed in the two central protoclusters, IRS1 and IRS2 by \cite{Peretto06} and \cite{Cunningham16}. 

As for the mass distribution of clumps, it strongly varies between  subregions (see \cref{fig:mass-reg}b). While they are similar in the northern and southern subregions, the mass distribution in the central subregion is strongly shifted toward higher masses. Out of the 29 clumps with masses above 4~$\Msol$, 24 are located in the central subregion. 

\subsection{Gravitational stability of clumps}
\label{sub:clump_vir}

\begin{figure*}
    \centering
    \includegraphics[width=0.47\hsize]{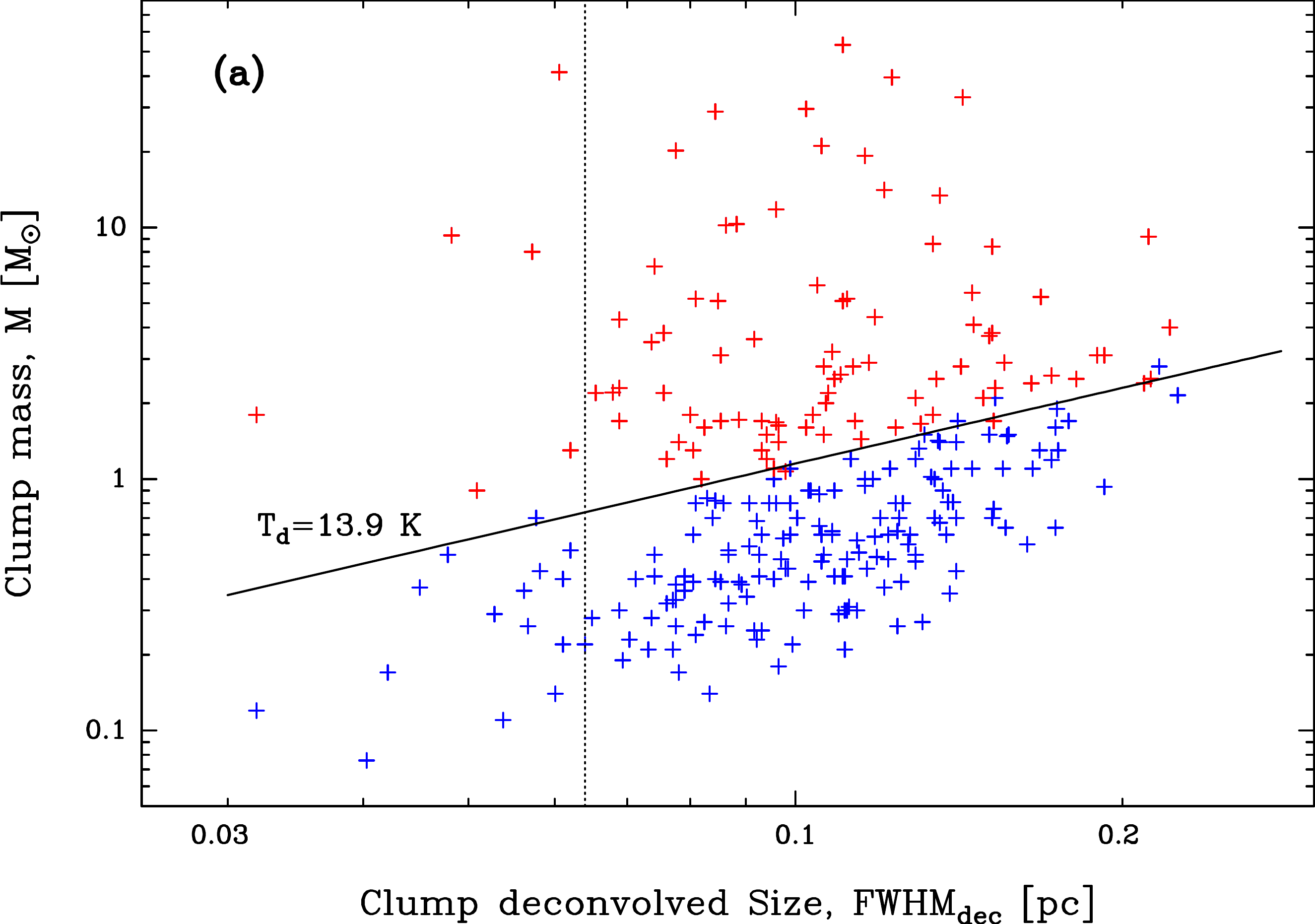}\hskip 0.7cm
    \includegraphics[width=0.47\hsize]{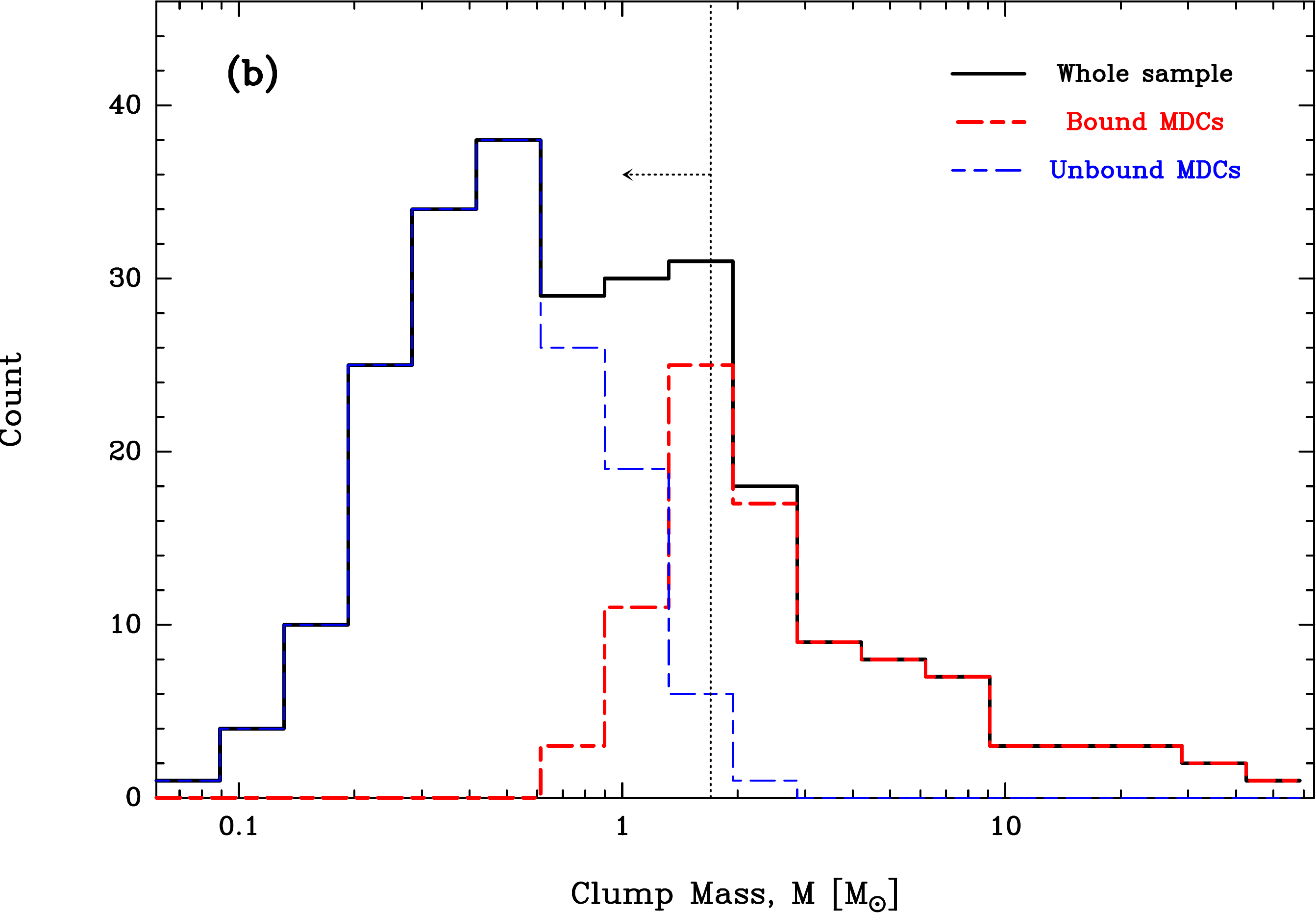}
    \caption{Mass versus size diagram (in \textbf{a}) and mass distribution (in \textbf{b}) for the 256 clumps detected in the NGC~2264 cloud. Gravitationally bound and unbound clumps, with respectively $\alpha_{\rm BE} < 2$ and $\alpha_{\rm BE} \geq 2$, are located with red (resp. blue) markers (in \textbf{a}) and sum up in a red (resp. blue) histogram (in \textbf{b}).
    \textbf{a)} The critical Bonnor-Ebert sphere model ($\alpha_{\rm BE} = 2$) at the median clump temperature, $T_{\rm dust}=13.9$~K, is plotted as a black solid line. The physical size of the beam at a distance of 723~pc is plotted as a vertical doted line.
    \textbf{b)} The global 90\% completeness level is indicated with a doted line.
    }
    \label{fig:MR}
\end{figure*}

In \emph{Herschel} studies,
the self-gravitating status of clumps is usually assessed using the Bonnor-Ebert ratio,  \(\alpha_{\rm BE} = R_{\rm dec}/R_{\rm BE}\), where $R_{\rm dec}$ is the deconvolved clump radius (here $FWHM_{\rm dec}$ and \(R_{\rm BE}\) the radius of a critical Bonnor-Ebert sphere of the same mass, $M$, and temperature, $T_{\rm dust}$ as the source, 
\begin{equation}
    R_{\rm BE}=\frac{M\,\mu_{\rm H_2}\,m_{\rm H}\, G}{2.4\times k\,T_{\rm dust}},
\end{equation}
where $G$ and $k$ are the gravitational and Boltzman constants.
This parameter estimates the ratio between the thermal support and gravitational force and clumps with  $\alpha_{\rm BE} < 2$ 
are considered as gravitationally bound. The Bonnor-Ebert ratio should however be considered as a rough estimator of the gravitational boundedness of the clump, since it does not include the effects of turbulence, magnetic fields, or external pressure.

Although $\alpha_{\rm BE}$ parameters theoretically depends on the mass, size, and temperature of the clump, the latter, which varies little within the sample ($13.9^{+3}_{-1}$~K,
see \cref{fig:temp-dist}), has a small influence here. As illustrated by the mass versus size diagram shown in \cref{fig:MR}a, the separation between bound and unbound clumps appears close to horizontal in this logarithmic representation. This implies that a threshold on the Bonnor-Ebert parameter approximately corresponds to a threshold in mass, $\alpha_{\rm BE} < 2 \Longleftrightarrow M \gtrsim$ 1-2~$\Msol$. 
This appears more clearly on the mass distributions shown in \cref{fig:MR}b. 
Bound clumps are significantly more massive than unbound clumps, with a median mass of 2.6~$\Msol$ vs 0.5~$\Msol$, respectively. 

In the whole NGC~2264 cloud and following the Bonnor-Ebert criterium, 36\% of the clumps are gravitationally bound. In more detail, half the clumps of the central subregion are bound while they are only $\sim$30\% in the northern and southern subregions (see \cref{tab:statMDC}). This result is in line with the variation of column-density background in the central subregion when compared with the southern and northern subregions (see \cref{tab:pow_spec_fit}).

\subsection{Spatial distribution of clumps}
\label{sub:clump_spacial}

\begin{figure}
    \centering
    \includegraphics[width=0.95\hsize]{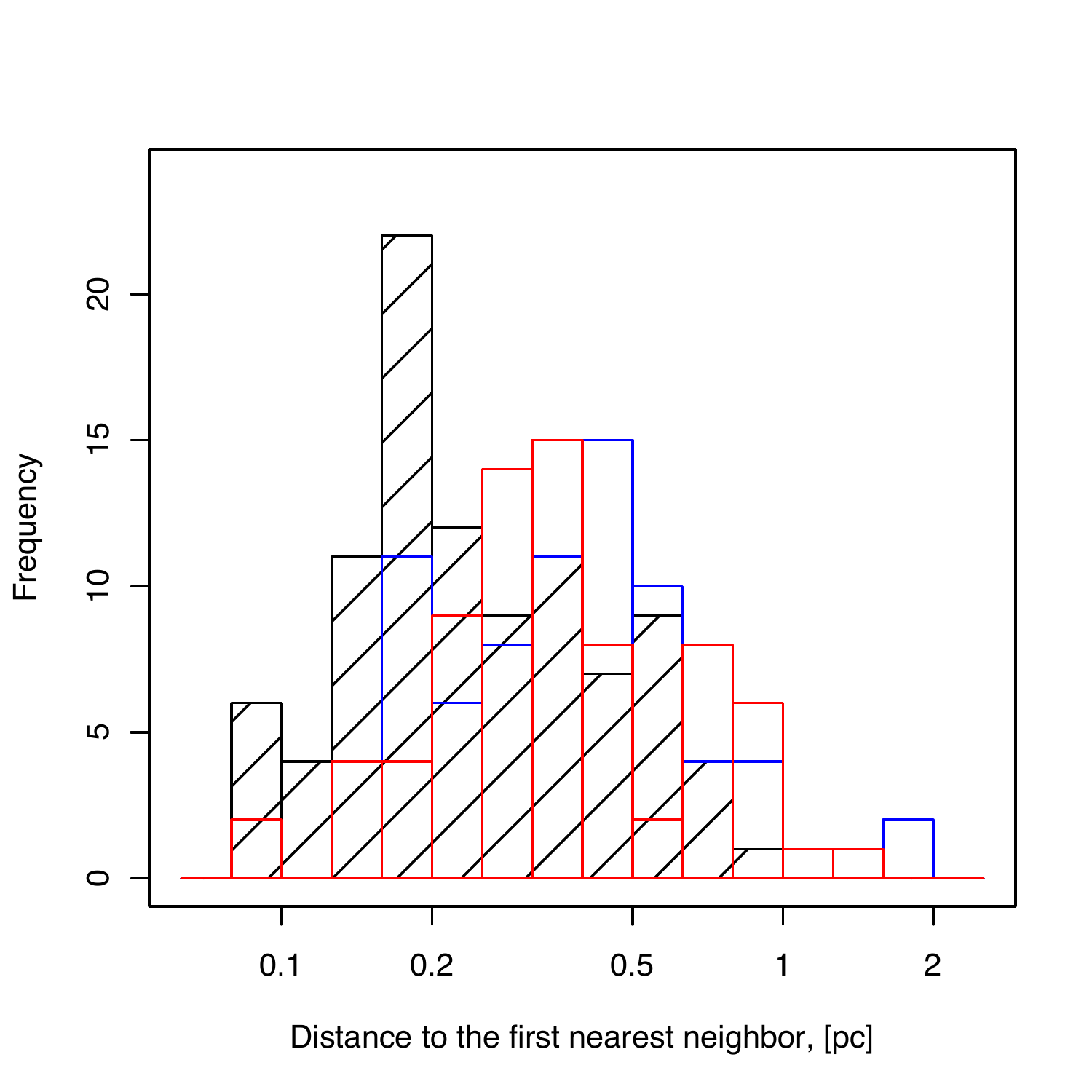}
    \caption{Distribution of the distance to the first nearest neighbor for clumps in the central (black hatched histogram), northern (red histogram), and southern (blue histogram) subregions of NGC~2264. Clumps are 1.6 times more closely packed in the central subregion than in the two others.}
    \label{fig:nndistMDC}
\end{figure}

Despite the different completeness limits of the three NGC~2264 subregions (see \cref{sub:compl}), clumps appear homogeneously distributed: 74, 97, and 85 in the northern, central, and southern subregions, respectively (see \cref{tab:statMDC}), of which $\sim$30\% lie above their 90\% completeness limit.
Note, however, that the density of clumps $\overline{\Sigma_{\rm clump}}$, defined as the ratio between the number of clumps and the subregion area, is about twice in the central subregion that in the northern and southern subregions (see \cref{tab:statMDC}).

To further investigate the spatial distribution of clumps, we first applied nearest neighbor statistics, using the distance from the center of each clump to the peak position of its nearest neighbor. 
The resulting distributions of nearest-neighbor clump separation are plotted in \Cref{fig:nndistMDC} for the three subregions of NGC~2264. 
In the northern and southern subregions, clumps have similar median distances, 0.34~pc and 0.36~pc respectively, and cannot be statistically distinguished, according to a Kolmogorov-Smirnov (KS) test.
The nearest neighbor distribution of the central subregion stands out with a large excess of clumps at short distance, $\sim$0.1--0.2~pc, and a median clump separation of $\sim$0.22~pc, which is 1.6 times smaller than in the northern and southern subregions. A KS test rejects, with a p-value below $10^{-4}$, the hypothesis that the central and northern $+$ southern subregions might have a common distance distribution. 
Interestingly, the clumps at the center of NGC~2264 are closely packed, with separations of about twice their median {\it FWHM} size, $\sim$0.11~pc (see \cref{sub:clumps_prop}), and thus once their  outer diameter.
They are mostly located in the two IRS1 and IRS2 protoclusters of NGC~2264 (see \cref{fig:map-MDC-z}).

We also computed the $Q$ parameter \citep{Cartwright04} for the NGC~2264 cloud and its three subregions (see \cref{tab:statMDC}).
The $Q$ parameter methods is based on the minimum spanning tree (MST), which is the set of straight lines (the edges) connecting a given set of points (here the clumps' center) without closed loops, such that the sum of all edges lengths is minimal.
$Q$ is defined as the ratio between the normalized mean 
edge length calculated by the MST, $\overline{l_{\rm edge}}$, and the mean clump separation, $\overline{s}$: 
\begin{equation}
    Q=\frac{\overline{l_{\rm edge}}}{\overline{s}}.
    \label{eq:Q-param}
\end{equation}

$Q$ values above 0.8 are associated with centrally concentrated spatial distribution, while lower $Q$ values indicate sub-clustering. 
With $Q\simeq0.7$, the NGC~2264 cloud overall displays only a moderate amount of sub-clustering for its clump population. 
Locally, the southern subregion appears 
to be more subclustered compared to the central and northern subregions ($Q \simeq 0.6$ versus $Q\simeq$\,0.8). On the column density map (see \cref{fig:map-MDC}), we indeed observed that its clumps are distributed either in a North-South filament joining the central subregion or in the south-western part of NGC~2264 .

\subsection{Mass segregation of clumps}
\label{sub:mass_seg}

\begin{figure*}
    \centering
    \vskip -1cm
    \includegraphics[width=0.45\hsize]{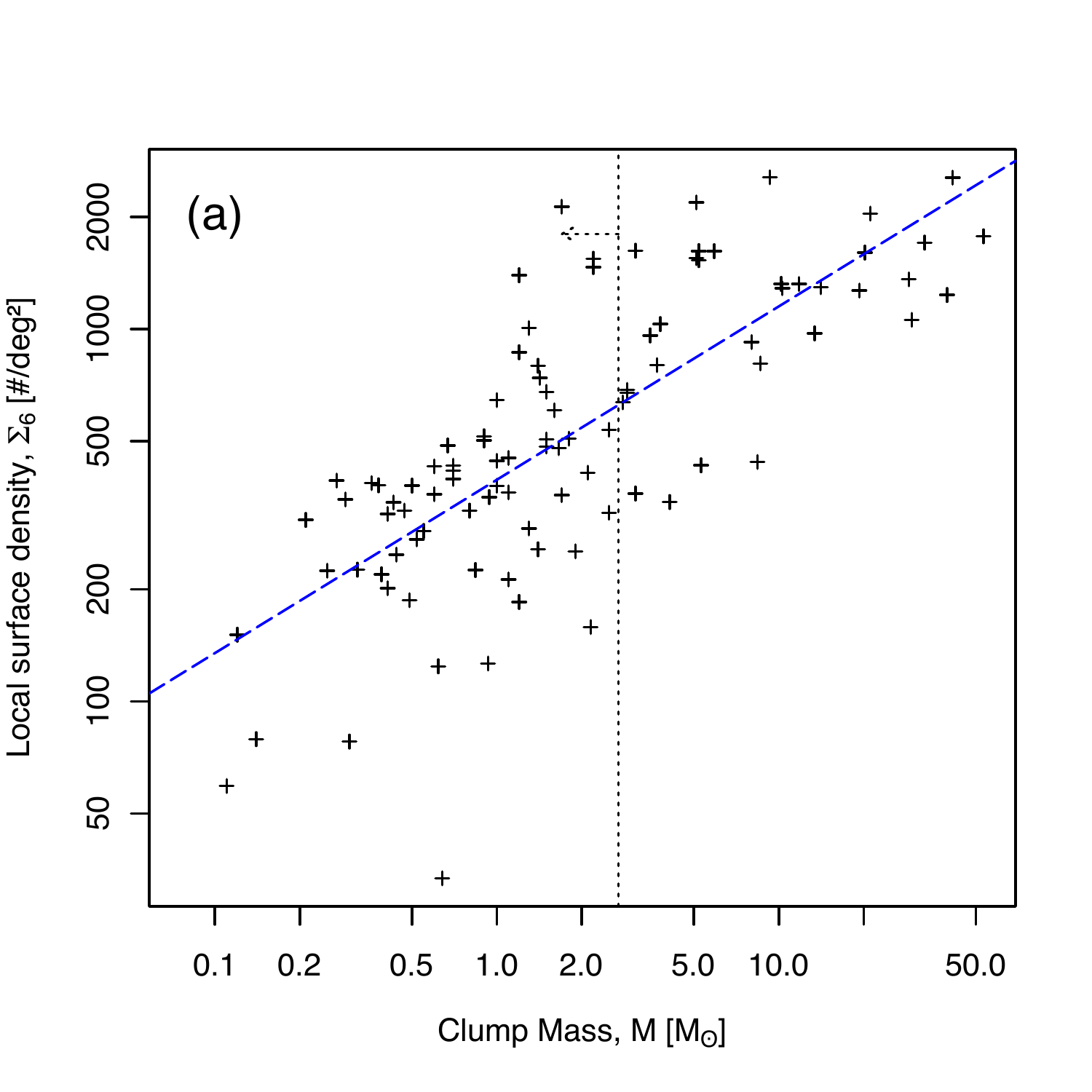}\hskip 0.5cm
    \includegraphics[width=0.45\hsize]{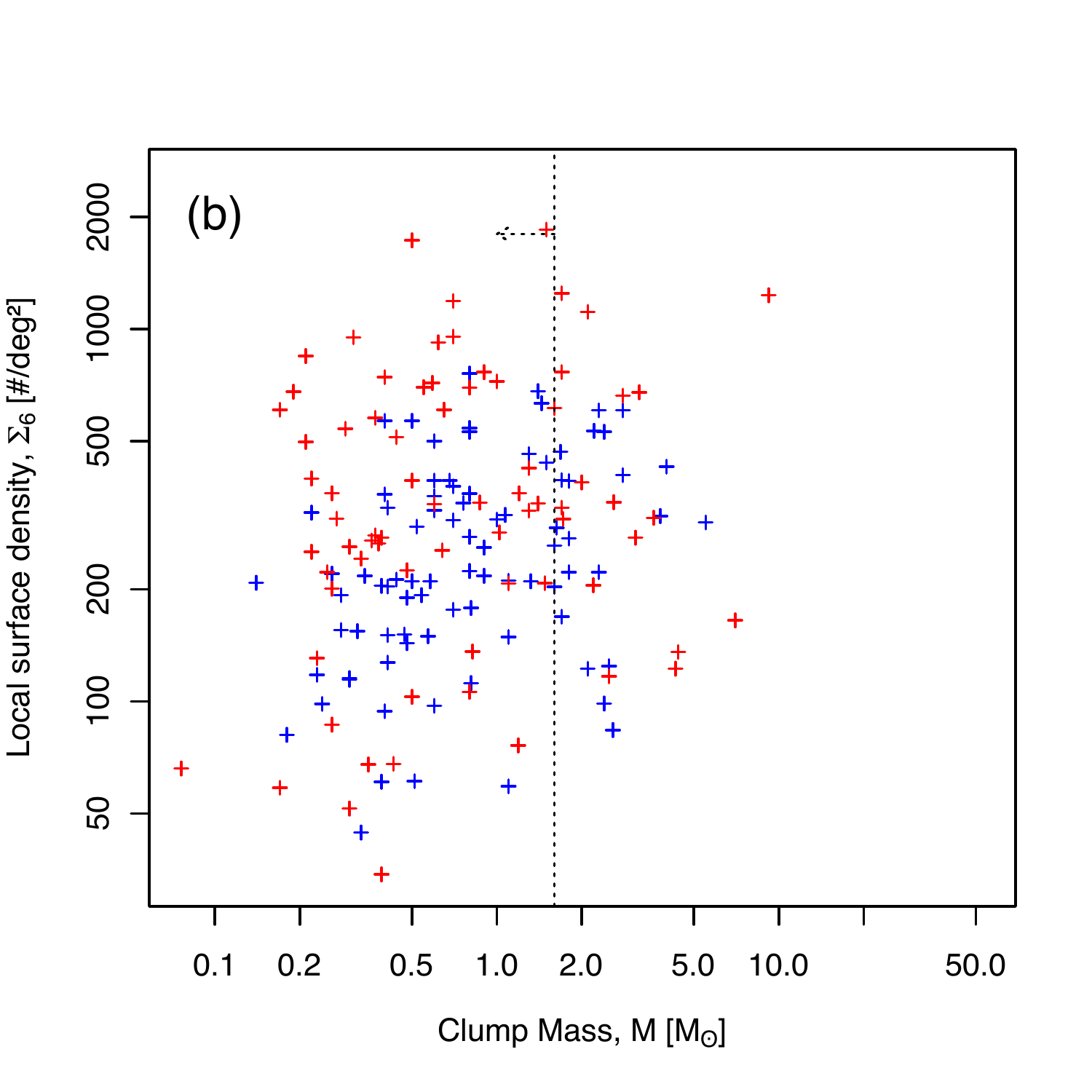}
    \vskip -0.3cm
    \caption{Local surface density $\Sigma_6$ as a function of the mass of the clump for the central subregion of NGC~2264 (in \textbf{a}, black markers) and in the northern and southern subregions (in \textbf{b}, red and blue markers, respectively). The potential correlation between $\Sigma_6$ and $M$ in \textbf{a}, $\Sigma_6 \propto M^{0.47}$, is represented by a dashed blue line, in both \textbf{a} and \textbf{b}.
    Completeness levels are indicated by vertical dotted lines.}
    \label{fig:sigma}
\end{figure*}
\begin{figure*}
    \centering
    \vskip -0.5cm    \includegraphics[width=0.47\hsize]{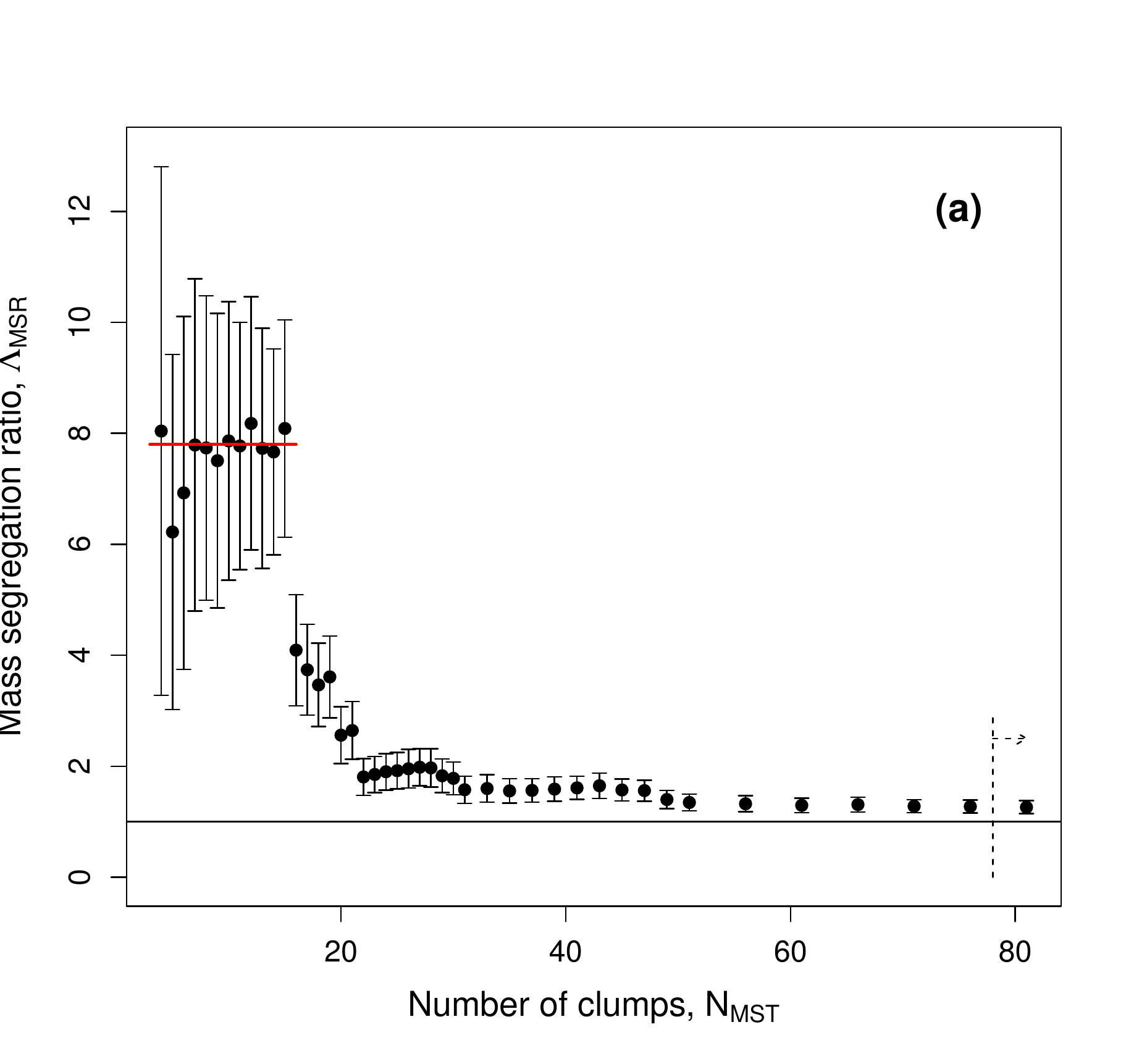}\hskip 0.5cm
    \includegraphics[width=0.47\hsize]{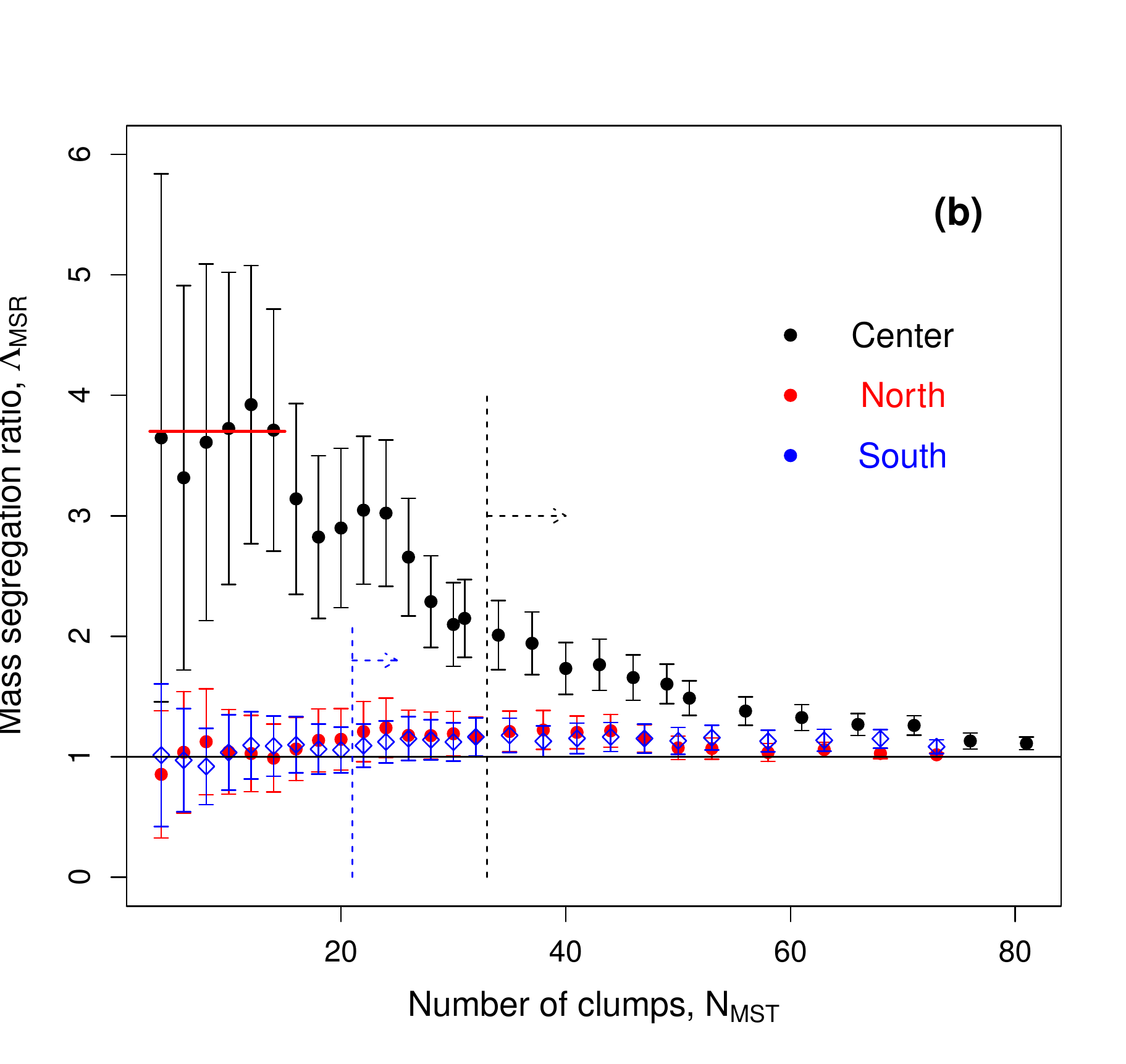}
    \vskip -0.3cm
    \caption{Mass segregation ratio, $\Lambda_{\rm MSR}$, as a function of the numbers of clumps, $N_{\rm MST}$, in the NGC~2264 cloud (in \textbf{a}) and in its three subregions (in \textbf{b}). Error bars represent the $\pm 2 \sigma$ uncertainties. Values above $\Lambda_{\rm MSR} \simeq 1$ (black solid line) suggest mass segregation. Completeness levels are indicated with vertical dashed lines.  
    \textbf{a)} The median segregation ratio for the 15 most massive clumps of NGC~2264, calculated over $N_{\rm MST}=4$ to 15, is $\Lambda_{\rm MSR}^{\rm med}=7.8$ (red segment). \textbf{b)} The $\Lambda_{\rm MSR}$ function of clumps in the central subregion (black symbols) is three to two times larger than the one in the northern and southern subregions (in red and blue, respectively), over the range of $N_{\rm MST}=4-33$.
    }
    \label{fig:lambda}
\end{figure*}

Mass segregation refers to a difference between the spatial distribution of massive objects compared to that of their lower-mass counterparts.
The methods used to quantify mass segregation have evolved with our current view on the subject and are associated to slightly different definitions, but in general, they all compare the distribution of high- and low-mass stars. Amongst them, the local surface density $\Sigma_{\rm j}$ \citep{Maschberger11}, and mass segregation ratio $\Lambda_{\rm MSR}$ \citep{Allison09}, do not require the specific definition of a cluster center and are the most widely used. They most probably also give the most robust results when combined \citep[e.g.,][]{Parker15}.
We applied both methods to measure the mass segregation of clumps in NGC~2264 and its subregions, using again the clumps peak position as a reference. 
The impact of the clumps spatial extent in these statistics will be investigated in future studies (Thomasson et al. in prep.).

The first method calculates the local surface density of source, $\Sigma_{\rm j}$, within an area encompassing the $j$th nearest neighbor, at a distance of $r_{\rm j}$:
\begin{equation}
    \Sigma_{\rm j} = \frac{j-1}{\pi r_{\rm j}^2}.
\end{equation}
As proposed by \cite{Maschberger11}, we took $j=6$,  as it constitutes a good compromise between estimating the local density and reducing low-numbers fluctuations \citep[see also][]{Casertano85}. 

Figures~\ref{fig:sigma}a-b present the $\Sigma_6$ versus $M$ diagrams that investigate if high-mass clumps are in denser groups than their lower-mass counterparts.
Since the clump populations of the northern and southern subregions exhibited similar behaviors in Sections~\ref{sec:clump_anal}.2--\ref{sec:clump_anal}.4., we assembled their clump samples. In contrast, we kept the central subregion alone since its clump sample had rather different characteristics.
In the central subregion, the local surface density correlates with the mass of the clumps, with a Pearson correlation coefficient of 0.75 (see \cref{fig:sigma}a). 
However, when accounting for the 90\% completeness level of 2.7~$\Msol$ (see \cref{sub:compl}), one can not exclude low-mass clumps in high-density groups have not been detected. 
Therefore, the most robust result of this $\Sigma_6$ analysis is that high-mass clumps in the central subregions are only found in high-density groups. The 15 most massive, $>$9.3~\Msol, clumps in NGC~2264, closely packed in the IRS1 and IRS2 protoclusters (see \cref{fig:map-MDC-z} and \cref{sub:clump_spacial}), have a median $\Sigma_6$ density about three times that of all clumps in NGC~2264 center. 
Conversely, the lower-mass clumps in the northern and southern subregions do not present a privileged location in high- or low-density groups according to their mass (see \cref{fig:sigma}b). \\

The second method to quantify mass segregation compares the MST of the most massive objects (here clumps) of a sample with the MST of randoms subsets of objects in this sample \citep[see][]{Allison09}. The mass segregation ratio, $\Lambda_{\rm MSR}$, is calculated as
\begin{equation}
    \Lambda_{\rm MSR}(N_{\rm MST}) = \frac{\overline{ l_{\rm random}}}{l_{\rm massive}} + \frac{\sigma_{\rm random}}{l_{\rm massive}},
\end{equation}
where $l_{\rm massive}$ is the length of MST for the $N_{\rm MST}$ most massive clumps and $\overline{ l_{\rm random}}$ is the average length of MST for $N_{\rm MST}$ random clumps. 
It was computed for 500 sets of $N_{\rm MST}$ random clumps along with the associated standard deviation, $\sigma_{\rm random}$. A value of $\Lambda_{\rm MSR} \approx 1$ indicates that the spatial distribution of massive clumps is comparable to that of other clumps and therefore that there is no mass segregation. Larger $\Lambda_{\rm MSR}$ values suggest that massive clumps are more concentrated than their lower-mass counterparts. 

Figures~\ref{fig:lambda}a-b display 
$\Lambda_{\rm MSR}$, for an increasing number, $N_{\rm MST}$, of the most massive clumps for the NGC~2264 cloud and its subregions.
The clump population of NGC2264 appears to be strongly mass segregated, with values of $\Lambda_{\rm MSR}$ around 7.8 for $N_{\rm MST} \leq 15$. If we consider only the central subregion, which contains the most massive clumps, segregation is still present, but the $\Lambda_{\rm MSR}$ values decreases to $\sim$3.7. This decrease is explained by smaller $l_{\rm random}$ values for clumps distributed in the central subregion than over the whole NGC~2264 cloud.
In \cref{fig:lambda}a, $\Lambda_{\rm MSR}$ drops to 4 for $N_{\rm MST}=16$, due to the fact that the 16th most massive clump is the first one located outside the IRS1 and IRS2 protoclusters. 
Mass segregation remain significant, with $\Lambda_{\rm MSR}$ values above 1 with more than 2 sigma uncertainty, up to $N_{\rm MST} \simeq 30$. 
In contrast, for the clump samples of the northern and southern subregions, \cref{fig:lambda}b do not reveal any mass segregation. The $\Lambda_{\rm MSR}$ plateau observed are listed in \cref{tab:statMDC}.

\section{Combined analysis of the YSO and clump populations in NGC~2264}
\label{sec:YSO-clumps}

In this section, we analyze the YSO content of the NGC~2264 subregions (\cref{sub:statYSO}) and quantify the link between the YSO and clump populations (\cref{sub:nnd}).

\subsection{Distribution in the NGC~2264 subregions}
\label{sub:statYSO}

\begin{figure*}[htpb]
    \centering
     \includegraphics[width=0.85\hsize]{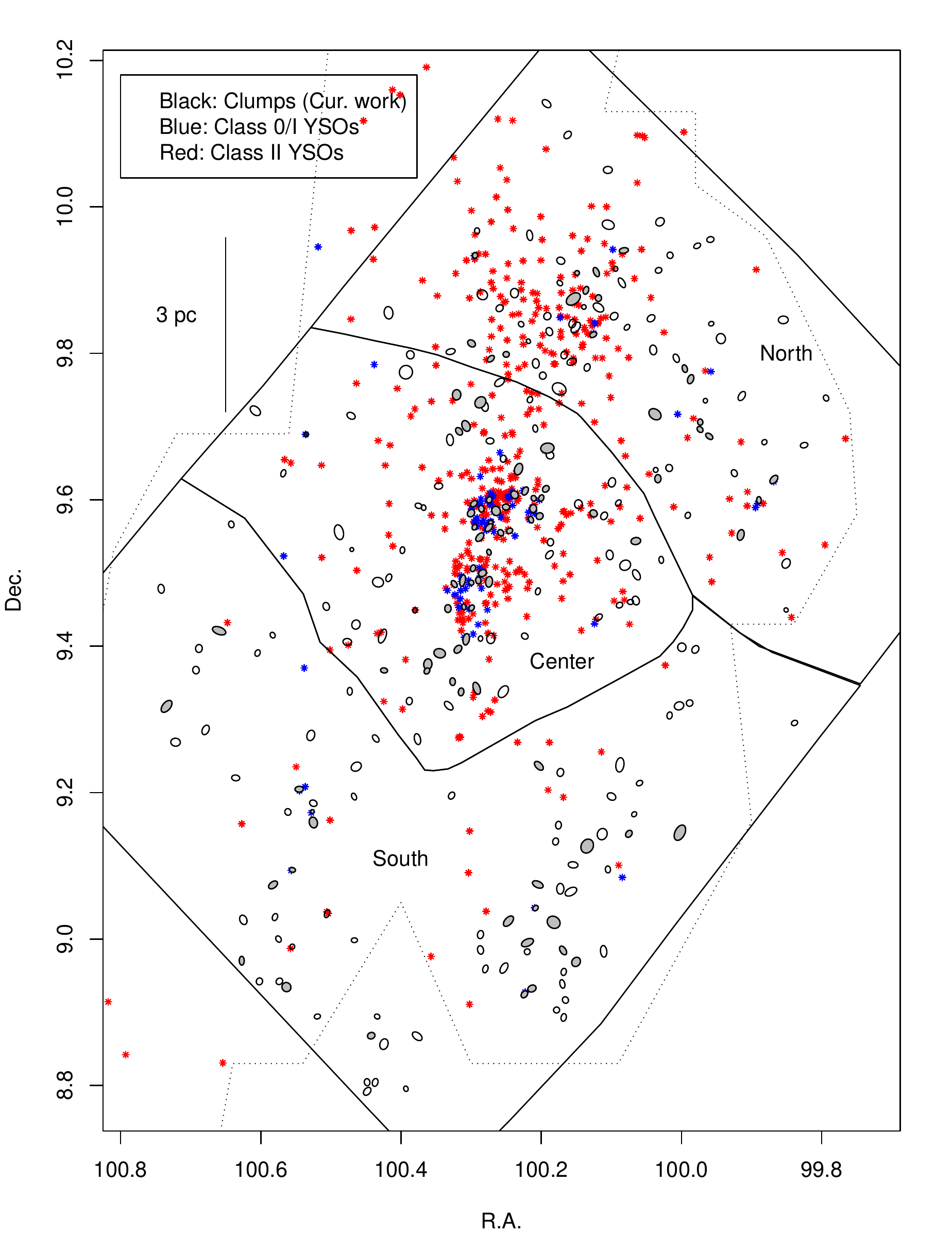}
    \caption{Spatial distribution of the clump (black ellipses) and YSO populations in the NGC~2264 cloud, as taken from the catalogs of \cref{tab:MDCs} and \cite{Rapson14}. 
    Bound clumps are represented with filled ellipses. 
    Class~0/I (blue star markers) and Class~II (red star markers) sources were identified in the area outlined by black dotted lines. The three subregions of NGC~2264 (see \cref{sub:3reg}) are outlined and labeled in black. Clumps are much more homogeneously distributed over the NGC~2264 cloud than YSOs.
    }
    \label{fig:MDC-YSO-map}
\end{figure*}

The spatial distribution of the clump and YSO populations is illustrated in \cref{fig:MDC-YSO-map}.
The clump population is rather homogeneously distributed over the whole \emph{Herschel} area (see \cref{sub:clump_spacial}) even if it obviously follows the gas concentration toward the two IRS1 and IRS2 protoclusters and the "Y"-shaped filament. The YSOs themselves cluster in three places: strongly around the IRS1 and IRS2 protoclusters and more distributed around the S~Mon massive star.

\Cref{tab:statMDC} lists the number of YSOs regardless of their class, the number of Class~0/I protostars and Class~II pre-main sequence stars found by \cite{Rapson14} in NGC~2264. We distributed this YSO population among the three subregions of the NGC~2264 cloud defined in \cref{sub:3reg}.
Unlike the spatial distribution of clumps in subregions, 
a very uneven distribution is observed for the YSOs.
The central subregion, which accounts for 22\% of the whole NGC~2264 cloud area and contains 38\% of the detected clumps (see \cref{tab:pow_spec_fit,tab:statMDC})
indeed gathers up to 62\% (302 out of 485) of the YSO population.
In contrast, the southern subregion, which covers 45\% of the whole cloud, contains 33\% of the clumps but only 6\% (27 out of 485) of the YSOs. This clear lack of YSOs in this subregion cannot be explained by the small area of the \emph{Herschel} image not investigated for YSOs by \cite{Rapson14} (see \cref{fig:MDC-YSO-map}).
The northen subregion show a more balanced distribution with 33\%, 29\%, and 32\% of the cloud area, clump and YSO populations. 

Regarding the evolutionary class of YSOs, the northern subregion is almost exclusively populated by Class~II sources (146 out of 156, 95\%, see \cref{tab:statMDC}). On the other hand, the distribution is more balanced in the central and southern subregions: 23\%-30\% of the YSOs are Class~0/I protostars and the complementary 70-77\% are Class~II sources.
Finally, we found that the 69 Class~0/I in the central subregion represent a large majority (79\%, see \cref{tab:statMDC}) of this younger YSO population in the whole NGC~2264 cloud.

\subsection{Link between the clump and YSO populations}
\label{sub:nnd}

\begin{figure*}[htbp]
    \centering
    \vskip -1.5cm
    \includegraphics[width=0.5\hsize]{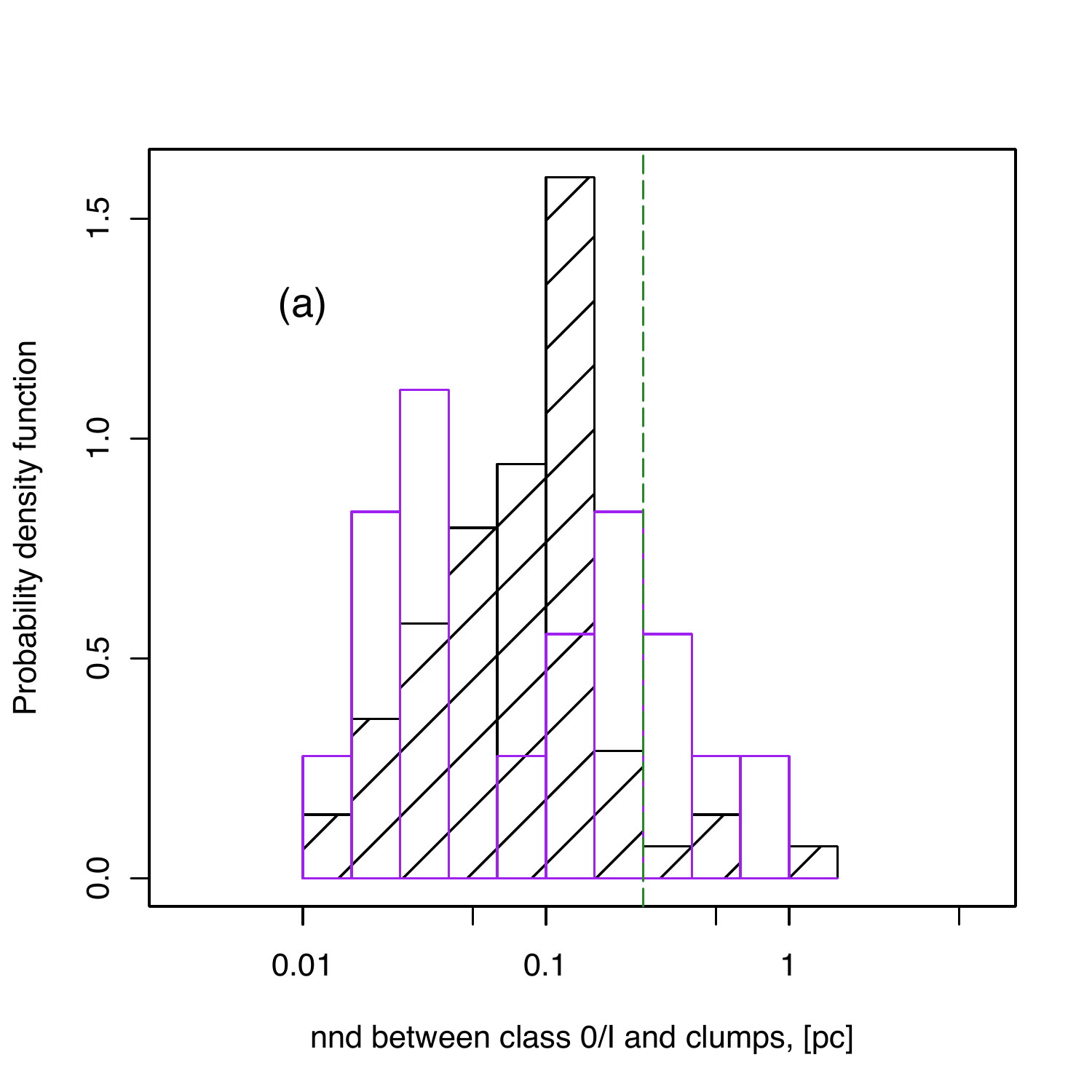}
    \hspace{-0.3cm}
    \includegraphics[width=0.5\hsize]{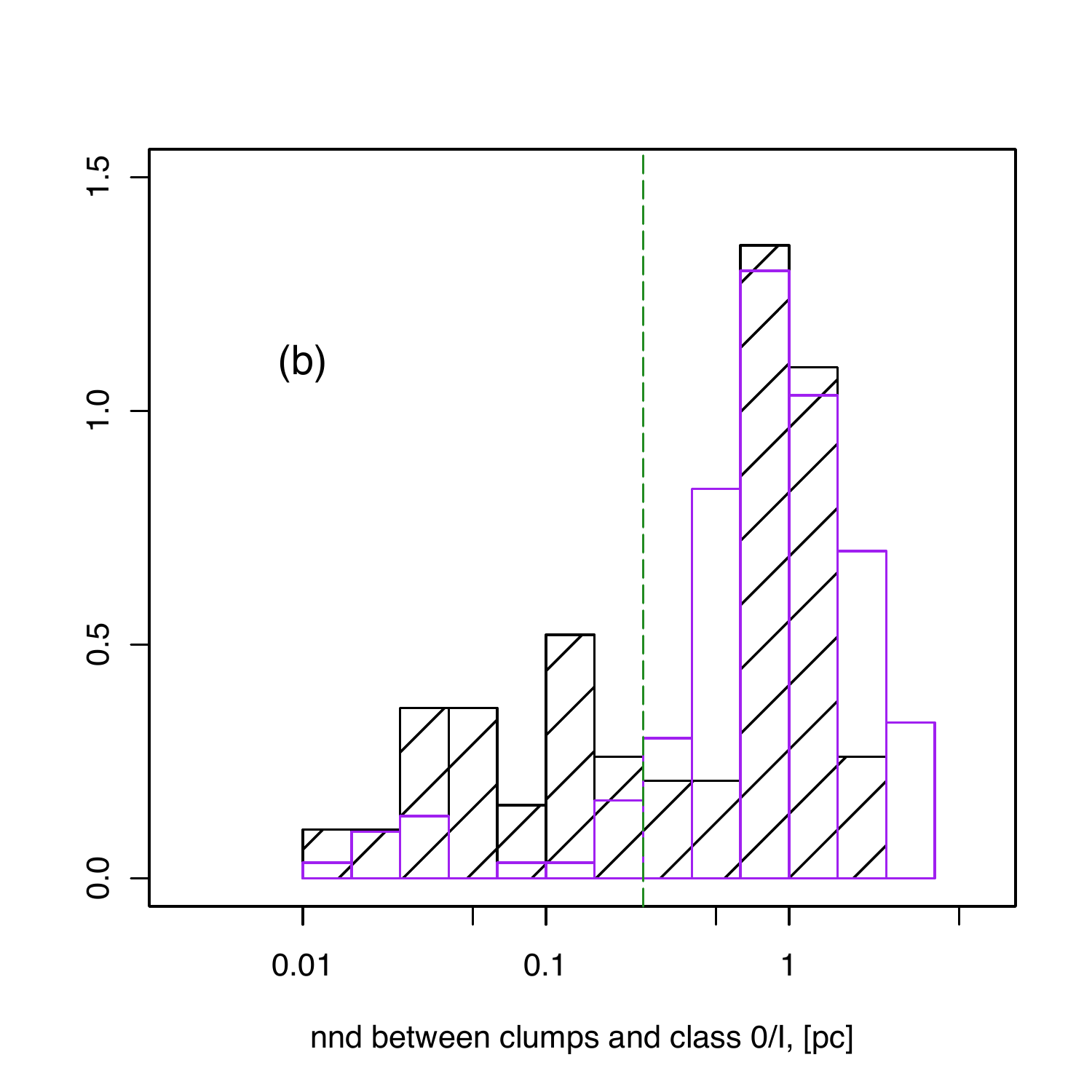}
    \vskip -0.4cm
    \includegraphics[width=0.5\hsize]{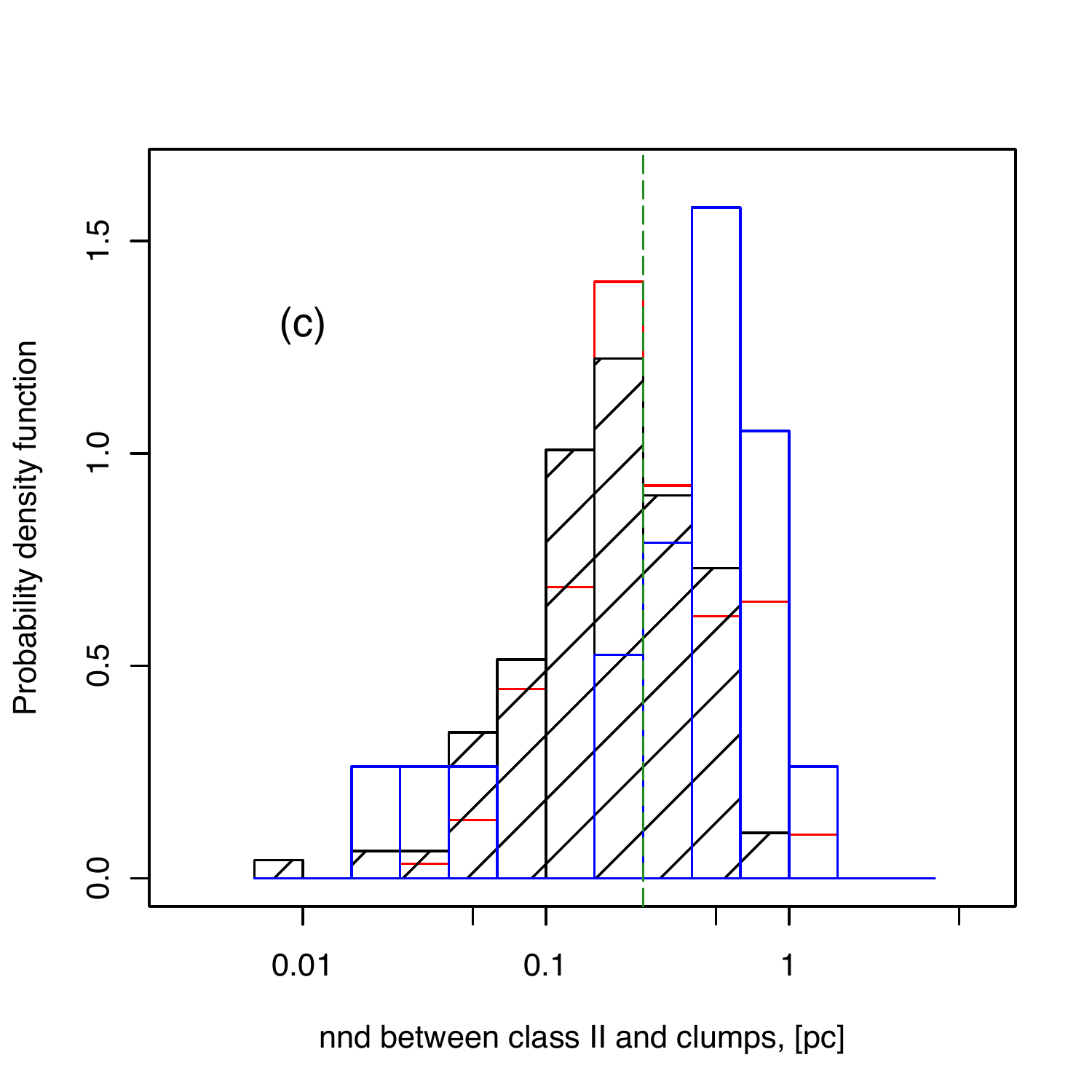}
    \hspace{-0.3cm}
    \includegraphics[width=0.5\hsize]{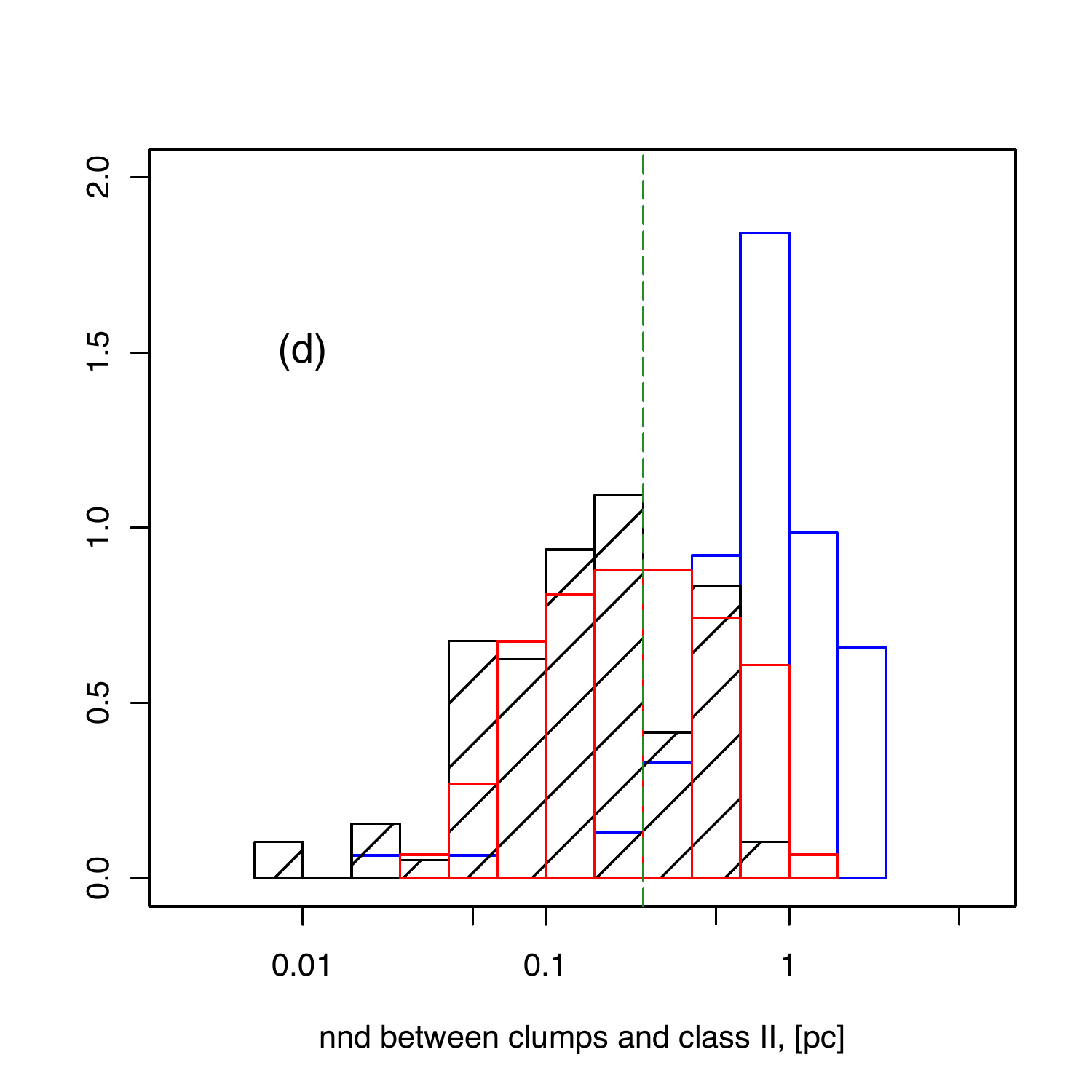}
    \caption{Distribution of the distance to the first nearest neighbor, $nnd$, from  YSOs to clumps (in \textbf{a} and \textbf{c}) and from clumps to YSOs (in \textbf{b} and \textbf{d}). 
    YSOs are Class 0/I in the upper panel (\textbf{a}-\textbf{b}), Class II in the lower panel (\textbf{c}-\textbf{d}). 
    Distributions are presented for the central subregion (black hatched histograms) and for the northern and southern subregions, combined in the upper panel (purple histogram) and separated in the lower panel (red and blue histograms, respectively). 
    Distributions have been normalised to facilitate the comparison involving small populations of YSOs (Class 0/I from northern and southern subregions in \textbf{a}, Class II from southern subregions in \textbf{c}).
    The red dashed vertical line indicates the median outer radius of clumps, calculated as twice the median FWHM: $R_{\rm out}^{\rm med}=2\,{FWHM}^{\rm med}=0.25$~pc.
    }
    \label{fig:nndist}
\end{figure*}

We investigated the spatial correlation between the clump and YSO populations found in the different subregions of NGC~2264 using the nearest neighbor statistics. Figures~\ref{fig:nndist}a-d show the distributions of the distance to the first nearest neighbor (nnd) from YSOs to clumps and vice versa, with YSO populations separated into Class~0/I protostars and Class~II pre-main sequence stars.
Class-0/I and clumps are similarly distributed in the northern and southern subregions (see \cref{tab:statMDC}  and \cref{fig:MDC-YSO-map}), in addition the nnd distributions between Class-0/I and clumps cannot be statistically distinguished, according to a KS test.
We therefore combined the northern and southern populations of Class~0/I protostars to increase the sample size (18 objects).
The 10 clumps detected in the \emph{Herschel} area not covered by \emph{Spitzer} (see \cref{fig:MDC-YSO-map} and \cref{tab:statMDC}) and thus not surveyed for YSOs by \citet{Rapson14} have been excluded from our analysis. 

The nearest neighbor distance, $nnd$, distribution between Class~0/I protostars and clumps first illustrates the strong link between Class~0/I protostars and their parental clump. Indeed, the large majority, $80-95\%$, 
of the Class~0s/Is of each of the NGC~2264 subregions lie at a distance smaller than the median outer radius of clumps, i.e. below 0.25~pc (see \cref{fig:nndist}a).
Conversely, the $nnd$ distribution of clumps with respect to Class~0/I protostars shows in \cref{fig:nndist}b that $\sim$40\% 
of the clumps in the central subregion host, within their outer radius, at least one Class~0/I protostar.
This fraction raises to 2/3 when considering only bound clumps in the central subregion.
In addition to illustrating this parental link, the $nnd$ distributions between clumps and Class~0s/Is presents, for both the central and northern $+$ southern samples a peak around 1~pc (see \cref{fig:nndist}b). 
This peak shows that aside the population of clumps tightly linked to Class 0/I, mostly clustered in the central region, there exists all over NGC 2264 a disperse population of clumps that are not associated to Class 0/I YSOs.   
As a matter of fact, most of the clumps that create this peak, $\sim$75\% whatever the subregion, are qualified as unbound. These unbound clumps could be more than transient cloud fragments because the external gas pressure associated with the global collapse is not taken into account in the calculation of gravitational boundedness. Excluding clumps presently qualified as unbound however, the $nnd$ distribution between clumps and Class~0s/Is still displays a peak, which is weaker and located at $\sim$0.5~pc.

Figures~\ref{fig:nndist}c-d display the $nnd$ distributions between clumps and Class~II pre-main sequence stars. Given the proper motions found by \cite{Buckner20} for the weakly embedded Class~II sources in the NGC~2264 cloud, $\sim$1~mas\,yr$^{-1}$, and their estimated age, $\sim\times 10^6$~yr \cite{Venuti18},
they could have dispersed by $\sim$6~pc from their original birth site. 
For these Class~II sources, we thus do not expect a direct parental link between Class~IIs and the presently observed clumps.
Other Class~II pre-main sequence stars could well  remain more tightly associated with their birth site, like it is probably the case in the NGC~2264 IRS1 and IRS2 protoclusters.
The histograms of Figs.~\ref{fig:nndist}c-d should thus reveal the parental link of a clustered population of Class~IIs plus
the average spatial distributions of a more extended population of Class~IIs, with no direct parental link with clumps.
If the small number, 19, of Class~II YSOs in the southern subregion makes any detailed characterization of the histogram shown in \cref{fig:nndist}c not relevant, that of \cref{fig:nndist}d should be robust to interpret.
While the $nnd$ distributions between clumps and Class~II sources are similar in the central and northern subregions, the one of the southern subregion clearly stands out as different (see \cref{fig:nndist}d).
The former are both broadly distributed around $\sim$0.2~pc.

In contrast, in the southern subregion, the $nnd$ distribution between clumps and Class~IIs presents a strong peak at $\sim$1~pc, which means at much further distances than in the other parts of the cloud.
This definitively is a consequence of the low density of YSOs observed in the southern subregion (see \cref{fig:MDC-YSO-map}).
One can question the association of Class~IIs with the southern subregion or even their membership in the NGC~2264 cloud.
Indeed, with proper motions like those measured by \cite{Buckner20}, the Class~IIs observed in the southern subregion could have formed in, and been ejected from, the central subregion of NGC~2264. As a guideline, \cite{Buckner20} also estimated that up to 30\% of the YSOs observed in the area covered by the NGC~2264 cloud may not be members of this star-forming region. A detailed comparison of the star and cloud velocities measured with ground-based radiotelescopes and the \emph{gaia} observatory would be necessary to evaluate the YSO membership in the southern subregion of NGC~2264.

The comparison of the $nnd$ distributions of clumps with respect to YSOs (see Figs.~\ref{fig:nndist}b and \ref{fig:nndist}d), in the central and northern subregions, leads to an apparent counter-intuitive result. On average, clumps are closer to Class~II pre-main sequence stars than to Class~0/I protostars: median $nnd$ of 0.2~pc versus $\simeq$0.8~pc. This reflects the differences in the spatial distribution visible in \cref{fig:MDC-YSO-map}: while Class 0s/Is are, for their large majority, clustered in the IRS1 and IRS2 protoclusters, Class~IIs and clumps display a more distributed population. 
This is totally inconsistent with scenarios of quiet cloud formation followed by a continuous process of star formation 
\citep[see, e.g.,][]{Shu87,Krumholz14}.



\section{Discussion}
\label{sec:discu}
We hereafter make the link between evidence for mass segregation observed for NGC~2264 clumps and the cloud structure (see \cref{sub:mass-disc}) and propose an updated scenario for the star and thus cloud formation history in NGC~2264  (see \cref{sub:SF-history}).

\subsection{Mass segregation of clumps and its relationship to cloud structure}
\label{sub:mass-disc}

Mass segregation has been studied extensively in stellar clusters for decades. 
%
More recently, mass segregation has been investigated for cloud fragments with 0.002~pc to 0.1~pc sizes, using tools developed for stellar clusters. In particular, \cite{Dib19} calculated the mass segregation ratio of cores in four star-forming regions: Taurus, Aquila, Corona Australis, and W43-MM1.
They found no mass segregation in Taurus but a significant level of mass segregation in the three other clouds: $\Lambda_{\rm MSR}$ up to 4-9 for the 6--14
most massive cores. \cite{Plunkett18} also reported that Serpens South is mass segregated, with a median $\Lambda_{\rm MSR} \simeq 4$ for the $N_{\rm MST} \leq 18$ most massive cores. 
\cite{Parker18} and \cite{Konyves20} in Orion~B, \cite{Roman19} in Orion~A, \cite{Sadaghiani20} in NGC 6334 also claimed to find mass segregation, although with lower values ($\Lambda_{\rm MSR} \simeq 2-3$) and/or involving a smaller number
of cores.
Besides, no significant mass segregation was found in 12 infrared-dark clouds by \cite{Sanhueza19}.
%
As for the surface density parameter, $\Sigma_{\rm j}$, sometimes used to quantify mass segregation, it was up to now 
only computed for a couple of star-forming regions. \cite{Lane16}, \cite{Parker18} and \cite{Dib19} notably reported tentative trends for the most massive cores in Orion-A, Orion~B, and Corona Australis to sit in areas of higher local surface density. In particular, \cite{Lane16} found in Orion~B a median $\Sigma_{\rm 10}$ value twice larger for the 10 most massive cores than for the whole source samples.

When comparing these published results with the mass segregation found in NGC~2264 (see \cref{sub:mass_seg}), the latter is among the strongest and affecting the largest numbers of objects. The $\Lambda_{\rm MSR}$ parameter is indeed measured to be $\Lambda_{\rm MSR} \simeq 8$ for the $N_{\rm MST}= 4$ to 15 most massive clumps (see \cref{tab:statMDC} and \cref{fig:lambda}a). 
Moreover, mass segregation remains significant for the $\sim$30 most massive clumps of the NGC~2264 cloud. 
When the mass segregation is measured by the local surface density, it has a median $\Sigma_{6}$ value for the 15 most massive clumps four times larger than that of the whole sample.
We refrained making more quantitative comparisons with published studies since the $\Lambda_{\rm MSR}$ values are measured with different methods (e.g., sliding window in \citealt{Roman19,Konyves20} instead of cumulative form here) and datasets are inhomogeneous. Sources were indeed extracted with different extraction tools (e.g., clumpfind in \citealt{Roman19}, FellWalker in \citealt{Parker18} versus \emph{getsf} here)
and thus have various definitions.
The cloud fragments considered also have different physical sizes, which range from $\sim$0.002~pc (or $\sim$400~AU) in \cite{Plunkett18}, 0.01-0.03~pc (or 2000-6000~AU) in \cite{Dib19} to $\sim$0.1~pc (or $\sim$2$\times 10^4$~AU) in the present study. Moreover, there are also issues, generally not taken into account, related to the incompleteness of samples and crowding of sources in measuring the mass segregation in dense environments like massive (proto)clusters \citep{Ascenso09}.
 
As shown in \cref{fig:lambda}b, this strong mass segregation is entirely due to the concentration of high-mass clumps in the  NGC~2264-IRS1 and IRS2 protoclusters, in the central subregion. We showed that these high-mass clumps all are gravitationally bound and lie at short distance from each other (see Sections~\ref{sub:clump_vir}-\ref{sub:clump_spacial} and \cref{fig:map-MDC-z}), thus forming a cluster of star-forming clumps.
The overabundance of high-mass clumps in the central subregion of NGC~2264 relative to the northern and southern subregions is consistent with its greater gas concentration.
After removing clumps from the column density image of \cref{fig:map-MDC}, the average column density of this background, at the location of clumps, is indeed almost twice higher in the central subregion than in the other NGC~2264 subregions (see Table~\ref{tab:pow_spec_fit}).
%
The relation between the mass of cloud fragments and their surrounding gas has been reported for several star-forming regions 
\citep[e.g.,][]{Konyves20}
and presents here a linear correlation between the mass of NGC~2264 clumps, $M$, and the column density of their surrounding background, $N_{\rm H_2}^{\rm back}$ (see \cref{fig:MNH2-reg}).
The incompleteness level of the clump sample could however partly explain this linear correlation and especially the lack of low-mass clumps over regions with high background level.

The overabundance of high-mass clumps in the central subregion is also in line with the fact that its hierarchical structure is different from that of the northern and southern subregions (see \cref{sub:MnGSeg}). Indeed, the power spectrum of the central subregion, which is hundred times stronger, in fact consists of the sum of a powerlaw function plus power excesses at the scales of cloud and clump mass reservoirs (see \cref{fig:gc-powspec}b and \cref{tab:pow_spec_fit}). While diffuse clouds and Gould Belt clouds generally have simple multi-fractal power spectra \citep{Robitaille19, Robitaille20}, those of high-mass star-forming regions are more complex and dominated, at given scales, by large gravity potentials such as hubs and ridges (Robitaille et al. in prep.).
Following the definition criteria set in HOBYS articles \citep[e.g.,][]{Hill11, Hennemann12, Nguyen13} and precursor papers \citep{Schneider10}, the central part of the central subregion of NGC~2264 indeed qualifies as a ridge.
In short, ridges are very dense, $>$10$^5$~cm$^{-3}$, $\sim$1~pc cloud structures actively forming clusters of intermediate- to high-mass stars \citep{Motte18a}.
The center of NGC~2264 indeed hosts a dense north-south filament,
which appears as a privileged site for intermediate- to high-mass star formation (see Figs.~\ref{fig:map-MDC-z} and \ref{fig:MDC-YSO-map}).
It contains a cluster of massive clumps hosting protostars, which could ultimately form a rich YSO cluster with at least a handful of high-mass stars.
The mass segregation observed for the NGC~2264 clumps could thus be at the origin of the mass segregation of future stars in, at least, the central stellar cluster of NGC~2264.

We postulate that this mass segregation could come from further afield, in the very way the mass of gas was assembled to form the molecular cloud in the NGC~2264 region.
The large-scale kinematics observed throughout the Monoceros cloud complex \citep{Loren77,
Montillaud19b} and the global infall discovered toward the NGC~2264-IRS1 protocluster \citep{Williams02, Peretto06} indeed suggest that a hierarchical collapse of the NGC~2264 cloud could have led to the formation of a ridge at its center. 
Further studies of the cloud kinematics are necessary to investigate whether the three filaments that appear converging toward the central ridge, drive material and therefore feed the  ridge and its protoclusters. 
This mode of cloud and thus star formation by competitive, inflowing gas is advocated in theories of hierarchical global collapse \citep[e.g.,][]{Vazquez19}, gravitationally-driven gas inflows \citep[e.g.,][]{Smith09,Hartmann12}, filament or conveyor-belt collapse \citep{Myers09,Krumholz20} and colliding flows \citep[e.g.,][]{Heitsch08}.

\subsection{Star formation history in NGC~2264}
\label{sub:SF-history}

\begin{figure}
    \centering
    \includegraphics[width=0.99\hsize]{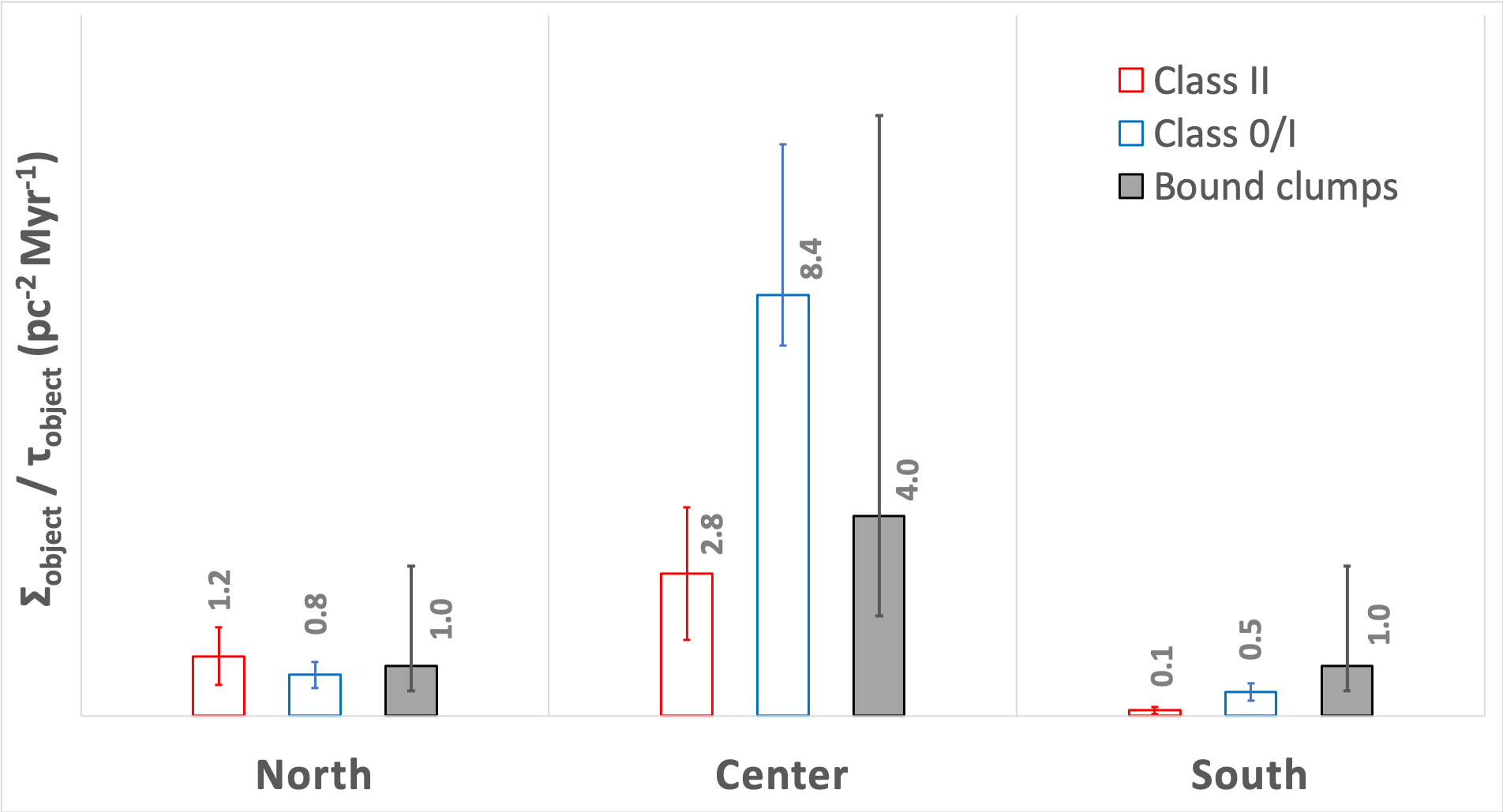}
    \caption{Surface density per lifetime, $\Sigma_{\rm object}/\tau_{\rm object}$, of Class~II YSOs, Class~0/I protostars, and clumps in the three NGC~2264 subregions. Statistical lifetimes of 2, 0.2 and 0.3~Myr were assumed for Class~IIs, Class~0/Is, and clumps, respectively. 
    Error bars take into account statistical uncertainties, contamination level of 30\% for Class~IIs, an incompleteness level of 25\% for the Class~0/I population in the central subregion, and a factor of two of uncertainty for the clumps.
   While the star formation activity of the northern subregion appears to be constant, it displays a burst in the central subregion and shows a tentative increase 
   in the southern subregion.}
    \label{fig:sigma-histo}
\end{figure}

Evidence for sequential star formation in NGC~2264 has been presented by several YSO studies \citep{Sung10,Teixeira12,Rapson14,Venuti18}. \cite{Venuti18} proposed that star formation began $\sim$5~Myr ago in the northern subregion and more precisely in the S~Mon area. Then, less than 1~Myr ago, star formation could have developed in the central subregion and especially in the NGC~2264-IRS1 and IRS2 protoclusters. 
In the present study, this age gradient is qualitatively confirmed when comparing YSO distributions of the northern and central subregions. The central subregion indeed concentrates about 80\% of the total Class~0/I protostars of the NGC~2264 cloud while the northern subregion is mainly populated by Class~II pre-main sequence stars (see \cref{tab:statMDC} and \cref{sub:statYSO}). As for the southern subregion of the NGC~2264, it has not been considered in YSO studies because it contains very few YSOs, only 6\% of the total population, suggesting that star formation has not yet been active there.
In order to more accurately compare subregions, \cref{fig:sigma-histo} displays, for each YSO (Class~II and Class~0/I) population, their surface density divided by their statistical lifetime.
Surface densities, $\Sigma_{\rm object}$, are calculated using the number of objects in each YSO class and the subregion surfaces listed in \cref{tab:pow_spec_fit,tab:statMDC}.
We assumed lifetimes, $\tau_{\rm object}$, of $2\times10^5$~yr and $2\times 10^6$~yr for Class~I protostars and Class~II pre-main sequence stars, respectively, in agreement with \cite{Evans09}, \cite{Venuti18} and references therein. 
Error bars on $\Sigma_{\rm object}/\tau_{\rm object}$ take into account statistical uncertainties, as well as an estimation of potentially undetected Class 0/Is in the central region and field contamination of Class II population \citep[see \cref{sub:nnd} and ][]{Buckner20}. 
The surface densities per lifetime of Class~IIs and Class~0/Is are similar in the northern subregion. 
This behavior is consistent with the idea that
a continuous star formation activity has developed in the northern part of NGC~2264 over the past few $10^6$ years. In contrast, in the central and southern subregions, the surface densities per lifetime are $3-5$ times larger for Class~0/I protostars than those measured for Class~IIs, suggesting an increase of the star formation activity over the past few $10^5$ years.

Identifying and characterizing the $\sim$0.1~pc clumps in NGC~2264, using its \emph{Herschel} column density image, allows, for the first time, to study the gas mass reservoir for current and future star formation in the NGC~2264 cloud. 
In particular, bound clumps represent structures that are forming or will likely form stars in the near future.
The clump lifetime is not as properly defined as the one of YSOs but, with a median density of $\sim\,8 \times 10^3$~cm$^{-3}$, they could live for one
free-fall time, i.e. $\sim\,3\times 10^5$~yrs. 
 With the goal to further characterize the relative star formation activity in the NGC~2264 cloud, \cref{fig:sigma-histo} displays the surface density per lifetime of bound clumps, as measured in the three subregions of NGC~2264.
We estimate that the $\Sigma_{\rm clump}/\tau_{\rm clump}$ values are uncertain by a factor of at least two.
Indeed, as the median densities of clumps vary by a factor of 2 depending on the subregions, the clump free-fall time is thus uncertain by a factor of $\sim\,50~\%$. Other sources of uncertainty
are the number of protostars that will form in each clump, here taken as one, and the number of free-fall times, here assumed to be one, required for these clumps to form stars.
With our current assumptions, the surface density per lifetime of clumps is similar to that of YSOs in the northern subregion, implying a continuous star formation activity. In contrast in the southern subregion, \cref{fig:sigma-histo} suggests
a tentative increase 
of the surface density per lifetime of clumps compared to that of the YSOs and therefore a possible increase of the star formation activity from $10^6$ years ago to the next few $10^5$ years. In the central subregion, uncertainties are too large to conclude that there is any trend for the future star formation activity relative to the current one.
All in one, the spatial distribution of YSO and clump populations suggests that the star formation activity will get
more and more intense as we go through the NGC~2264 cloud, from north to south, in the coming $\sim3 \times 10^5$ years.
Star formation first developed mainly in the northern subregion and continues today. Star formation is now extremely active in the central subregion while it was less so in the past.
Star formation now begins in the southern subregion and could get more active
while it was almost nonexistent in the past. 
This time sequence, based on the study of clump and YSO populations, agrees and extends the evolutionary sequence proposed by, e.g., \cite{Venuti18}.

We propose here that this sequence of star formation is related to a prior sequence of cloud formation, first in the north, now in the center and in the future in the south of NGC~2264. With its large mass reservoir at pc scale (see \cref{tab:pow_spec_fit}), the central subregion, and especially its ridge, has a very large potential for new events of star formation. The cloud power spectrum of the central subregion, and especially its coherent component associated to star formation, also implies that the cloud reservoir for star formation will remain important or will even increase in the coming $10^6$ years (see \cref{fig:gc-powspec}b). 
Indeed, the coherent component of the cloud contains almost all the power and thus most of the gas mass reservoir in the central subregion (see Figs.~\ref{fig:gc-powspec}a-b, \cref{fig:sigma-histo} and \cref{tab:pow_spec_fit}). Interestingly, this fraction of coherent cloud structures supposedly associated with star formation is orders of magnitude
larger in the central subregion of NGC~2264 than in the northern and southern subregions.
The southern subregion has similar properties with respect to its gas distribution (column density and power spectrum) than the northern subregion (see \cref{fig:gc-powspec}b and \cref{tab:statMDC}).
These observations of the gas thus show that the northern and southern subregions should have the same 
potential for future star formation events, i.e. similar numbers of stars that could form in the coming $\sim\,10^6$ years. This is also consistent with the similarity observed in term of mass, boundedness, and surface density (see \cref{fig:sigma-histo} and \cref{tab:statMDC}).
We therefore propose that the NGC~2264 cloud initially concentrated in the northern region several Myrs ago. 
The cloud distribution in the central subregion makes a strong case for extremely active cloud formation activity now and possibly in the coming $10^6$ years. 
As for the southern subregion, cloud is not yet highly concentrated but it could be enhanced in the close future.
Kinematical studies of the the NGC~2264 cloud are obviously necessary to confirm this last assertion.
 
Over the past few Myr, the NGC~2264 cloud has thus undergone several episodes of cloud and thus star formation.
We showed in \cref{sub:nnd} that a large fraction of clumps in the NGC~2264-IRS1 and IRS2 protoclusters are the parental cloud structures from which the Class 0/I protostars currently gather their mass. 
In addition, a population of clumps is observed at large, $\sim$1~pc, distances from the current sites of star-formation activity (see \cref{fig:MDC-YSO-map} and \cref{fig:nndist}b).
These clumps are smaller in mass and mostly unbound. They thus need to gather more mass and concentrate themselves before they could be considered as the cloud structures that will form new protostars. 
Future star-formation episodes could develop either aside the current star-formation sites or clumps must first be driven toward them. 
The population of clumps also appears to be spatially well correlated with Class~II pre-main sequence stars, surprisingly better than with Class~0/I protostars.
These elements argue for a dynamical scenario of star formation in NGC~2264.
We speculate that, first and simultaneously, the cloud undergoes global collapse and clumps converge toward the central gravity potentials, called ridges, reducing their distance to star-forming sites.
Protostellar collapse would then form a cluster of stars with a definite mass segregation at birth. The last phase would then be for at least part of the population of pre-main sequence stars to disperse from their original birth sites and spread over the entire extent of the NGC~2264 cloud, thus simulating a tight spatial correlation with a new generation of clumps. Whether the most massive and most clustered protostars, embedded within cloud ridges will become closely-packed main sequence stars will define if mass segregation is determined at birth in NGC~2264 and potentially in other dynamical clouds.
This scenario is consistent with the finding of \cite{Buckner20} that YSOs with increasing evolutionary stages tend to be less clustered. It would also qualitatively agree with numerical simulations of cloud formation through, e.g., the hierarchical global collapse \citep{Vazquez19} and stellar cluster evolution with the ejection of many YSOs members \citep{Oh16}.

\section{Conclusions}
\label{sec:conclu}
We used an \emph{Herschel} column density image of the NGC~2264 cloud 
to identify and characterize its clumps, study their spatial distribution, and make the link of these clumps with the YSO population and cloud structure. Our main results can be summarized as follows:

\begin{enumerate}
    \item We applied \textit{getsf} to the 18.2$\arcsec$-angular resolution column density image. We found a population of 256 clumps with $\sim$0.1~pc sizes, $\sim$14~K temperature and masses ranging from 0.08~$\Msol$ to 53~$\Msol$. 36\% of the NGC~2264 clumps are gravitationally bound following the Bonnor-Ebert criterium.
    
    \item After analyzing the cloud structure, we divided the NGC~2264 cloud into three subregions (northern, central, and southern) of roughly similar areas.
    The MnGseg technique revealed that the coherent cloud component associated to star formation dominates over the Gaussian component, especially for the central subregion.
    We qualified as a ridge the cloud structure at the center of the central subregion of NGC~2264 because it is massive and hosts the NGC~2264-IRS1 and IRS2 protoclusters.
 
    \item 
    We then applied the $nnd$ statistics to characterize the spatial distribution of clumps in the NGC~2264 cloud.
    The majority of the clumps are low-mass, unbound structures distributed all over the cloud. In addition, we found within the NGC~2264 ridge a cluster of higher-mass, bound clumps. 
    The typical distance between clumps in this second population is half that found elsewhere.

    \item 
    We quantified the mass segregation of NGC~2264 clumps using two complementary methods. 
    The 15 most massive clumps, with $M = 9.3-53~\Msol$ and all located in the central protoclusters, have a median local surface density, $\Sigma_{6}$, three times that of the entire clump population. This high degree of mass segregation is also indicated by the strong mass segregation ratio: $\Lambda_{\rm MSR} \simeq 8$ 
    for the $N_{\rm MST} \leq 15$ most massive clumps. We propose that this mass segregation is due to the presence and therefore to the formation of a ridge by the 
    global collapse of clouds in the central subregion of NGC~2264.
    
    \item 
    Except for those in the ridge, most of the clumps in the NGC~2264 cloud are not associated with Class 0/I protostars.
    At odd with what a parental link would suggest, $nnd$ distributions notably show that 
    the spatial distribution of clumps better correlates with that of Class~II pre-main sequence stars.
    This argues for a dynamical scenario where clumps are driven to the central parts of the NGC~2264, protostars are formed in situ, and some Class~IIs are ejected from their birth site.
    
    \item 
    Our combined study of clump and YSO populations in NGC~2264 agrees and extends the sequential star formation sequence from north to south, which was initially proposed by YSO studies. Star formation first developed in the northern subregion. It is now extremely active in the central subregion and should soon begin in the southern subregion. We propose to explain it by a sequence of cloud formation from north to south.
    
\end{enumerate}

The strong mass segregation observed for the NGC~2264 clumps, which is most likely due to the formation of a favorable cloud structure, the ridge, could itself be at the origin of future mass segregation in the NGC~2264 stellar cluster. Detailed studies of the cloud and YSO kinematics are mandatory to confirm our current interpretation and predictions. 

 
\begin{acknowledgements}
This project has received funding from the European Union’s Horizon 2020 research and innovation programs StarFormMapper under grant agreement No 687528 and Filaments-to-stars
under the Marie Skłodowska-Curie Grant Agreement No. 750920. This project was also supported by the Programme National de Physique Stellaire and Physique et Chimie du Milieu Interstellaire (PNPS and PCMI) of CNRS/INSU (with INC/INP/IN2P3) co-funded by CEA and CNES. This research has made use of data from the Herschel imaging survey of OB young stellar objects (HOBYS) project (http:// hobys-herschel.cea.fr). HOBYS is a \emph{Herschel} Key Program jointly carried out by SPIRE Specialist Astronomy Group 3 (SAG 3), scientists of several institutes in the PACS Consortium (LAM Marseille and CEA Saclay), and scientists of the Herschel Science Center (HSC).
\end{acknowledgements}

\bibliographystyle{aa}
\bibliography{NGC2264}   

\begin{appendix}

\renewcommand{\thefigure}{A\arabic{figure}}

\setcounter{figure}{0}
\vspace{4cm }

\begin{figure*}
    \includegraphics[width=1\hsize]{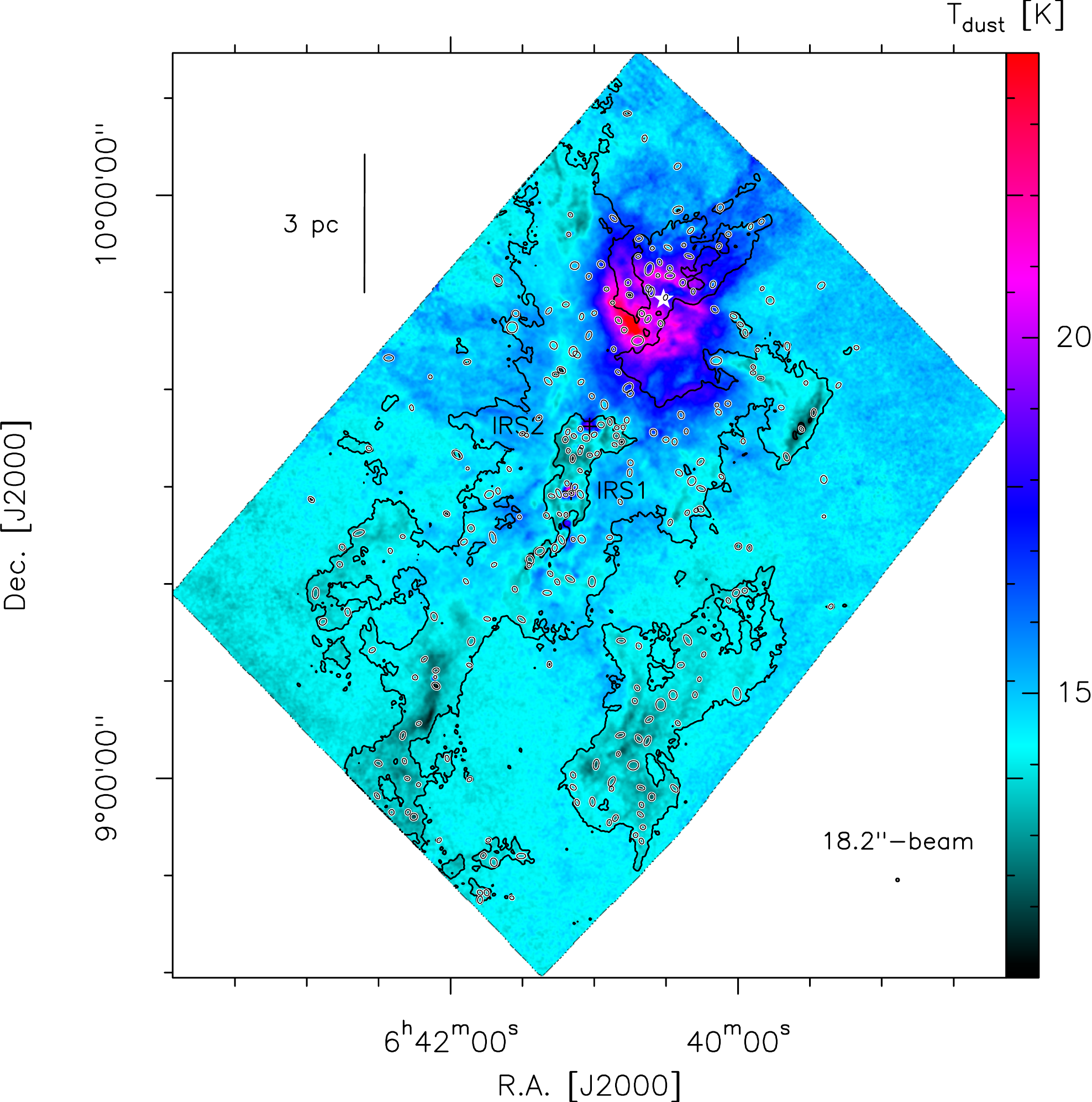}
    \caption{\emph{Herschel} dust temperature image of the NGC~2264 cloud, with $3.4\times 10^{21}$~H$_2$\,cm$^{-2}$ and $1.6\times 10^{22}$~H$_2$\,cm$^{-2}$ contours taken from the column density image of \cref{fig:map-MDC}. 
    The resolution of the map, 18.2\arcsec, is shown in the lower-right corner and a scale bar is given in the upper-left corner.
    A white star locates the OB star cluster which heats its surrounding cloud over several parsecs. The IRS1 and IRS2 protoclusters are also indicated.
    }
    \label{fig:map-temp}
\end{figure*}

\begin{figure*}
    \centering
    \includegraphics[width=0.525\hsize]{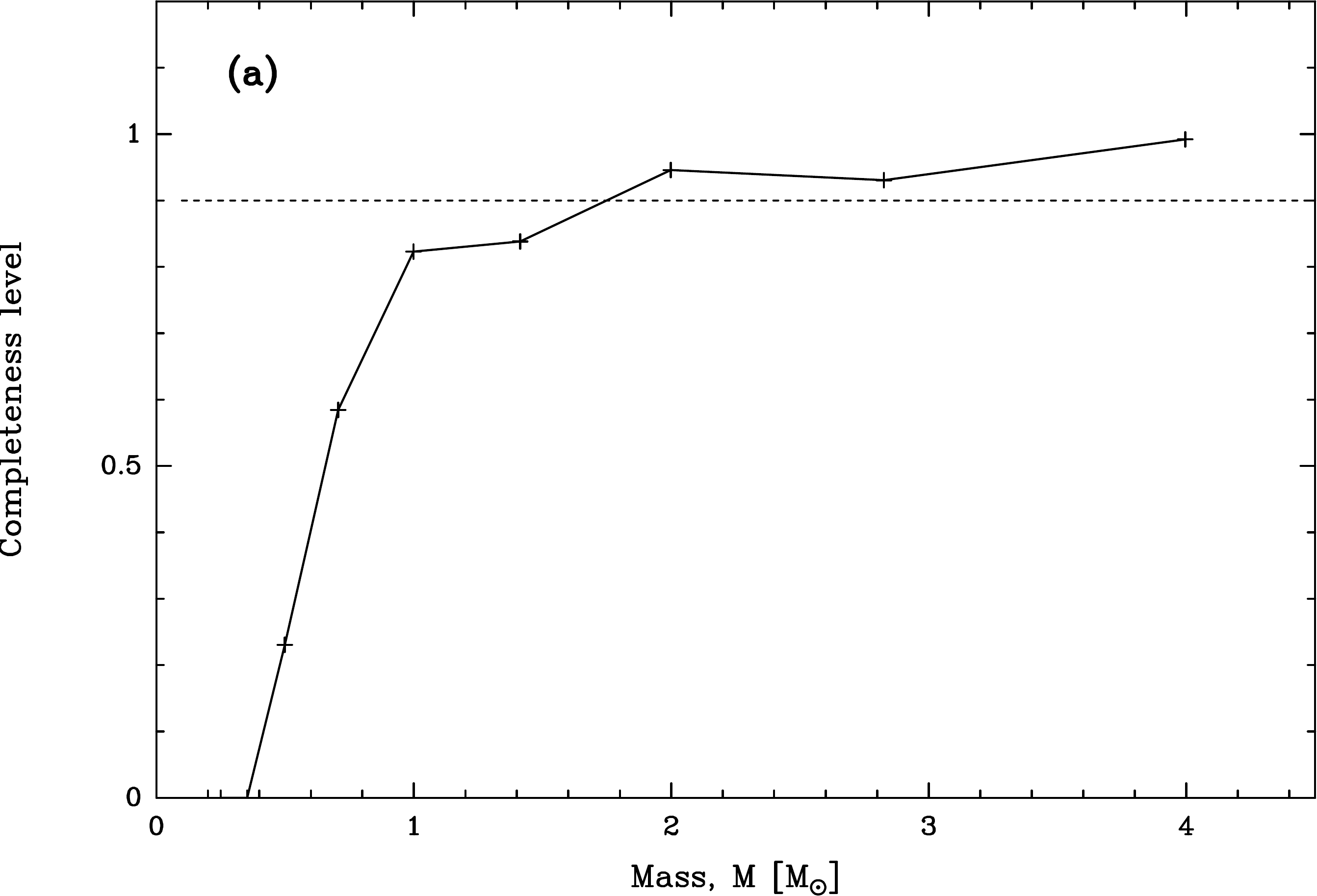}
    \includegraphics[width=0.47\hsize]{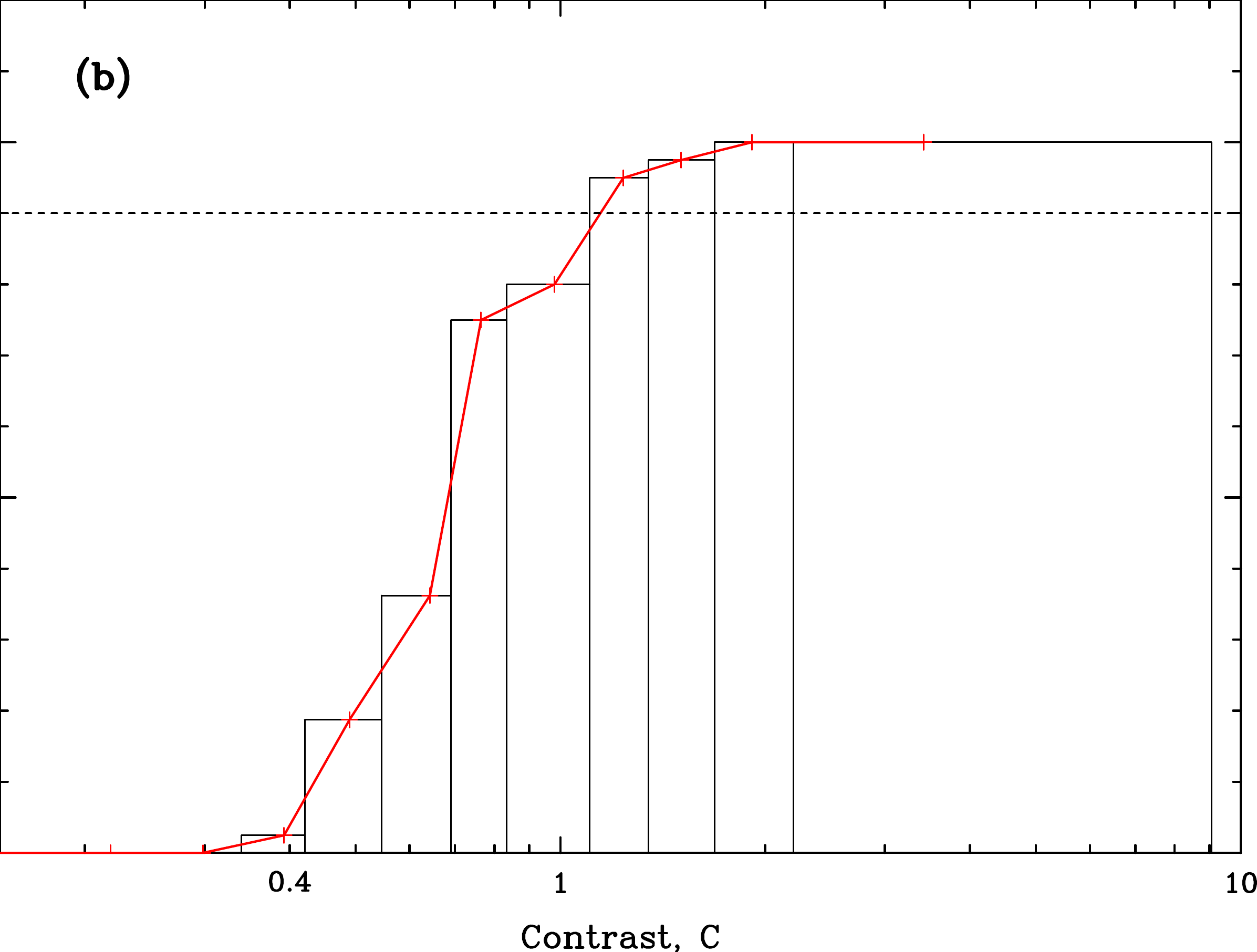}
    \caption{
    Completeness level of the clumps extraction 
    as a function of the clump mass (in \textbf{a}) and of the contrast of the clump column density over its local background 
    (in \textbf{b}). The 90\% completeness level is indicated with a dashed horizontal line. 
    \textbf{a)} The clump sample is 90\% complete above 1.7~\Msol.
    \textbf{b)} The histogram has been constructed with a constant number of 80 synthetic sources in the 11 first bins, plus a bin containing the remaining 290 sources which have all been detected. The red curve connects the median contrast value in each bin. 90\% of the clump are detected for a contrast larger than 1.1.
    }
    \label{fig:comp}
\end{figure*} 

\begin{figure*}
    \centering
    \includegraphics[width=0.47\hsize]{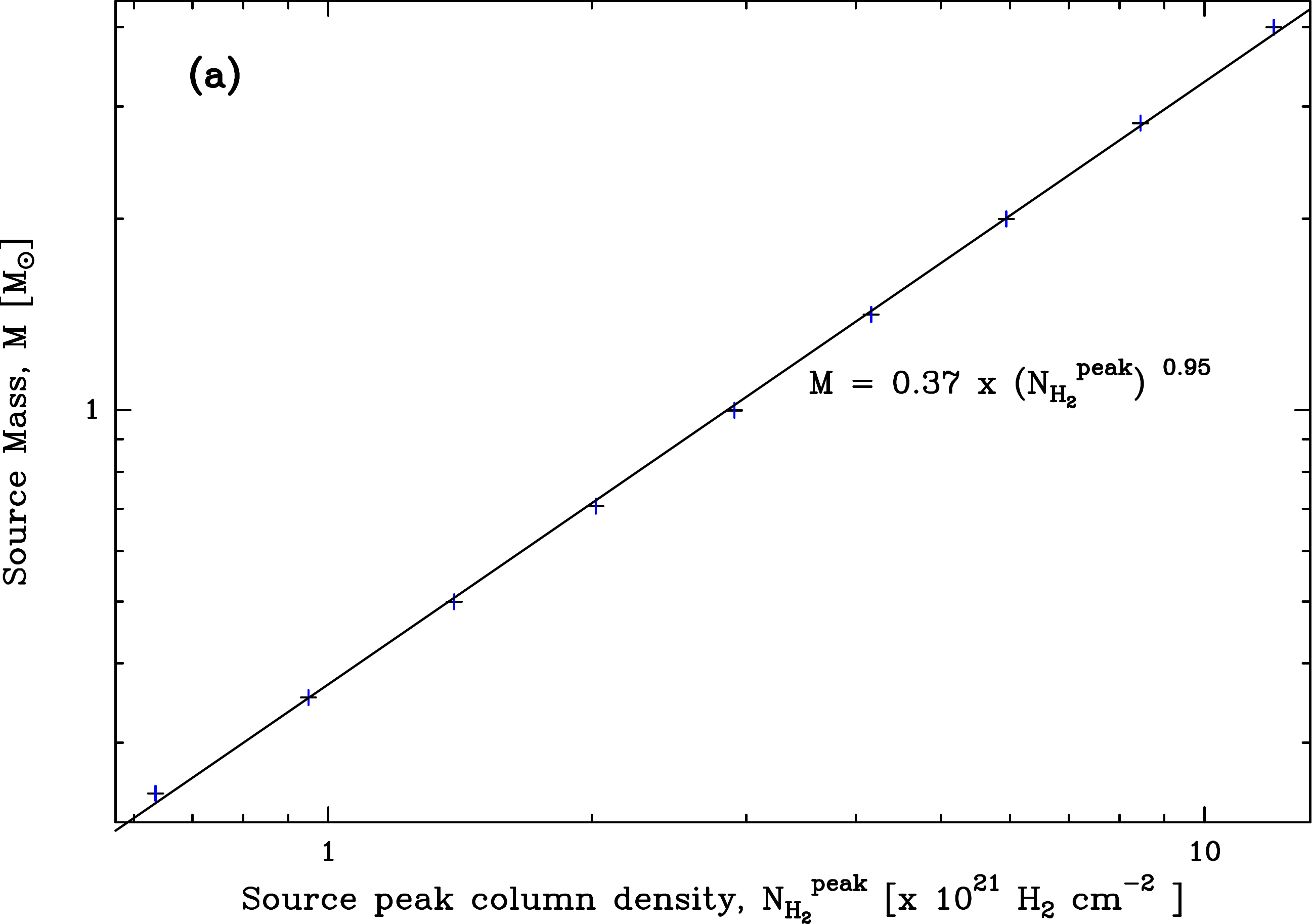}\hskip 0.9cm
    \includegraphics[width=0.47\hsize]{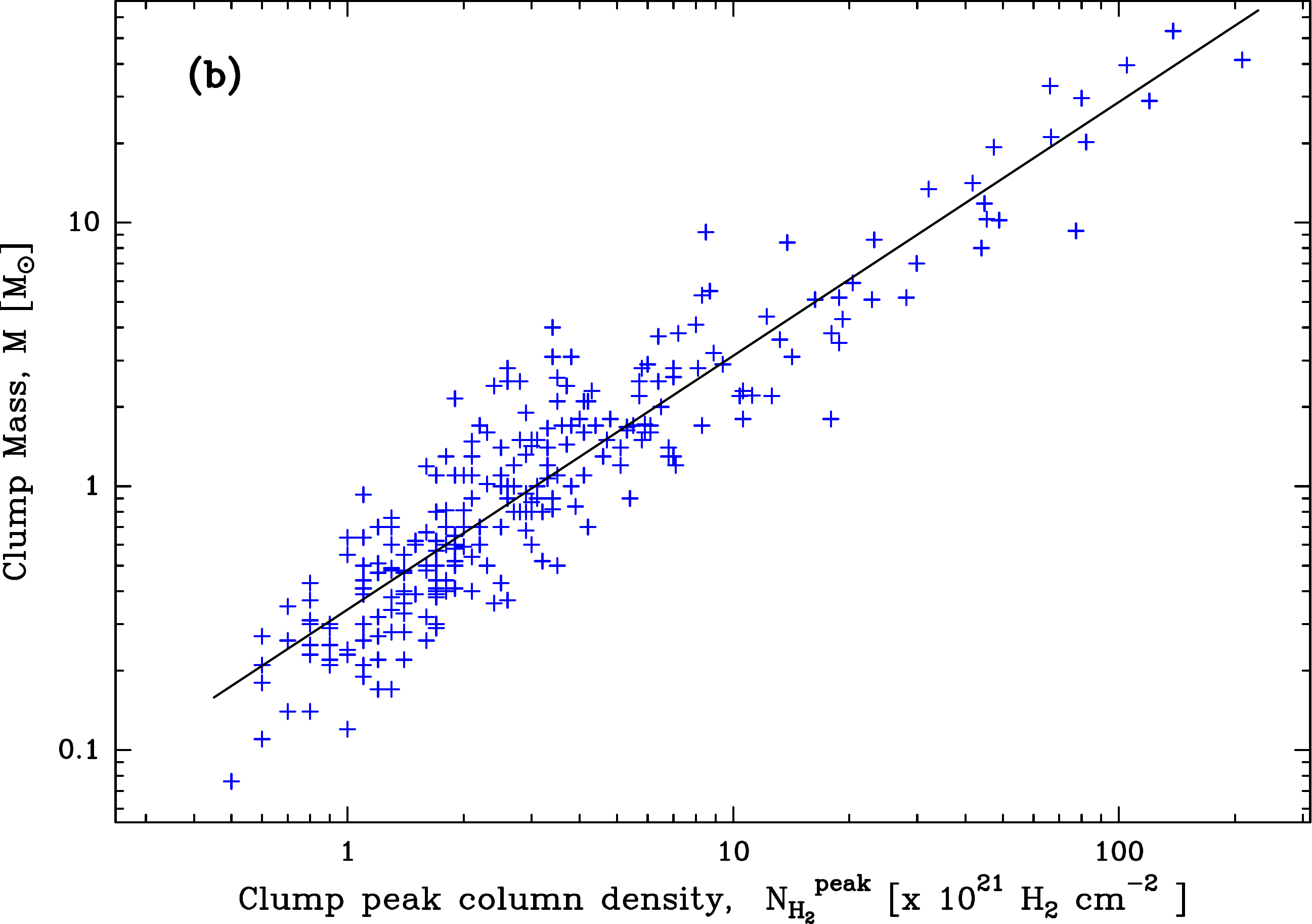}
    \caption{
    In relation with the intensity profile of clumps,
    clump masses are plotted against the peak column density \textbf{a)} for synthetic sources and \textbf{b)} for observed clumps. A simple regression gives the relation $M = \beta (N_{\rm H2}^{\rm peak})^{0.95}$ with $\beta= 0.37$ in \textbf{a} over a [$0.6-12\times 10^{21}$~cm$^{-2}$] range, $\beta= 0.34$ in \textbf{b} over a [$0.5-200\times 10^{21}$~cm$^{-2}$].
    The similarity in the fitted parameters confirms that the synthetic sources correctly represent the observed clumps.
    }
    \label{fig:Mpic}
\end{figure*}

\begin{figure}[h!]
    \centering
    \includegraphics[width=0.9\hsize]{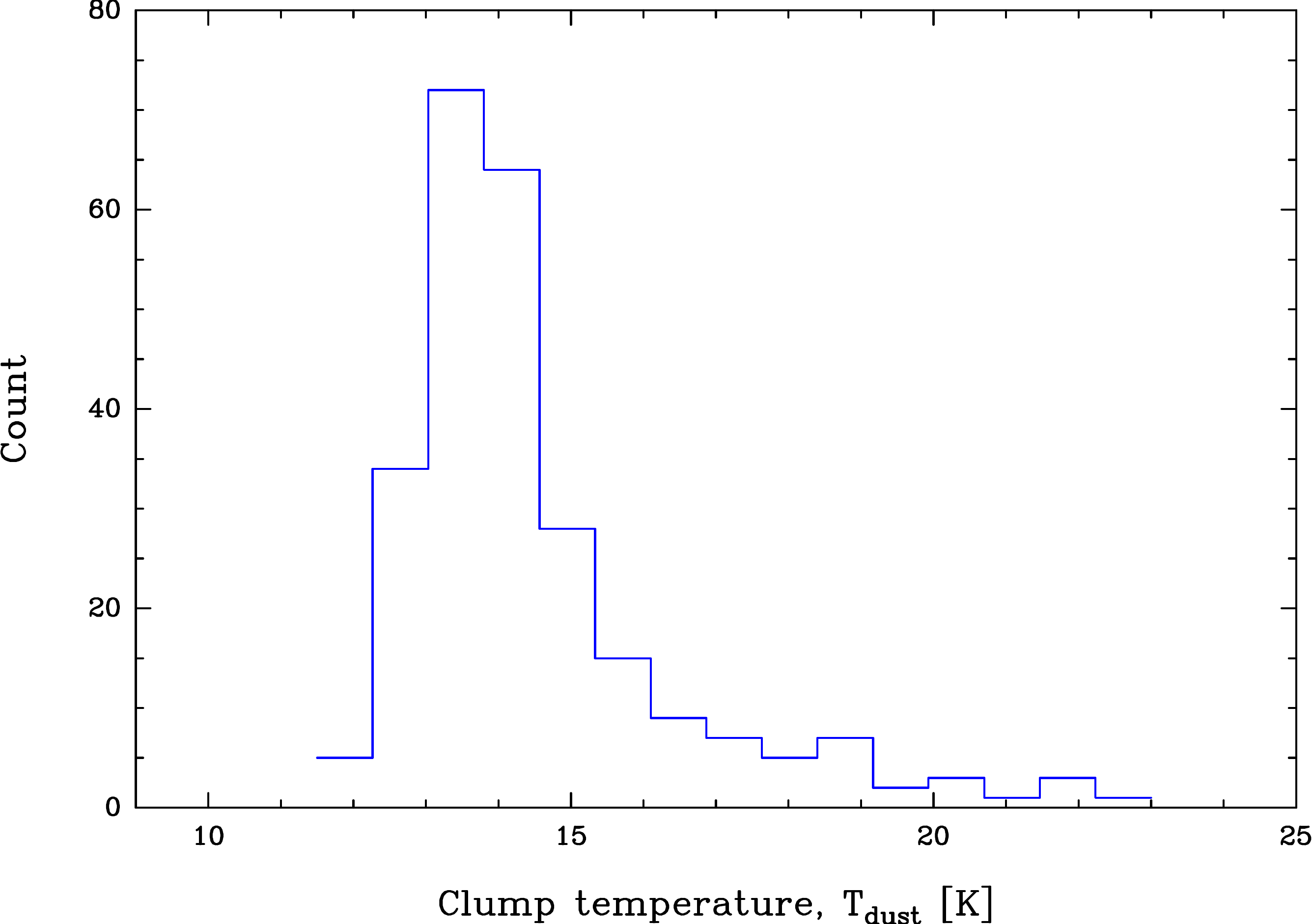}
    \caption{Temperature distribution of the 256 clumps detected in NGC 2264. The temperature dispersion between clumps in small, 80\% of the clumps having a temperature between 
    12.7 and 16.9~K}
    \label{fig:temp-dist}
\end{figure}

\begin{figure}[h!]
    \centering
    \includegraphics[width=0.95\hsize]{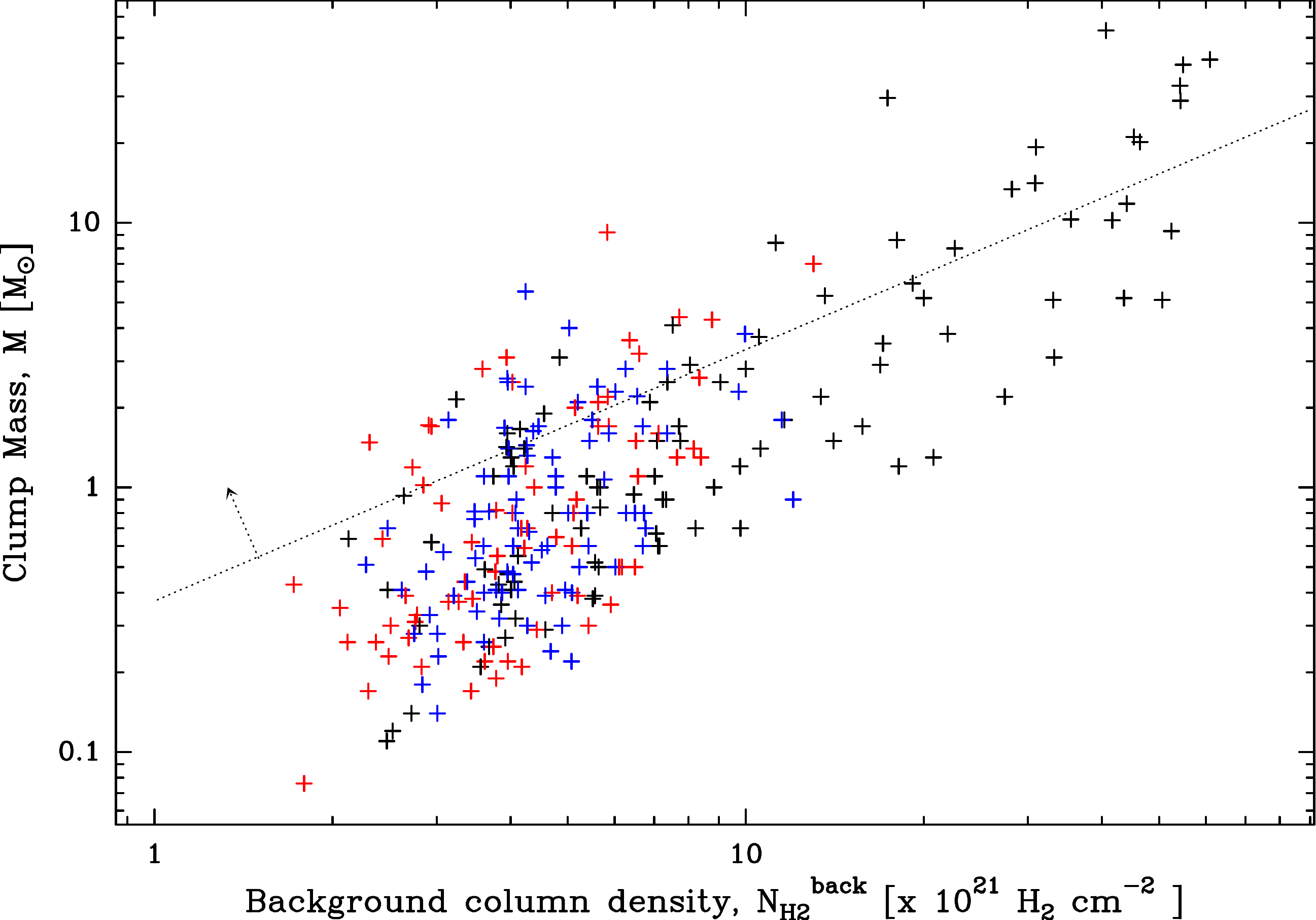}
    \caption{A correlation is observed between the clump mass and its background column density. The color code highlighting the spatial location of the clumps is the same as in \cref{fig:mass-reg}. The completeness curve in dotted line is calculated using a 90\% completeness contrast of 1.1 and the M-$N_{\rm H2}^{\rm peak}$ relation (see \cref{fig:comp,fig:Mpic}): $M=0.34\,(1.1 \times N_{\rm H2}^{\rm back})^{0.95}$  }
    \label{fig:MNH2-reg}
\end{figure}

\renewcommand{\thetable}{A\arabic{table}}
\begin{table*}[h!]
	\small	
   \caption[]{Catalog of clumps.}
   \makebox[\textwidth][c]{
\begin{tabular}{|c|c|c|c|c|c|c|c|c|c|c|c|c|} 
\hline    
\#    &     Alp  &   Del  &    $A \times B$     &     PA        &      Sig    &   $N^{\rm peak}_{\rm H_2}$  &   $N^{\rm int}_{\rm H_2}$  &    FWHM$_{\rm dec}\tablefootmark{b} $       &   $T_{\rm dust}$  &    $ M$   &   $\alpha_{\rm BE}$  &   $N^{\rm back}_{\rm H_2}$\\
          &  [J2000]         &  [J2000]            &       [$" \times "$]     &  [$\degree$]  &       &   [Av units]\tablefootmark{a} & [Av units]\tablefootmark{a} & [pc]      &   [K]    &  [$\Msol$]&      &  [Av units]\tablefootmark{a} \\
\hline    
1  &      100.5263   &   9.1742   &     $18.2 \times 18.2$   &    18.3   &     80.5   &    $18 \pm 2$    &    $14 \pm 2$   &     0.03   &    $20 \pm 2$    &    $1.8 \pm 0.2$    &   0.3   &  $11.5 \pm 0.2$  \\
2  &      100.2023   &   9.5775   &     $36.6 \times 32.4$   &    152.2   &     174.4   &    $80 \pm 6$    &    $286 \pm 10$   &     0.1   &    $11.8 \pm 0.3$    &    $30 \pm 1$    &   0.03   &  $17.4 \pm 0.6$  \\
3  &      100.2944   &   9.9336   &     $27.4 \times 26.2$   &    24.5   &     104.8   &    $19 \pm 2$    &    $41 \pm 2$   &     0.07   &    $15.7 \pm 0.7$    &    $4.3 \pm 0.2$    &   0.21   &  $8.8 \pm 0.2$  \\
4  &      99.8894   &   9.5995   &     $33.6 \times 23.2$   &    44.3   &     94.8   &    $30 \pm 3$    &    $68 \pm 4$   &     0.07   &    $11.5 \pm 0.4$    &    $7.0 \pm 0.4$    &   0.1   &  $13.0 \pm 0.3$  \\
5  &      100.2081   &   9.0423   &     $30.1 \times 23.4$   &    168.7   &     76.2   &    $11.2 \pm 0.8$    &    $21.0 \pm 0.9$   &     0.07   &    $12.3 \pm 0.2$    &    $2.2 \pm 0.1$    &   0.31   &  $6.5 \pm 0.1$  \\
6  &      100.2997   &   9.4868   &     $26.5 \times 23.8$   &    93.5   &     64.4   &    $209 \pm 12$    &    $400 \pm 15$   &     0.06   &    $17.3 \pm 0.7$    &    $41 \pm 2$    &   0.02   &  $61 \pm 1$  \\
7  &      100.2741   &   9.5998   &     $31.3 \times 26.3$   &    7.2   &     56.5   &    $82 \pm 11$    &    $195 \pm 14$   &     0.08   &    $15.0 \pm 0.4$    &    $20 \pm 1$    &   0.05   &  $46 \pm 3$  \\
8  &      100.272   &   9.5686   &     $22.8 \times 22.8$   &    22.6   &     47.7   &    $77 \pm 11$    &    $90 \pm 13$   &     0.05   &    $15.4 \pm 0.7$    &    $9 \pm 1$    &   0.07   &  $52 \pm 2$  \\
9  &      100.3104   &   9.4536   &     $29.5 \times 20.2$   &    48.4   &     50.1   &    $44 \pm 8$    &    $77 \pm 9$   &     0.06   &    $13.3 \pm 0.4$    &    $8.0 \pm 0.9$    &   0.08   &  $22.6 \pm 0.6$  \\
10  &      100.5549   &   9.0945   &     $31.8 \times 22.5$   &    115.5   &     42.3   &    $11 \pm 2$    &    $22 \pm 3$   &     0.07   &    $12.6 \pm 0.4$    &    $2.3 \pm 0.3$    &   0.31   &  $9.7 \pm 0.5$  \\
11  &      99.9603   &   9.6866   &     $40.8 \times 24.9$   &    93.2   &     45.8   &    $13 \pm 1$    &    $35 \pm 1$   &     0.09   &    $12.2 \pm 0.2$    &    $3.6 \pm 0.1$    &   0.26   &  $6.4 \pm 0.2$  \\
12  &      99.9953   &   9.779   &     $38.2 \times 25.4$   &    56.7   &     46.9   &    $5.9 \pm 0.2$    &    $17.0 \pm 0.3$   &     0.09   &    $13.5 \pm 0.3$    &    $1.7 \pm 0.0$    &   0.58   &  $2.9 \pm 0.0$  \\
13  &      100.2838   &   9.4998   &     $41.0 \times 32.3$   &    153.3   &     48.6   &    $138 \pm 13$    &    $514 \pm 21$   &     0.11   &    $16 \pm 2$    &    $53 \pm 2$    &   0.03   &  $41 \pm 3$  \\
14  &      100.25   &   9.7985   &     $31.9 \times 27.8$   &    129.0   &     40.5   &    $5.9 \pm 0.9$    &    $15 \pm 1$   &     0.08   &    $15.1 \pm 0.3$    &    $1.6 \pm 0.1$    &   0.65   &  $7.1 \pm 0.2$  \\
15  &      100.224   &   8.9245   &     $38.6 \times 28.0$   &    174.2   &     44.6   &    $5.3 \pm 0.4$    &    $16.0 \pm 0.6$   &     0.1   &    $13.0 \pm 0.2$    &    $1.7 \pm 0.1$    &   0.61   &  $3.9 \pm 0.1$  \\
16  &      100.2746   &   9.5619   &     $46.7 \times 26.6$   &    174.2   &     32.2   &    $67 \pm 11$    &    $204 \pm 15$   &     0.11   &    $13.7 \pm 0.7$    &    $21 \pm 2$    &   0.06   &  $45 \pm 3$  \\
17  &      100.2961   &   9.5946   &     $37.4 \times 25.9$   &    166.2   &     31.2   &    $46 \pm 13$    &    $100 \pm 13$   &     0.09   &    $14.0 \pm 0.9$    &    $10 \pm 1$    &   0.1   &  $35 \pm 3$  \\
18  &      100.1488   &   9.9097   &     $23.8 \times 22.8$   &    113.5   &     31.8   &    $5 \pm 1$    &    $9 \pm 2$   &     0.05   &    $15.8 \pm 0.4$    &    $0.9 \pm 0.2$    &   0.74   &  $5.2 \pm 0.3$  \\
19  &      100.1259   &   9.8254   &     $37.1 \times 24.9$   &    155.3   &     39.4   &    $6 \pm 1$    &    $16 \pm 2$   &     0.09   &    $16.5 \pm 0.3$    &    $1.7 \pm 0.2$    &   0.69   &  $5.9 \pm 0.2$  \\
20  &      100.3121   &   9.4895   &     $57.4 \times 27.1$   &    25.2   &     36.5   &    $105 \pm 13$    &    $382 \pm 19$   &     0.12   &    $14.3 \pm 0.4$    &    $40 \pm 2$    &   0.04   &  $55 \pm 3$  \\
21  &      99.9869   &   9.7646   &     $46.5 \times 29.6$   &    15.4   &     36.3   &    $4.4 \pm 0.3$    &    $16.0 \pm 0.5$   &     0.11   &    $13.3 \pm 0.3$    &    $1.7 \pm 0.1$    &   0.73   &  $2.9 \pm 0.1$  \\
22  &      100.2898   &   9.4898   &     $33.9 \times 26.9$   &    47.0   &     26.1   &    $120 \pm 13$    &    $280 \pm 14$   &     0.08   &    $22 \pm 1$    &    $29 \pm 2$    &   0.05   &  $54 \pm 2$  \\
23  &      100.1215   &   9.9109   &     $48.7 \times 26.4$   &    63.9   &     45.7   &    $9 \pm 1$    &    $31 \pm 2$   &     0.11   &    $15.0 \pm 0.2$    &    $3.2 \pm 0.2$    &   0.42   &  $6.6 \pm 0.2$  \\
24$^*$  &      100.4425   &   8.868   &     $38.6 \times 31.3$   &    143.8   &     47.3   &    $4.8 \pm 0.4$    &    $17.0 \pm 0.8$   &     0.1   &    $12.6 \pm 0.2$    &    $1.8 \pm 0.1$    &   0.6   &  $3.1 \pm 0.1$  \\
25  &      100.2921   &   9.3425   &     $61.6 \times 33.6$   &    59.7   &     54.6   &    $8 \pm 1$    &    $39 \pm 2$   &     0.15   &    $13.4 \pm 0.1$    &    $4.1 \pm 0.2$    &   0.4   &  $7.5 \pm 0.3$  \\
26  &      100.238   &   9.6068   &     $40.9 \times 34.9$   &    82.4   &     38.8   &    $47 \pm 13$    &    $186 \pm 21$   &     0.12   &    $13.6 \pm 0.6$    &    $19 \pm 2$    &   0.07   &  $31 \pm 4$  \\
27  &      100.3241   &   9.4849   &     $38.9 \times 24.1$   &    75.9   &     28.5   &    $49 \pm 13$    &    $98 \pm 15$   &     0.09   &    $13.4 \pm 0.3$    &    $10 \pm 2$    &   0.09   &  $42 \pm 4$  \\
28  &      100.3172   &   9.6934   &     $42.1 \times 29.6$   &    80.3   &     34.0   &    $8 \pm 2$    &    $27 \pm 3$   &     0.11   &    $13.9 \pm 0.2$    &    $2.8 \pm 0.3$    &   0.44   &  $10.0 \pm 0.3$  \\
29  &      100.1688   &   8.9845   &     $48.3 \times 28.3$   &    73.7   &     40.7   &    $7.0 \pm 0.9$    &    $27 \pm 1$   &     0.11   &    $12.4 \pm 0.1$    &    $2.8 \pm 0.1$    &   0.41   &  $7.4 \pm 0.1$  \\
30  &      100.3801   &   9.449   &     $26.5 \times 22.8$   &    25.5   &     25.6   &    $2.5 \pm 0.8$    &    $4.0 \pm 0.9$   &     0.06   &    $15.6 \pm 0.3$    &    $0.4 \pm 0.1$    &   1.7   &  $3.8 \pm 0.1$  \\
31  &      100.0337   &   9.6287   &     $39.7 \times 31.8$   &    36.3   &     38.8   &    $5.7 \pm 0.9$    &    $21 \pm 1$   &     0.11   &    $14.7 \pm 0.2$    &    $2.2 \pm 0.1$    &   0.59   &  $5.8 \pm 0.1$  \\
32  &      99.9659   &   9.7352   &     $22.7 \times 21.9$   &    22.4   &     25.5   &    $2.6 \pm 0.8$    &    $4.0 \pm 0.9$   &     0.05   &    $13.8 \pm 0.2$    &    $0.4 \pm 0.1$    &   1.4   &  $3.3 \pm 0.1$  \\
33  &      100.5634   &   8.9344   &     $46.5 \times 44.0$   &    95.3   &     50.8   &    $8.7 \pm 0.7$    &    $53 \pm 2$   &     0.15   &    $12.4 \pm 0.2$    &    $5.5 \pm 0.2$    &   0.27   &  $4.2 \pm 0.2$  \\
34  &      100.1253   &   9.5813   &     $38.9 \times 33.2$   &    98.1   &     34.3   &    $6.4 \pm 0.8$    &    $24 \pm 1$   &     0.11   &    $15.1 \pm 0.3$    &    $2.5 \pm 0.1$    &   0.54   &  $7.4 \pm 0.1$  \\
35  &      100.0074   &   9.7932   &     $43.4 \times 28.3$   &    65.5   &     27.8   &    $3.0 \pm 0.2$    &    $8.0 \pm 0.2$   &     0.11   &    $14.2 \pm 0.3$    &    $0.9 \pm 0.0$    &   1.4   &  $3.1 \pm 0.1$  \\
36  &      100.2145   &   9.5727   &     $33.2 \times 24.0$   &    61.2   &     22.6   &    $18 \pm 7$    &    $36 \pm 8$   &     0.08   &    $13.1 \pm 0.3$    &    $3.8 \pm 0.8$    &   0.22   &  $21.9 \pm 0.6$  \\
37  &      100.1184   &   9.8753   &     $33.0 \times 31.5$   &    18.7   &     28.9   &    $6 \pm 1$    &    $16 \pm 2$   &     0.09   &    $17.8 \pm 0.5$    &    $1.7 \pm 0.2$    &   0.81   &  $5.6 \pm 0.2$  \\
38  &      100.2994   &   9.8134   &     $42.3 \times 31.2$   &    96.0   &     28.2   &    $7 \pm 1$    &    $25 \pm 2$   &     0.11   &    $13.8 \pm 0.2$    &    $2.6 \pm 0.2$    &   0.48   &  $8.3 \pm 0.2$  \\
39  &      100.2034   &   9.2366   &     $54.4 \times 31.1$   &    86.2   &     38.4   &    $4.2 \pm 0.5$    &    $20.0 \pm 0.9$   &     0.13   &    $13.0 \pm 0.2$    &    $2.1 \pm 0.1$    &   0.66   &  $5.2 \pm 0.1$  \\
40  &      100.2644   &   9.5847   &     $47.0 \times 42.3$   &    75.8   &     27.5   &    $66 \pm 13$    &    $318 \pm 19$   &     0.14   &    $13.8 \pm 0.4$    &    $33 \pm 2$    &   0.05   &  $54 \pm 3$  \\
41  &      100.6272   &   8.9705   &     $43.1 \times 25.2$   &    38.8   &     34.5   &    $5.3 \pm 0.5$    &    $16.0 \pm 0.6$   &     0.1   &    $12.6 \pm 0.2$    &    $1.6 \pm 0.1$    &   0.62   &  $4.4 \pm 0.2$  \\
42  &      100.4702   &   9.532   &     $19.4 \times 19.4$   &    4.4   &     19.8   &    $1.0 \pm 0.1$    &    $1.0 \pm 0.1$   &     0.03   &    $13.8 \pm 0.2$    &    $0.1 \pm 0.0$    &   3.0   &  $2.5 \pm 0.0$  \\
43  &      100.0651   &   9.5435   &     $49.8 \times 34.5$   &    135.0   &     34.2   &    $3.3 \pm 0.5$    &    $16.0 \pm 0.8$   &     0.13   &    $13.5 \pm 0.2$    &    $1.7 \pm 0.1$    &   0.88   &  $4.2 \pm 0.0$  \\
44  &      100.0434   &   9.8478   &     $42.8 \times 31.7$   &    161.1   &     24.5   &    $3.3 \pm 0.9$    &    $12 \pm 1$   &     0.11   &    $16.2 \pm 0.3$    &    $1.2 \pm 0.1$    &   1.3   &  $4.2 \pm 0.1$  \\
45  &      100.075   &   9.1437   &     $38.5 \times 28.8$   &    175.8   &     23.3   &    $3.3 \pm 0.6$    &    $10.0 \pm 0.9$   &     0.1   &    $13.1 \pm 0.1$    &    $1.1 \pm 0.1$    &   0.99   &  $5.8 \pm 0.1$  \\
46  &      100.2876   &   9.5489   &     $50.5 \times 30.1$   &    173.9   &     21.5   &    $42 \pm 6$    &    $137 \pm 8$   &     0.12   &    $11.8 \pm 0.2$    &    $14.1 \pm 0.8$    &   0.08   &  $31 \pm 1$  \\
47  &      100.1391   &   9.8632   &     $22.8 \times 22.8$   &    116.8   &     18.3   &    $4 \pm 1$    &    $5 \pm 1$   &     0.05   &    $19.8 \pm 0.3$    &    $0.5 \pm 0.1$    &   1.6   &  $6.5 \pm 0.2$  \\
48  &      100.2183   &   9.8165   &     $23.6 \times 20.1$   &    70.1   &     17.4   &    $1.3 \pm 0.6$    &    $2.0 \pm 0.6$   &     0.04   &    $22.7 \pm 0.7$    &    $0.2 \pm 0.1$    &   4.7   &  $3.4 \pm 0.1$  \\
49  &      100.0822   &   9.9404   &     $47.0 \times 26.7$   &    136.8   &     20.7   &    $7 \pm 1$    &    $19 \pm 1$   &     0.11   &    $14.7 \pm 0.2$    &    $2.0 \pm 0.1$    &   0.65   &  $5.1 \pm 0.1$  \\
50  &      100.2993   &   9.4523   &     $31.3 \times 24.7$   &    175.5   &     18.4   &    $19 \pm 5$    &    $34 \pm 5$   &     0.07   &    $13.2 \pm 0.4$    &    $3.5 \pm 0.6$    &   0.23   &  $17.1 \pm 0.7$  \\
51  &      100.1999   &   9.6023   &     $32.3 \times 24.7$   &    12.9   &     18.1   &    $10 \pm 3$    &    $21 \pm 3$   &     0.08   &    $13.8 \pm 0.4$    &    $2.2 \pm 0.3$    &   0.39   &  $13.4 \pm 0.4$  \\
52  &      100.2532   &   8.9604   &     $57.3 \times 31.9$   &    4.4   &     20.2   &    $3.3 \pm 0.6$    &    $14.0 \pm 0.8$   &     0.14   &    $13.3 \pm 0.2$    &    $1.4 \pm 0.1$    &   1.1   &  $4.0 \pm 0.2$  \\
53  &      100.1503   &   8.9688   &     $48.3 \times 41.0$   &    176.3   &     21.7   &    $6 \pm 1$    &    $27 \pm 2$   &     0.14   &    $12.6 \pm 0.1$    &    $2.8 \pm 0.2$    &   0.53   &  $6.3 \pm 0.3$  \\
54  &      100.1049   &   9.0951   &     $34.1 \times 26.5$   &    43.2   &     20.4   &    $2.5 \pm 0.8$    &    $7 \pm 1$   &     0.08   &    $13.2 \pm 0.1$    &    $0.7 \pm 0.1$    &   1.3   &  $4.1 \pm 0.2$  \\
55  &      100.3751   &   9.5917   &     $34.9 \times 25.5$   &    111.1   &     17.1   &    $3.9 \pm 0.7$    &    $8.0 \pm 0.7$   &     0.08   &    $14.5 \pm 0.2$    &    $0.8 \pm 0.1$    &   1.2   &  $5.7 \pm 0.0$  \\
56  &      100.118   &   9.4395   &     $41.1 \times 27.4$   &    162.4   &     17.6   &    $4 \pm 1$    &    $10 \pm 2$   &     0.1   &    $13.5 \pm 0.3$    &    $1.1 \pm 0.2$    &   1.0   &  $5.4 \pm 0.1$  \\
57  &      100.3073   &   9.7007   &     $55.8 \times 39.1$   &    66.0   &     23.9   &    $6 \pm 2$    &    $36 \pm 3$   &     0.15   &    $14.0 \pm 0.2$    &    $3.7 \pm 0.3$    &   0.47   &  $10.5 \pm 0.4$  \\
58  &      100.2719   &   9.7925   &     $40.7 \times 25.5$   &    168.4   &     19.7   &    $4.6 \pm 0.8$    &    $12 \pm 1$   &     0.09   &    $14.5 \pm 0.1$    &    $1.3 \pm 0.1$    &   0.86   &  $8.4 \pm 0.2$  \\
59  &      100.2287   &   9.8397   &     $29.6 \times 25.1$   &    92.0   &     21.3   &    $1.8 \pm 0.7$    &    $4 \pm 1$   &     0.07   &    $19.1 \pm 0.5$    &    $0.4 \pm 0.1$    &   2.8   &  $4.7 \pm 0.1$  \\
60  &      100.5452   &   9.2047   &     $42.5 \times 27.9$   &    129.7   &     17.7   &    $6 \pm 3$    &    $16 \pm 3$   &     0.1   &    $12.7 \pm 0.3$    &    $1.6 \pm 0.3$    &   0.67   &  $7.4 \pm 0.6$  \\
61  &      100.4242   &   9.3669   &     $35.7 \times 24.6$   &    119.9   &     18.9   &    $4 \pm 1$    &    $9 \pm 1$   &     0.08   &    $14.1 \pm 0.1$    &    $1.0 \pm 0.1$    &   0.96   &  $5.7 \pm 0.1$  \\
62  &      100.3618   &   9.3753   &     $52.8 \times 41.7$   &    30.6   &     29.5   &    $14 \pm 2$    &    $81 \pm 4$   &     0.15   &    $12.7 \pm 0.2$    &    $8.4 \pm 0.4$    &   0.19   &  $11.2 \pm 0.3$  \\
63  &      100.0883   &   9.4561   &     $34.2 \times 27.1$   &    151.5   &     20.9   &    $3 \pm 1$    &    $8 \pm 2$   &     0.09   &    $13.4 \pm 0.2$    &    $0.8 \pm 0.2$    &   1.2   &  $4.7 \pm 0.2$  \\
64  &      100.2161   &   9.7366   &     $31.7 \times 26.4$   &    7.5   &     21.2   &    $1.4 \pm 0.4$    &    $4.0 \pm 0.6$   &     0.08   &    $17.5 \pm 0.2$    &    $0.4 \pm 0.1$    &   3.2   &  $3.9 \pm 0.1$  \\
65  &      99.9776   &   9.9467   &     $37.4 \times 24.4$   &    153.7   &     18.3   &    $3.4 \pm 0.8$    &    $8.0 \pm 0.9$   &     0.08   &    $14.5 \pm 0.2$    &    $0.8 \pm 0.1$    &   1.2   &  $3.8 \pm 0.1$  \\
\hline   
\end{tabular} 
 } \\
   \label{tab:clumps-long}
 \end{table*}

   \setcounter{table}{0}
   \begin{table*}
      \small	
   \caption[]{Catalog of clumps (continued).}
   \makebox[\textwidth][c]{
\begin{tabular}{|c|c|c|c|c|c|c|c|c|c|c|c|c|} 
\hline    
\#    &     Alp  &   Del  &    $A \times B$     &     PA        &      Sig    &   $N^{\rm peak}_{\rm H_2}$  &   $N^{\rm int}_{\rm H_2}$  &    FWHM$_{\rm dec}\tablefootmark{b} $       &   $T_{\rm dust}$  &    $ M$   &   $\alpha_{\rm BE}$  &   $N^{\rm back}_{\rm H_2}$\\
          &  [J2000]         &  [J2000]            &       [$" \times "$]     &  [$\degree$]  &       &   [Av units]\tablefootmark{a} & [Av units]\tablefootmark{a} & [pc]      &   [K]    &  [$\Msol$]&      &  [Av units]\tablefootmark{a} \\
\hline    
66  &      99.991   &   9.5787   &     $31.2 \times 26.4$   &    8.5   &     20.1   &    $1.4 \pm 0.3$    &    $3.0 \pm 0.3$   &     0.08   &    $14.2 \pm 0.2$    &    $0.3 \pm 0.0$    &   2.8   &  $2.8 \pm 0.1$  \\
67  &      99.9575   &   9.8404   &     $27.4 \times 21.7$   &    145.0   &     15.9   &    $1.6 \pm 0.2$    &    $3.0 \pm 0.2$   &     0.06   &    $13.9 \pm 0.2$    &    $0.3 \pm 0.0$    &   2.5   &  $2.4 \pm 0.0$  \\
68  &      100.2132   &   8.9324   &     $43.1 \times 32.7$   &    148.7   &     20.8   &    $3.7 \pm 0.6$    &    $14.0 \pm 0.9$   &     0.11   &    $13.0 \pm 0.2$    &    $1.4 \pm 0.1$    &   0.86   &  $4.3 \pm 0.1$  \\
69  &      99.9438   &   9.82   &     $50.0 \times 46.5$   &    57.5   &     30.9   &    $2.1 \pm 0.2$    &    $14.0 \pm 0.5$   &     0.16   &    $13.9 \pm 0.2$    &    $1.5 \pm 0.1$    &   1.2   &  $2.3 \pm 0.0$  \\
70  &      100.3765   &   9.2731   &     $54.6 \times 30.9$   &    52.7   &     28.4   &    $2.7 \pm 0.6$    &    $12.0 \pm 0.9$   &     0.13   &    $13.4 \pm 0.1$    &    $1.2 \pm 0.1$    &   1.2   &  $4.0 \pm 0.2$  \\
71  &      100.3143   &   9.3375   &     $38.3 \times 28.0$   &    28.8   &     20.0   &    $4 \pm 1$    &    $10 \pm 2$   &     0.1   &    $13.5 \pm 0.2$    &    $1.1 \pm 0.2$    &   0.97   &  $7.0 \pm 0.4$  \\
72  &      100.2324   &   9.6418   &     $58.2 \times 39.5$   &    15.8   &     21.7   &    $6 \pm 2$    &    $28 \pm 4$   &     0.16   &    $15.1 \pm 0.4$    &    $2.9 \pm 0.4$    &   0.67   &  $8.1 \pm 0.6$  \\
73$^*$  &      100.4484   &   8.8045   &     $34.9 \times 28.7$   &    47.8   &     17.4   &    $2.1 \pm 0.3$    &    $5.0 \pm 0.4$   &     0.09   &    $13.6 \pm 0.2$    &    $0.5 \pm 0.0$    &   1.9   &  $3.5 \pm 0.1$  \\
74  &      100.22   &   8.9826   &     $30.1 \times 28.7$   &    136.3   &     17.6   &    $3 \pm 1$    &    $8 \pm 1$   &     0.08   &    $13.3 \pm 0.1$    &    $0.8 \pm 0.1$    &   1.1   &  $5.4 \pm 0.2$  \\
75  &      100.6594   &   9.4211   &     $71.2 \times 35.8$   &    108.8   &     31.6   &    $3.7 \pm 0.8$    &    $23 \pm 2$   &     0.16   &    $13.1 \pm 0.2$    &    $2.4 \pm 0.2$    &   0.75   &  $4.2 \pm 0.3$  \\
76  &      100.2049   &   9.0746   &     $57.5 \times 31.1$   &    105.4   &     19.2   &    $4.0 \pm 0.7$    &    $17 \pm 1$   &     0.13   &    $12.5 \pm 0.2$    &    $1.8 \pm 0.1$    &   0.77   &  $5.5 \pm 0.1$  \\
77  &      100.2746   &   9.5283   &     $29.5 \times 21.9$   &    56.9   &     16.8   &    $7 \pm 4$    &    $13 \pm 4$   &     0.06   &    $13.2 \pm 0.1$    &    $1.3 \pm 0.4$    &   0.52   &  $20.8 \pm 0.3$  \\
78  &      99.9735   &   9.6959   &     $33.6 \times 25.5$   &    88.5   &     15.4   &    $7 \pm 2$    &    $13 \pm 2$   &     0.08   &    $13.0 \pm 0.2$    &    $1.3 \pm 0.2$    &   0.67   &  $7.7 \pm 0.3$  \\
79  &      99.8682   &   9.6277   &     $49.2 \times 29.9$   &    173.2   &     20.6   &    $12 \pm 4$    &    $43 \pm 5$   &     0.12   &    $11.9 \pm 0.1$    &    $4.4 \pm 0.5$    &   0.27   &  $7.7 \pm 0.8$  \\
80  &      100.1543   &   9.8742   &     $79.4 \times 50.0$   &    161.9   &     36.8   &    $9 \pm 2$    &    $89 \pm 4$   &     0.21   &    $18.5 \pm 0.7$    &    $9.2 \pm 0.4$    &   0.35   &  $5.8 \pm 0.4$  \\
81  &      100.085   &   9.8963   &     $54.0 \times 27.2$   &    71.2   &     18.8   &    $2.0 \pm 0.4$    &    $6.0 \pm 0.5$   &     0.12   &    $17.7 \pm 0.5$    &    $0.6 \pm 0.1$    &   2.9   &  $4.2 \pm 0.1$  \\
82  &      100.0651   &   9.4623   &     $43.9 \times 35.3$   &    62.1   &     19.6   &    $2.5 \pm 0.7$    &    $10 \pm 1$   &     0.12   &    $13.6 \pm 0.2$    &    $1.1 \pm 0.1$    &   1.3   &  $3.7 \pm 0.1$  \\
83$^*$  &      99.8388   &   9.2951   &     $32.8 \times 24.9$   &    151.4   &     17.0   &    $1.4 \pm 0.4$    &    $3.0 \pm 0.5$   &     0.08   &    $13.8 \pm 0.2$    &    $0.3 \pm 0.1$    &   2.7   &  $2.9 \pm 0.0$  \\
84  &      100.5825   &   9.0739   &     $50.9 \times 31.0$   &    162.1   &     19.2   &    $4 \pm 1$    &    $16 \pm 2$   &     0.12   &    $12.3 \pm 0.1$    &    $1.6 \pm 0.2$    &   0.79   &  $5.9 \pm 0.1$  \\
85  &      100.0642   &   9.4975   &     $55.7 \times 32.6$   &    69.7   &     20.0   &    $3.0 \pm 0.4$    &    $14.0 \pm 0.7$   &     0.14   &    $13.6 \pm 0.2$    &    $1.4 \pm 0.1$    &   1.1   &  $3.9 \pm 0.1$  \\
86  &      100.1978   &   9.7694   &     $35.2 \times 35.1$   &    24.4   &     18.8   &    $1.9 \pm 0.7$    &    $6 \pm 1$   &     0.11   &    $22.0 \pm 0.5$    &    $0.7 \pm 0.1$    &   3.0   &  $4.8 \pm 0.2$  \\
87  &      99.9733   &   9.706   &     $30.7 \times 26.9$   &    48.4   &     14.7   &    $7 \pm 3$    &    $14 \pm 3$   &     0.08   &    $12.9 \pm 0.2$    &    $1.4 \pm 0.3$    &   0.59   &  $8.2 \pm 0.3$  \\
88  &      100.1362   &   9.886   &     $39.5 \times 26.6$   &    15.5   &     16.8   &    $6 \pm 2$    &    $14 \pm 2$   &     0.09   &    $17.7 \pm 0.7$    &    $1.5 \pm 0.2$    &   0.92   &  $6.5 \pm 0.4$  \\
89  &      100.3631   &   9.3662   &     $32.6 \times 26.1$   &    175.5   &     12.0   &    $11 \pm 2$    &    $18 \pm 2$   &     0.08   &    $12.6 \pm 0.1$    &    $1.8 \pm 0.2$    &   0.46   &  $11.6 \pm 0.6$  \\
90  &      100.5747   &   9.0   &     $33.8 \times 25.5$   &    74.7   &     14.3   &    $3.0 \pm 0.9$    &    $6 \pm 1$   &     0.08   &    $12.5 \pm 0.1$    &    $0.6 \pm 0.1$    &   1.4   &  $5.4 \pm 0.2$  \\
91  &      100.2115   &   9.5879   &     $38.6 \times 34.9$   &    27.1   &     13.7   &    $19 \pm 7$    &    $51 \pm 8$   &     0.11   &    $12.9 \pm 0.2$    &    $5.2 \pm 0.8$    &   0.23   &  $20 \pm 1$  \\
92  &      100.1875   &   9.6607   &     $39.5 \times 36.2$   &    112.1   &     13.0   &    $3 \pm 1$    &    $10 \pm 2$   &     0.12   &    $16.8 \pm 0.3$    &    $1.0 \pm 0.2$    &   1.6   &  $5.6 \pm 0.2$  \\
93  &      100.0378   &   9.7167   &     $66.5 \times 48.9$   &    95.0   &     29.8   &    $4 \pm 1$    &    $30 \pm 2$   &     0.19   &    $15.0 \pm 0.4$    &    $3.1 \pm 0.2$    &   0.76   &  $3.9 \pm 0.1$  \\
94  &      100.1673   &   9.7977   &     $38.5 \times 33.2$   &    19.4   &     20.2   &    $1.7 \pm 0.5$    &    $6.0 \pm 0.8$   &     0.11   &    $20.6 \pm 0.2$    &    $0.6 \pm 0.1$    &   3.0   &  $3.4 \pm 0.1$  \\
95  &      100.2196   &   8.9951   &     $64.8 \times 34.5$   &    150.7   &     19.5   &    $4.3 \pm 0.9$    &    $22 \pm 2$   &     0.15   &    $13.1 \pm 0.1$    &    $2.3 \pm 0.2$    &   0.72   &  $6.0 \pm 0.2$  \\
96  &      100.6249   &   9.0262   &     $48.3 \times 35.3$   &    63.9   &     21.1   &    $2.9 \pm 0.4$    &    $13.0 \pm 0.6$   &     0.13   &    $12.8 \pm 0.1$    &    $1.3 \pm 0.1$    &   1.0   &  $4.3 \pm 0.0$  \\
97  &      100.5551   &   8.9896   &     $25.9 \times 24.6$   &    164.2   &     13.3   &    $2 \pm 1$    &    $3 \pm 1$   &     0.06   &    $13.3 \pm 0.3$    &    $0.4 \pm 0.1$    &   1.7   &  $5.1 \pm 0.2$  \\
98  &      100.7344   &   9.3178   &     $69.8 \times 39.3$   &    175.9   &     27.8   &    $3.5 \pm 0.4$    &    $25.0 \pm 0.8$   &     0.17   &    $13.1 \pm 0.1$    &    $2.6 \pm 0.1$    &   0.72   &  $4.0 \pm 0.1$  \\
99  &      100.3448   &   9.3903   &     $59.3 \times 44.4$   &    116.7   &     20.2   &    $8 \pm 3$    &    $51 \pm 5$   &     0.17   &    $13.5 \pm 0.1$    &    $5.3 \pm 0.5$    &   0.35   &  $13.6 \pm 0.6$  \\
100  &      100.096   &   9.8506   &     $42.0 \times 34.8$   &    162.8   &     20.8   &    $3 \pm 1$    &    $10 \pm 2$   &     0.12   &    $19.0 \pm 0.4$    &    $1.0 \pm 0.2$    &   1.9   &  $4.4 \pm 0.4$  \\
101  &      100.2797   &   8.9593   &     $39.6 \times 25.9$   &    17.2   &     14.6   &    $1.7 \pm 0.4$    &    $4.0 \pm 0.5$   &     0.09   &    $13.6 \pm 0.1$    &    $0.4 \pm 0.1$    &   2.5   &  $3.8 \pm 0.1$  \\
102  &      100.1655   &   8.9167   &     $33.1 \times 30.1$   &    76.7   &     16.0   &    $2.8 \pm 0.9$    &    $8 \pm 1$   &     0.09   &    $13.2 \pm 0.1$    &    $0.8 \pm 0.1$    &   1.2   &  $5.0 \pm 0.2$  \\
103  &      100.2451   &   9.5575   &     $32.0 \times 25.2$   &    151.1   &     13.8   &    $7 \pm 5$    &    $12 \pm 6$   &     0.08   &    $14.2 \pm 0.2$    &    $1.2 \pm 0.6$    &   0.75   &  $18 \pm 1$  \\
104  &      100.1744   &   9.7514   &     $73.3 \times 56.5$   &    101.7   &     32.5   &    $2.6 \pm 0.6$    &    $27 \pm 2$   &     0.22   &    $21.2 \pm 0.8$    &    $2.8 \pm 0.2$    &   1.4   &  $3.6 \pm 0.2$  \\
105  &      100.2541   &   9.3378   &     $65.8 \times 42.6$   &    179.1   &     19.8   &    $2.9 \pm 0.6$    &    $18 \pm 1$   &     0.17   &    $13.5 \pm 0.2$    &    $1.9 \pm 0.1$    &   1.0   &  $4.6 \pm 0.1$  \\
106  &      100.2503   &   9.59   &     $37.3 \times 29.1$   &    125.1   &     13.9   &    $45 \pm 11$    &    $114 \pm 12$   &     0.1   &    $14.8 \pm 0.5$    &    $12 \pm 1$    &   0.1   &  $44.1 \pm 0.6$  \\
107  &      100.0727   &   9.4603   &     $31.7 \times 24.6$   &    129.4   &     9.9   &    $1.9 \pm 0.8$    &    $4.0 \pm 0.9$   &     0.07   &    $13.5 \pm 0.1$    &    $0.4 \pm 0.1$    &   2.0   &  $4.0 \pm 0.2$  \\
108$^*$  &      100.434   &   8.8942   &     $28.9 \times 23.4$   &    103.2   &     14.4   &    $1.4 \pm 0.5$    &    $3.0 \pm 0.6$   &     0.06   &    $13.5 \pm 0.2$    &    $0.3 \pm 0.1$    &   2.6   &  $3.0 \pm 0.1$  \\
109  &      100.2467   &   9.0244   &     $63.2 \times 35.1$   &    172.2   &     21.4   &    $4 \pm 1$    &    $16 \pm 2$   &     0.15   &    $12.5 \pm 0.2$    &    $1.7 \pm 0.2$    &   0.93   &  $6.7 \pm 0.7$  \\
110  &      100.0598   &   9.213   &     $34.0 \times 24.7$   &    154.0   &     15.3   &    $1.7 \pm 0.3$    &    $4.0 \pm 0.4$   &     0.08   &    $13.4 \pm 0.1$    &    $0.4 \pm 0.0$    &   2.1   &  $5.0 \pm 0.1$  \\
111  &      100.2915   &   9.9672   &     $29.9 \times 25.9$   &    42.8   &     14.5   &    $2.3 \pm 0.9$    &    $5 \pm 1$   &     0.07   &    $13.3 \pm 0.1$    &    $0.5 \pm 0.1$    &   1.6   &  $6.2 \pm 0.1$  \\
112  &      100.094   &   9.9152   &     $23.7 \times 23.7$   &    149.1   &     13.4   &    $1.7 \pm 0.6$    &    $3.0 \pm 0.7$   &     0.05   &    $16.4 \pm 0.2$    &    $0.3 \pm 0.1$    &   2.5   &  $4.4 \pm 0.0$  \\
113  &      100.1795   &   9.1332   &     $35.3 \times 30.0$   &    20.3   &     14.0   &    $2.7 \pm 0.8$    &    $7 \pm 1$   &     0.09   &    $13.2 \pm 0.1$    &    $0.8 \pm 0.1$    &   1.3   &  $6.7 \pm 0.0$  \\
114  &      100.1754   &   9.1555   &     $37.9 \times 29.6$   &    32.7   &     14.2   &    $3 \pm 1$    &    $8 \pm 1$   &     0.1   &    $13.0 \pm 0.1$    &    $0.8 \pm 0.1$    &   1.3   &  $6.5 \pm 0.2$  \\
115  &      100.2806   &   9.7276   &     $50.4 \times 35.6$   &    53.6   &     12.1   &    $3 \pm 2$    &    $10 \pm 3$   &     0.13   &    $14.6 \pm 0.1$    &    $1.0 \pm 0.3$    &   1.6   &  $8.8 \pm 0.4$  \\
116  &      100.1912   &   9.6705   &     $67.9 \times 49.3$   &    133.8   &     24.3   &    $3 \pm 1$    &    $29 \pm 3$   &     0.19   &    $17.2 \pm 0.5$    &    $3.1 \pm 0.3$    &   0.88   &  $4.8 \pm 0.3$  \\
117  &      99.9587   &   9.9553   &     $37.3 \times 28.3$   &    138.8   &     13.7   &    $2.9 \pm 0.9$    &    $7 \pm 1$   &     0.09   &    $15.7 \pm 0.2$    &    $0.8 \pm 0.1$    &   1.5   &  $4.0 \pm 0.2$  \\
118  &      100.4639   &   9.2354   &     $54.6 \times 41.8$   &    155.9   &     21.4   &    $1.9 \pm 0.9$    &    $11 \pm 2$   &     0.16   &    $13.5 \pm 0.1$    &    $1.1 \pm 0.2$    &   1.6   &  $4.8 \pm 0.1$  \\
119  &      100.7421   &   9.4783   &     $44.0 \times 31.9$   &    47.3   &     18.3   &    $1.2 \pm 0.1$    &    $5.0 \pm 0.2$   &     0.11   &    $14.3 \pm 0.2$    &    $0.5 \pm 0.0$    &   2.7   &  $2.3 \pm 0.0$  \\
120  &      100.4643   &   9.4295   &     $46.0 \times 42.1$   &    11.5   &     24.5   &    $2.5 \pm 0.6$    &    $14 \pm 1$   &     0.14   &    $13.7 \pm 0.2$    &    $1.4 \pm 0.1$    &   1.1   &  $4.2 \pm 0.2$  \\
121  &      100.1103   &   9.5779   &     $40.7 \times 31.8$   &    0.8   &     13.9   &    $2.6 \pm 0.8$    &    $8 \pm 1$   &     0.11   &    $15.0 \pm 0.1$    &    $0.9 \pm 0.1$    &   1.5   &  $7.2 \pm 0.1$  \\
122  &      100.1735   &   9.8958   &     $30.9 \times 23.3$   &    93.5   &     13.3   &    $1.1 \pm 0.5$    &    $2.0 \pm 0.5$   &     0.07   &    $17.7 \pm 0.2$    &    $0.2 \pm 0.1$    &   5.3   &  $3.8 \pm 0.1$  \\
123  &      100.3207   &   9.7433   &     $49.9 \times 42.8$   &    35.8   &     20.8   &    $4 \pm 2$    &    $20 \pm 3$   &     0.15   &    $14.5 \pm 0.1$    &    $2.1 \pm 0.3$    &   0.85   &  $6.9 \pm 0.5$  \\
124  &      100.642   &   9.5663   &     $33.6 \times 27.6$   &    56.7   &     14.1   &    $1.5 \pm 0.4$    &    $4.0 \pm 0.5$   &     0.09   &    $13.8 \pm 0.1$    &    $0.4 \pm 0.1$    &   2.5   &  $3.2 \pm 0.1$  \\
125  &      100.2854   &   9.5877   &     $33.2 \times 27.5$   &    52.6   &     12.5   &    $23 \pm 13$    &    $49 \pm 14$   &     0.08   &    $13.4 \pm 0.3$    &    $5 \pm 1$    &   0.18   &  $51 \pm 2$  \\
126  &      100.1556   &   9.8406   &     $44.1 \times 34.0$   &    15.8   &     12.6   &    $2 \pm 1$    &    $7 \pm 2$   &     0.12   &    $18.5 \pm 0.4$    &    $0.7 \pm 0.2$    &   2.6   &  $4.2 \pm 0.3$  \\
127  &      100.4369   &   8.8042   &     $39.6 \times 28.4$   &    19.0   &     12.9   &    $1.8 \pm 0.2$    &    $4.0 \pm 0.3$   &     0.1   &    $13.6 \pm 0.2$    &    $0.4 \pm 0.0$    &   2.5   &  $3.4 \pm 0.0$  \\
128$^*$  &      100.3264   &   9.3515   &     $37.5 \times 28.1$   &    42.0   &     12.1   &    $5 \pm 2$    &    $11 \pm 2$   &     0.09   &    $13.5 \pm 0.2$    &    $1.2 \pm 0.2$    &   0.88   &  $9.8 \pm 0.5$  \\
129  &      100.1184   &   9.8625   &     $59.1 \times 37.7$   &    96.4   &     13.4   &    $4 \pm 1$    &    $21 \pm 2$   &     0.15   &    $19.7 \pm 0.5$    &    $2.1 \pm 0.2$    &   1.2   &  $5.6 \pm 0.3$  \\
\hline
\end{tabular} 
 } \\
 \end{table*}

    \setcounter{table}{0}
   \begin{table*}
      \small	
   \caption[]{Catalog of clumps (continued).}
   \makebox[\textwidth][c]{
\begin{tabular}{|c|c|c|c|c|c|c|c|c|c|c|c|c|} 
\hline    
\#    &     Alp  &   Del  &    $A \times B$     &     PA        &      Sig    &   $N^{\rm peak}_{\rm H_2}$  &   $N^{\rm int}_{\rm H_2}$  &    FWHM$_{\rm dec}\tablefootmark{b} $       &   $T_{\rm dust}$  &    $ M$   &   $\alpha_{\rm BE}$  &   $N^{\rm back}_{\rm H_2}$\\
          &  [J2000]         &  [J2000]            &       [$" \times "$]     &  [$\degree$]  &       &   [Av units]\tablefootmark{a} & [Av units]\tablefootmark{a} & [pc]      &   [K]    &  [$\Msol$]&      &  [Av units]\tablefootmark{a} \\
\hline    
130  &      100.4482   &   8.792   &     $45.3 \times 30.7$   &    173.2   &     14.2   &    $1.7 \pm 0.3$    &    $6.0 \pm 0.4$   &     0.11   &    $13.6 \pm 0.2$    &    $0.6 \pm 0.1$    &   2.3   &  $3.1 \pm 0.1$  \\
131$^*$  &      100.1699   &   8.9387   &     $43.1 \times 25.1$   &    53.0   &     14.0   &    $2.7 \pm 0.5$    &    $8.0 \pm 0.8$   &     0.1   &    $13.1 \pm 0.1$    &    $0.8 \pm 0.1$    &   1.3   &  $6.3 \pm 0.2$  \\
132  &      100.583   &   9.415   &     $29.9 \times 24.0$   &    73.2   &     11.9   &    $1.7 \pm 0.8$    &    $3.0 \pm 0.9$   &     0.07   &    $13.4 \pm 0.1$    &    $0.3 \pm 0.1$    &   2.5   &  $4.9 \pm 0.2$  \\
133  &      100.0878   &   9.2379   &     $70.8 \times 41.2$   &    29.5   &     20.6   &    $2.2 \pm 0.5$    &    $16 \pm 1$   &     0.18   &    $13.6 \pm 0.1$    &    $1.7 \pm 0.1$    &   1.2   &  $4.5 \pm 0.0$  \\
134  &      100.2048   &   9.4749   &     $34.9 \times 24.6$   &    19.8   &     12.4   &    $1.7 \pm 0.6$    &    $4.0 \pm 0.7$   &     0.08   &    $14.9 \pm 0.1$    &    $0.4 \pm 0.1$    &   2.6   &  $5.6 \pm 0.1$  \\
135  &      100.3867   &   9.7979   &     $40.7 \times 38.9$   &    67.4   &     15.3   &    $1.5 \pm 0.3$    &    $6.0 \pm 0.4$   &     0.12   &    $15.1 \pm 0.2$    &    $0.6 \pm 0.0$    &   2.5   &  $2.9 \pm 0.1$  \\
136  &      100.436   &   9.2739   &     $42.4 \times 25.8$   &    79.5   &     13.3   &    $1.6 \pm 0.4$    &    $5.0 \pm 0.6$   &     0.1   &    $14.0 \pm 0.1$    &    $0.5 \pm 0.1$    &   2.3   &  $4.0 \pm 0.1$  \\
137  &      100.2108   &   9.6019   &     $28.3 \times 25.4$   &    72.3   &     12.2   &    $8 \pm 3$    &    $16 \pm 3$   &     0.07   &    $13.5 \pm 0.2$    &    $1.7 \pm 0.3$    &   0.45   &  $15.7 \pm 0.5$  \\
138  &      100.09   &   9.5957   &     $33.9 \times 24.3$   &    174.1   &     13.1   &    $1.7 \pm 0.6$    &    $4.0 \pm 0.7$   &     0.08   &    $15.8 \pm 0.1$    &    $0.4 \pm 0.1$    &   2.7   &  $5.5 \pm 0.1$  \\
139  &      100.0368   &   9.6407   &     $24.3 \times 24.2$   &    47.6   &     11.4   &    $2.4 \pm 0.6$    &    $3.0 \pm 0.6$   &     0.06   &    $16.1 \pm 0.2$    &    $0.4 \pm 0.1$    &   2.1   &  $5.9 \pm 0.0$  \\
140  &      99.9155   &   9.5519   &     $53.3 \times 34.0$   &    22.3   &     14.3   &    $6 \pm 2$    &    $24 \pm 2$   &     0.13   &    $12.8 \pm 0.2$    &    $2.5 \pm 0.2$    &   0.57   &  $4.0 \pm 0.4$  \\
141  &      100.1577   &   9.0647   &     $64.2 \times 36.4$   &    154.7   &     15.8   &    $3 \pm 1$    &    $14 \pm 2$   &     0.16   &    $12.6 \pm 0.2$    &    $1.5 \pm 0.2$    &   1.1   &  $5.4 \pm 0.2$  \\
142  &      100.0998   &   9.1946   &     $46.0 \times 34.1$   &    103.0   &     16.3   &    $1.7 \pm 0.6$    &    $8 \pm 1$   &     0.12   &    $13.6 \pm 0.1$    &    $0.8 \pm 0.1$    &   1.7   &  $4.1 \pm 0.1$  \\
143  &      100.2731   &   9.6803   &     $39.9 \times 29.9$   &    127.7   &     12.9   &    $3 \pm 1$    &    $9 \pm 2$   &     0.1   &    $14.5 \pm 0.2$    &    $0.9 \pm 0.2$    &   1.4   &  $7.3 \pm 0.4$  \\
144  &      100.4178   &   9.8553   &     $61.1 \times 44.9$   &    46.0   &     18.3   &    $1.6 \pm 0.2$    &    $11.0 \pm 0.4$   &     0.17   &    $13.8 \pm 0.2$    &    $1.2 \pm 0.0$    &   1.7   &  $2.7 \pm 0.0$  \\
145  &      100.2594   &   9.8615   &     $38.6 \times 32.2$   &    159.2   &     13.2   &    $1.6 \pm 0.8$    &    $5 \pm 1$   &     0.11   &    $15.9 \pm 0.2$    &    $0.5 \pm 0.1$    &   2.8   &  $6.1 \pm 0.3$  \\
146  &      100.0022   &   9.1453   &     $81.3 \times 49.2$   &    8.8   &     20.2   &    $2.6 \pm 0.7$    &    $25 \pm 2$   &     0.21   &    $13.8 \pm 0.1$    &    $2.5 \pm 0.2$    &   0.97   &  $4.0 \pm 0.2$  \\
147  &      100.3188   &   9.3958   &     $45.5 \times 31.7$   &    171.5   &     11.9   &    $9 \pm 2$    &    $28 \pm 3$   &     0.12   &    $14.2 \pm 0.2$    &    $2.9 \pm 0.3$    &   0.0   &  $16.9 \pm 0.6$  \\
148  &      100.2951   &   9.5625   &     $29.1 \times 23.3$   &    2.6   &     11.4   &    $13 \pm 6$    &    $22 \pm 7$   &     0.07   &    $12.6 \pm 0.2$    &    $2.2 \pm 0.7$    &   0.31   &  $27.4 \pm 0.3$  \\
149  &      100.1934   &   9.6148   &     $37.1 \times 29.3$   &    167.1   &     11.4   &    $5 \pm 2$    &    $13 \pm 2$   &     0.1   &    $14.5 \pm 0.2$    &    $1.4 \pm 0.2$    &   0.83   &  $10.6 \pm 0.3$  \\
150  &      100.2077   &   9.7861   &     $46.5 \times 34.8$   &    43.1   &     14.7   &    $1.7 \pm 0.7$    &    $7 \pm 1$   &     0.13   &    $21.6 \pm 0.5$    &    $0.8 \pm 0.1$    &   2.8   &  $5.1 \pm 0.1$  \\
151  &      100.0163   &   9.9338   &     $31.8 \times 30.7$   &    77.5   &     13.0   &    $1.3 \pm 0.3$    &    $4.0 \pm 0.4$   &     0.09   &    $16.0 \pm 0.1$    &    $0.4 \pm 0.1$    &   3.1   &  $3.5 \pm 0.1$  \\
152  &      100.6882   &   9.3968   &     $39.1 \times 34.3$   &    31.1   &     15.7   &    $1.4 \pm 0.6$    &    $5 \pm 1$   &     0.11   &    $13.4 \pm 0.1$    &    $0.5 \pm 0.1$    &   2.6   &  $2.9 \pm 0.1$  \\
153  &      100.5352   &   9.6892   &     $24.5 \times 23.1$   &    25.0   &     11.5   &    $0.6 \pm 0.2$    &    $1.0 \pm 0.2$   &     0.05   &    $15.0 \pm 0.1$    &    $0.1 \pm 0.0$    &   6.1   &  $2.5 \pm 0.0$  \\
154  &      99.7937   &   9.7389   &     $31.5 \times 26.2$   &    73.7   &     11.6   &    $1.2 \pm 0.1$    &    $2.0 \pm 0.1$   &     0.08   &    $14.3 \pm 0.1$    &    $0.2 \pm 0.0$    &   5.4   &  $2.3 \pm 0.1$  \\
155  &      100.5735   &   8.9423   &     $35.2 \times 30.5$   &    162.6   &     10.3   &    $3.1 \pm 0.7$    &    $9 \pm 1$   &     0.1   &    $12.7 \pm 0.1$    &    $1.0 \pm 0.1$    &   1.0   &  $4.8 \pm 0.2$  \\
156  &      100.1781   &   8.9031   &     $31.2 \times 29.2$   &    138.6   &     12.5   &    $1.7 \pm 0.7$    &    $4 \pm 1$   &     0.08   &    $13.5 \pm 0.1$    &    $0.4 \pm 0.1$    &   2.4   &  $3.9 \pm 0.2$  \\
157  &      100.398   &   9.5311   &     $33.2 \times 24.1$   &    30.6   &     10.8   &    $1.6 \pm 0.4$    &    $3.0 \pm 0.4$   &     0.08   &    $13.9 \pm 0.1$    &    $0.3 \pm 0.0$    &   2.7   &  $4.1 \pm 0.0$  \\
158  &      100.1566   &   9.7885   &     $50.2 \times 32.7$   &    149.7   &     13.0   &    $1.4 \pm 0.4$    &    $5.0 \pm 0.5$   &     0.13   &    $20.2 \pm 0.2$    &    $0.6 \pm 0.1$    &   3.9   &  $3.8 \pm 0.1$  \\
159  &      100.3769   &   8.8673   &     $56.4 \times 33.4$   &    92.8   &     13.2   &    $2.0 \pm 0.4$    &    $8.0 \pm 0.6$   &     0.14   &    $13.7 \pm 0.1$    &    $0.8 \pm 0.1$    &   1.9   &  $3.7 \pm 0.0$  \\
160$^*$  &      100.5798   &   9.03   &     $36.2 \times 26.0$   &    14.2   &     12.1   &    $1.9 \pm 0.7$    &    $5.0 \pm 0.9$   &     0.09   &    $12.9 \pm 0.2$    &    $0.5 \pm 0.1$    &   1.8   &  $4.3 \pm 0.1$  \\
161  &      100.467   &   9.1946   &     $35.8 \times 23.0$   &    66.2   &     12.4   &    $1.1 \pm 0.6$    &    $3.0 \pm 0.8$   &     0.08   &    $13.9 \pm 0.1$    &    $0.3 \pm 0.1$    &   3.4   &  $3.6 \pm 0.1$  \\
162  &      100.5067   &   9.4543   &     $39.9 \times 32.4$   &    59.8   &     13.2   &    $1.1 \pm 0.2$    &    $4.0 \pm 0.3$   &     0.11   &    $13.9 \pm 0.1$    &    $0.4 \pm 0.0$    &   3.0   &  $2.5 \pm 0.0$  \\
163  &      100.2866   &   9.4104   &     $26.2 \times 23.0$   &    27.2   &     11.1   &    $4 \pm 2$    &    $7 \pm 2$   &     0.06   &    $16.4 \pm 0.3$    &    $0.7 \pm 0.2$    &   1.1   &  $9.8 \pm 0.3$  \\
164  &      100.2973   &   9.5064   &     $34.1 \times 25.5$   &    17.6   &     10.8   &    $28 \pm 7$    &    $50 \pm 7$   &     0.08   &    $14.1 \pm 0.7$    &    $5.2 \pm 0.8$    &   0.18   &  $44 \pm 3$  \\
165  &      100.1823   &   9.0229   &     $68.1 \times 57.3$   &    99.1   &     14.6   &    $2 \pm 1$    &    $23 \pm 3$   &     0.21   &    $13.1 \pm 0.1$    &    $2.4 \pm 0.3$    &   0.95   &  $5.6 \pm 0.5$  \\
166  &      100.1345   &   9.127   &     $71.8 \times 60.0$   &    3.2   &     20.6   &    $3.4 \pm 0.9$    &    $39 \pm 2$   &     0.22   &    $13.3 \pm 0.1$    &    $4.0 \pm 0.3$    &   0.61   &  $5.0 \pm 0.2$  \\
167  &      100.3678   &   9.5892   &     $27.1 \times 23.7$   &    50.0   &     10.1   &    $3.2 \pm 0.6$    &    $5.0 \pm 0.6$   &     0.06   &    $14.6 \pm 0.2$    &    $0.5 \pm 0.1$    &   1.4   &  $5.6 \pm 0.0$  \\
168  &      100.1722   &   9.9264   &     $42.3 \times 34.0$   &    106.6   &     12.4   &    $1.1 \pm 0.4$    &    $4.0 \pm 0.7$   &     0.12   &    $16.7 \pm 0.1$    &    $0.4 \pm 0.1$    &   3.7   &  $3.4 \pm 0.1$  \\
169  &      100.3882   &   9.325   &     $36.5 \times 31.0$   &    31.4   &     10.8   &    $2 \pm 1$    &    $6 \pm 1$   &     0.1   &    $13.4 \pm 0.1$    &    $0.6 \pm 0.1$    &   1.8   &  $7.1 \pm 0.1$  \\
170  &      100.2085   &   9.6972   &     $35.8 \times 24.6$   &    49.7   &     10.8   &    $1.2 \pm 0.5$    &    $3.0 \pm 0.6$   &     0.08   &    $17.5 \pm 0.2$    &    $0.3 \pm 0.1$    &   4.4   &  $3.9 \pm 0.1$  \\
171  &      100.1884   &   9.8296   &     $41.4 \times 32.7$   &    64.5   &     15.5   &    $0.8 \pm 0.2$    &    $3.0 \pm 0.3$   &     0.11   &    $18.8 \pm 0.4$    &    $0.3 \pm 0.0$    &   5.6   &  $2.8 \pm 0.0$  \\
172  &      99.8817   &   9.6786   &     $27.6 \times 24.2$   &    165.0   &     10.9   &    $1.2 \pm 0.3$    &    $2.0 \pm 0.4$   &     0.06   &    $13.8 \pm 0.1$    &    $0.2 \pm 0.0$    &   3.3   &  $4.0 \pm 0.0$  \\
173  &      100.5193   &   8.8943   &     $32.0 \times 24.2$   &    139.5   &     10.9   &    $1.3 \pm 0.3$    &    $3.0 \pm 0.4$   &     0.07   &    $13.5 \pm 0.2$    &    $0.3 \pm 0.0$    &   2.9   &  $2.8 \pm 0.0$  \\
174  &      100.111   &   8.9833   &     $57.2 \times 32.3$   &    44.0   &     13.6   &    $2.1 \pm 0.8$    &    $9 \pm 1$   &     0.14   &    $13.3 \pm 0.1$    &    $0.9 \pm 0.1$    &   1.7   &  $4.1 \pm 0.2$  \\
175  &      100.4885   &   9.5556   &     $74.0 \times 45.2$   &    52.5   &     13.8   &    $1.1 \pm 0.2$    &    $9.0 \pm 0.5$   &     0.19   &    $14.0 \pm 0.2$    &    $0.9 \pm 0.1$    &   2.4   &  $2.6 \pm 0.1$  \\
176  &      99.9884   &   9.3223   &     $34.1 \times 31.4$   &    143.5   &     10.7   &    $1.4 \pm 0.2$    &    $4.0 \pm 0.3$   &     0.1   &    $13.4 \pm 0.1$    &    $0.4 \pm 0.0$    &   2.6   &  $3.6 \pm 0.0$  \\
177  &      100.2578   &   9.5549   &     $33.4 \times 27.7$   &    82.4   &     10.1   &    $14 \pm 9$    &    $30 \pm 9$   &     0.09   &    $12.7 \pm 0.1$    &    $3.1 \pm 0.9$    &   0.29   &  $33 \pm 4$  \\
178  &      99.9995   &   9.3985   &     $42.9 \times 41.9$   &    123.5   &     15.6   &    $1.3 \pm 0.2$    &    $7.0 \pm 0.3$   &     0.13   &    $13.8 \pm 0.1$    &    $0.7 \pm 0.0$    &   2.2   &  $2.5 \pm 0.0$  \\
179  &      100.2595   &   9.7602   &     $55.4 \times 33.0$   &    160.7   &     12.0   &    $1.6 \pm 0.6$    &    $6 \pm 1$   &     0.14   &    $15.4 \pm 0.1$    &    $0.7 \pm 0.1$    &   2.6   &  $7.0 \pm 0.1$  \\
180  &      100.2866   &   9.0061   &     $40.1 \times 30.9$   &    24.0   &     11.7   &    $1.9 \pm 0.8$    &    $6 \pm 1$   &     0.11   &    $13.5 \pm 0.1$    &    $0.6 \pm 0.1$    &   2.0   &  $4.6 \pm 0.2$  \\
181  &      100.113   &   9.1433   &     $56.7 \times 49.6$   &    12.3   &     13.7   &    $1.8 \pm 0.6$    &    $12 \pm 1$   &     0.17   &    $13.6 \pm 0.1$    &    $1.3 \pm 0.1$    &   1.5   &  $4.7 \pm 0.1$  \\
182  &      100.4264   &   9.4137   &     $72.1 \times 34.7$   &    16.4   &     13.3   &    $1.0 \pm 0.4$    &    $5.0 \pm 0.7$   &     0.16   &    $15.1 \pm 0.1$    &    $0.6 \pm 0.1$    &   3.7   &  $4.1 \pm 0.1$  \\
183  &      100.1023   &   9.5246   &     $47.2 \times 35.8$   &    48.0   &     12.3   &    $1.2 \pm 0.4$    &    $5.0 \pm 0.6$   &     0.13   &    $14.4 \pm 0.2$    &    $0.5 \pm 0.1$    &   3.3   &  $4.0 \pm 0.1$  \\
184  &      100.2541   &   9.656   &     $34.5 \times 30.1$   &    30.6   &     14.3   &    $2 \pm 1$    &    $6 \pm 1$   &     0.09   &    $15.5 \pm 0.1$    &    $0.6 \pm 0.1$    &   2.0   &  $7.1 \pm 0.2$  \\
185  &      100.5251   &   9.1592   &     $53.8 \times 41.0$   &    54.7   &     12.1   &    $7 \pm 2$    &    $37 \pm 3$   &     0.15   &    $12.3 \pm 0.3$    &    $3.8 \pm 0.3$    &   0.41   &  $10.0 \pm 0.7$  \\
186  &      100.7216   &   9.2688   &     $48.9 \times 39.1$   &    123.7   &     13.3   &    $2.1 \pm 0.6$    &    $10 \pm 1$   &     0.14   &    $13.5 \pm 0.1$    &    $1.1 \pm 0.1$    &   1.4   &  $3.6 \pm 0.1$  \\
187  &      100.6924   &   9.3673   &     $37.0 \times 33.5$   &    51.6   &     12.1   &    $1.4 \pm 0.5$    &    $5.0 \pm 0.8$   &     0.11   &    $13.4 \pm 0.1$    &    $0.5 \pm 0.1$    &   2.5   &  $4.0 \pm 0.0$  \\
188  &      100.4766   &   9.4058   &     $44.4 \times 33.3$   &    122.5   &     11.4   &    $1.3 \pm 0.5$    &    $5.0 \pm 0.8$   &     0.12   &    $14.6 \pm 0.1$    &    $0.5 \pm 0.1$    &   2.9   &  $3.6 \pm 0.1$  \\
189  &      100.0175   &   9.6433   &     $35.3 \times 33.8$   &    84.8   &     10.7   &    $1.4 \pm 0.5$    &    $4.0 \pm 0.7$   &     0.1   &    $15.8 \pm 0.1$    &    $0.4 \pm 0.1$    &   3.5   &  $5.2 \pm 0.2$  \\
190  &      100.4665   &   8.9984   &     $31.6 \times 23.2$   &    128.7   &     10.6   &    $1.0 \pm 0.2$    &    $2.0 \pm 0.3$   &     0.07   &    $13.7 \pm 0.1$    &    $0.2 \pm 0.0$    &   3.5   &  $3.0 \pm 0.0$  \\
191  &      100.2866   &   8.9851   &     $42.1 \times 30.4$   &    51.5   &     11.3   &    $1.8 \pm 0.6$    &    $6.0 \pm 0.8$   &     0.11   &    $13.8 \pm 0.1$    &    $0.6 \pm 0.1$    &   2.1   &  $4.0 \pm 0.1$  \\
192  &      100.332   &   9.3187   &     $53.7 \times 29.8$   &    85.9   &     12.4   &    $1.8 \pm 0.9$    &    $6 \pm 1$   &     0.12   &    $14.4 \pm 0.1$    &    $0.7 \pm 0.1$    &   2.1   &  $5.3 \pm 0.2$  \\
193  &      100.0675   &   9.1705   &     $28.1 \times 22.7$   &    166.9   &     9.8   &    $1.4 \pm 0.4$    &    $2.0 \pm 0.5$   &     0.06   &    $13.4 \pm 0.1$    &    $0.2 \pm 0.1$    &   3.1   &  $5.1 \pm 0.0$  \\
194  &      100.307   &   9.4094   &     $56.4 \times 31.8$   &    26.6   &     9.7   &    $23 \pm 3$    &    $83 \pm 5$   &     0.13   &    $13.8 \pm 0.3$    &    $8.6 \pm 0.5$    &   0.18   &  $18 \pm 1$  \\
195  &      100.1469   &   9.5938   &     $43.4 \times 40.0$   &    40.8   &     12.9   &    $3 \pm 1$    &    $14 \pm 3$   &     0.13   &    $16.7 \pm 0.3$    &    $1.5 \pm 0.3$    &   1.2   &  $7.1 \pm 0.3$  \\

\hline
\end{tabular} 
 } \\
   \label{table}
 \end{table*} 
 
     \setcounter{table}{0}
    \begin{table*}
      \small	
   \caption[]{Catalog of clumps (continued).}
   \makebox[\textwidth][c]{
\begin{tabular}{|c|c|c|c|c|c|c|c|c|c|c|c|c|} 
\hline    
\#    &     Alp  &   Del  &    $A \times B$     &     PA        &      Sig    &   $N^{\rm peak}_{\rm H_2}$  &   $N^{\rm int}_{\rm H_2}$  &    FWHM$_{\rm dec}\tablefootmark{b} $       &   $T_{\rm dust}$  &    $ M$   &   $\alpha_{\rm BE}$  &   $N^{\rm back}_{\rm H_2}$\\
          &  [J2000]         &  [J2000]            &       [$" \times "$]     &  [$\degree$]  &       &   [Av units]\tablefootmark{a} & [Av units]\tablefootmark{a} & [pc]     &   [K]    &  [$\Msol$]&      &  [Av units]\tablefootmark{a} \\
\hline   
196  &      100.3293   &   9.8023   &     $47.4 \times 28.1$   &    140.5   &     11.3   &    $1 \pm 1$    &    $3 \pm 1$   &     0.11   &    $14.8 \pm 0.2$    &    $0.3 \pm 0.1$    &   4.5   &  $5.4 \pm 0.4$  \\
197  &      100.0306   &   9.9795   &     $46.5 \times 38.3$   &    161.4   &     11.9   &    $2.3 \pm 0.4$    &    $10.0 \pm 0.6$   &     0.13   &    $15.8 \pm 0.2$    &    $1.0 \pm 0.1$    &   1.7   &  $2.9 \pm 0.1$  \\
198  &      100.5287   &   9.2781   &     $52.4 \times 39.1$   &    18.0   &     14.0   &    $2.0 \pm 0.6$    &    $10 \pm 1$   &     0.15   &    $13.6 \pm 0.1$    &    $1.1 \pm 0.1$    &   1.5   &  $4.0 \pm 0.1$  \\
199  &      100.4731   &   9.3384   &     $37.7 \times 25.7$   &    38.9   &     9.9   &    $1.7 \pm 0.3$    &    $4.0 \pm 0.3$   &     0.09   &    $14.3 \pm 0.1$    &    $0.4 \pm 0.0$    &   2.7   &  $4.6 \pm 0.2$  \\
200  &      100.4175   &   9.4934   &     $39.8 \times 27.9$   &    9.9   &     10.2   &    $1.7 \pm 0.5$    &    $4.0 \pm 0.6$   &     0.1   &    $14.2 \pm 0.2$    &    $0.4 \pm 0.1$    &   2.6   &  $4.1 \pm 0.0$  \\
201  &      100.5678   &   9.6361   &     $37.4 \times 23.9$   &    15.1   &     10.0   &    $0.7 \pm 0.2$    &    $1.0 \pm 0.3$   &     0.08   &    $14.3 \pm 0.1$    &    $0.1 \pm 0.0$    &   7.0   &  $2.7 \pm 0.0$  \\
202  &      100.1883   &   9.5249   &     $48.1 \times 29.7$   &    173.9   &     10.2   &    $2.9 \pm 0.7$    &    $9.0 \pm 0.9$   &     0.12   &    $15.0 \pm 0.1$    &    $0.9 \pm 0.1$    &   1.5   &  $6.5 \pm 0.1$  \\
203  &      100.2228   &   9.6119   &     $43.5 \times 28.1$   &    7.9   &     9.7   &    $20 \pm 6$    &    $57 \pm 7$   &     0.1   &    $12.9 \pm 0.2$    &    $5.9 \pm 0.7$    &   0.19   &  $19 \pm 2$  \\
204  &      100.6019   &   8.9422   &     $34.6 \times 31.9$   &    12.9   &     11.5   &    $1.9 \pm 0.6$    &    $6.0 \pm 0.9$   &     0.1   &    $13.1 \pm 0.1$    &    $0.6 \pm 0.1$    &   1.8   &  $4.5 \pm 0.1$  \\
205  &      100.3301   &   9.6818   &     $56.2 \times 34.8$   &    43.8   &     12.3   &    $4 \pm 2$    &    $17 \pm 3$   &     0.14   &    $14.5 \pm 0.1$    &    $1.7 \pm 0.3$    &   1.0   &  $7.7 \pm 0.6$  \\
206  &      100.1509   &   9.8352   &     $61.0 \times 36.1$   &    178.2   &     11.7   &    $1 \pm 1$    &    $6 \pm 2$   &     0.15   &    $18.3 \pm 0.4$    &    $0.7 \pm 0.2$    &   3.3   &  $4.3 \pm 0.3$  \\
207  &      100.0777   &   9.8355   &     $33.5 \times 22.9$   &    171.4   &     10.2   &    $1.1 \pm 0.7$    &    $2.0 \pm 0.8$   &     0.07   &    $18.4 \pm 0.2$    &    $0.2 \pm 0.1$    &   5.3   &  $4.2 \pm 0.2$  \\
208  &      100.393   &   8.7956   &     $27.3 \times 22.9$   &    61.5   &     9.9   &    $0.8 \pm 0.2$    &    $1.0 \pm 0.3$   &     0.06   &    $13.9 \pm 0.1$    &    $0.1 \pm 0.0$    &   4.9   &  $3.0 \pm 0.0$  \\
209  &      100.6358   &   9.2201   &     $41.5 \times 28.3$   &    121.7   &     10.3   &    $1.1 \pm 0.4$    &    $3.0 \pm 0.5$   &     0.1   &    $13.8 \pm 0.1$    &    $0.3 \pm 0.1$    &   3.9   &  $4.3 \pm 0.1$  \\
210$^*$  &      100.1728   &   9.0687   &     $52.5 \times 36.9$   &    37.7   &     10.6   &    $1.8 \pm 0.8$    &    $7 \pm 1$   &     0.14   &    $12.7 \pm 0.1$    &    $0.7 \pm 0.1$    &   2.1   &  $6.8 \pm 0.2$  \\
211  &      99.8251   &   9.6745   &     $39.1 \times 26.2$   &    141.1   &     11.2   &    $0.8 \pm 0.3$    &    $2.0 \pm 0.5$   &     0.09   &    $14.3 \pm 0.1$    &    $0.2 \pm 0.1$    &   4.8   &  $2.5 \pm 0.1$  \\
212  &      100.4328   &   9.4874   &     $56.5 \times 46.5$   &    116.7   &     13.3   &    $2.1 \pm 0.4$    &    $13.0 \pm 0.7$   &     0.17   &    $14.2 \pm 0.1$    &    $1.3 \pm 0.1$    &   1.5   &  $4.0 \pm 0.1$  \\
213  &      100.1546   &   9.1013   &     $50.2 \times 30.6$   &    122.3   &     10.7   &    $1.8 \pm 0.8$    &    $6 \pm 1$   &     0.12   &    $12.8 \pm 0.1$    &    $0.6 \pm 0.1$    &   2.2   &  $6.7 \pm 0.4$  \\
214  &      99.8508   &   9.513   &     $49.7 \times 39.1$   &    4.2   &     18.1   &    $0.8 \pm 0.2$    &    $4.0 \pm 0.3$   &     0.14   &    $14.1 \pm 0.1$    &    $0.4 \pm 0.0$    &   3.8   &  $1.7 \pm 0.0$  \\
215  &      100.2164   &   9.9611   &     $51.6 \times 29.7$   &    47.7   &     11.0   &    $1.3 \pm 0.5$    &    $5.0 \pm 0.8$   &     0.12   &    $16.5 \pm 0.1$    &    $0.5 \pm 0.1$    &   3.5   &  $3.8 \pm 0.1$  \\
216  &      100.1868   &   9.5419   &     $36.1 \times 31.9$   &    29.6   &     10.9   &    $2 \pm 1$    &    $7 \pm 2$   &     0.1   &    $15.1 \pm 0.1$    &    $0.7 \pm 0.2$    &   1.8   &  $8.2 \pm 0.5$  \\
217  &      100.393   &   9.7742   &     $67.7 \times 65.9$   &    157.0   &     19.8   &    $1.9 \pm 0.4$    &    $21.0 \pm 0.9$   &     0.22   &    $14.5 \pm 0.2$    &    $2.1 \pm 0.1$    &   1.3   &  $3.2 \pm 0.1$  \\
218  &      100.3331   &   9.4513   &     $38.3 \times 32.5$   &    39.8   &     10.5   &    $5 \pm 2$    &    $15 \pm 3$   &     0.11   &    $14.2 \pm 0.2$    &    $1.5 \pm 0.3$    &   0.83   &  $14.1 \pm 0.6$  \\
219  &      100.0169   &   9.3053   &     $35.0 \times 26.9$   &    26.4   &     9.8   &    $1.2 \pm 0.3$    &    $3.0 \pm 0.4$   &     0.09   &    $13.4 \pm 0.1$    &    $0.3 \pm 0.0$    &   3.0   &  $3.8 \pm 0.1$  \\
220  &      100.0327   &   9.9133   &     $34.0 \times 27.6$   &    144.4   &     9.8   &    $1.1 \pm 0.3$    &    $3.0 \pm 0.4$   &     0.09   &    $16.9 \pm 0.2$    &    $0.3 \pm 0.0$    &   4.7   &  $3.3 \pm 0.1$  \\
221  &      100.2234   &   9.4137   &     $35.1 \times 29.0$   &    101.4   &     9.7   &    $0.8 \pm 0.4$    &    $2.0 \pm 0.6$   &     0.09   &    $14.8 \pm 0.1$    &    $0.3 \pm 0.1$    &   4.5   &  $3.7 \pm 0.1$  \\
222  &      100.1727   &   9.2276   &     $39.8 \times 33.6$   &    23.4   &     10.6   &    $1.1 \pm 0.3$    &    $4.0 \pm 0.5$   &     0.11   &    $13.9 \pm 0.1$    &    $0.4 \pm 0.1$    &   3.1   &  $4.1 \pm 0.0$  \\
223  &      100.0034   &   9.3185   &     $48.9 \times 39.3$   &    142.6   &     10.5   &    $1.8 \pm 0.4$    &    $8.0 \pm 0.6$   &     0.14   &    $13.4 \pm 0.1$    &    $0.8 \pm 0.1$    &   1.9   &  $3.5 \pm 0.1$  \\
224  &      99.9801   &   9.3956   &     $41.0 \times 32.4$   &    177.0   &     11.2   &    $1.1 \pm 0.2$    &    $4.0 \pm 0.4$   &     0.11   &    $13.7 \pm 0.1$    &    $0.4 \pm 0.0$    &   3.1   &  $2.6 \pm 0.0$  \\
225  &      100.168   &   8.8929   &     $38.3 \times 26.7$   &    23.5   &     9.4   &    $2.9 \pm 0.7$    &    $7.0 \pm 0.8$   &     0.09   &    $13.3 \pm 0.1$    &    $0.7 \pm 0.1$    &   1.5   &  $4.3 \pm 0.1$  \\
226  &      100.1345   &   9.7806   &     $38.2 \times 29.7$   &    13.0   &     9.5   &    $0.9 \pm 0.6$    &    $2.0 \pm 0.7$   &     0.1   &    $20.0 \pm 0.1$    &    $0.2 \pm 0.1$    &   7.5   &  $3.6 \pm 0.1$  \\
227  &      100.1679   &   8.9551   &     $35.4 \times 26.7$   &    22.8   &     9.5   &    $1.9 \pm 0.6$    &    $5.0 \pm 0.8$   &     0.09   &    $13.1 \pm 0.1$    &    $0.5 \pm 0.1$    &   1.9   &  $6.0 \pm 0.1$  \\
228  &      100.2744   &   9.4879   &     $52.5 \times 34.8$   &    29.2   &     9.8   &    $32 \pm 10$    &    $130 \pm 15$   &     0.14   &    $14.4 \pm 0.6$    &    $13 \pm 2$    &   0.12   &  $28 \pm 4$  \\
229  &      100.4712   &   9.7146   &     $44.5 \times 31.1$   &    101.7   &     10.8   &    $0.8 \pm 0.3$    &    $3.0 \pm 0.5$   &     0.11   &    $14.4 \pm 0.1$    &    $0.3 \pm 0.1$    &   4.5   &  $2.8 \pm 0.1$  \\
230  &      100.5055   &   9.0346   &     $40.9 \times 24.4$   &    13.9   &     9.4   &    $1.3 \pm 0.2$    &    $3.0 \pm 0.3$   &     0.09   &    $13.6 \pm 0.1$    &    $0.3 \pm 0.0$    &   3.0   &  $3.5 \pm 0.1$  \\
231  &      100.6787   &   9.2856   &     $50.0 \times 33.1$   &    9.4   &     10.6   &    $1.5 \pm 0.4$    &    $6.0 \pm 0.6$   &     0.13   &    $13.7 \pm 0.1$    &    $0.6 \pm 0.1$    &   2.4   &  $3.6 \pm 0.1$  \\
232  &      100.2841   &   9.8799   &     $54.2 \times 47.1$   &    104.2   &     13.7   &    $1.7 \pm 0.8$    &    $11 \pm 2$   &     0.17   &    $15.0 \pm 0.1$    &    $1.1 \pm 0.2$    &   1.9   &  $6.6 \pm 0.2$  \\
233  &      100.0403   &   9.8045   &     $48.1 \times 35.8$   &    45.5   &     11.7   &    $0.6 \pm 0.3$    &    $3.0 \pm 0.5$   &     0.13   &    $16.9 \pm 0.3$    &    $0.3 \pm 0.1$    &   6.8   &  $2.7 \pm 0.1$  \\
234  &      99.8508   &   9.4494   &     $22.7 \times 20.4$   &    94.0   &     10.1   &    $0.5 \pm 0.1$    &    $1.0 \pm 0.1$   &     0.04   &    $14.0 \pm 0.1$    &    $0.1 \pm 0.0$    &   6.1   &  $1.8 \pm 0.0$  \\
235  &      100.5251   &   9.1855   &     $39.1 \times 30.7$   &    109.6   &     9.5   &    $3 \pm 1$    &    $9 \pm 2$   &     0.1   &    $12.4 \pm 0.1$    &    $0.9 \pm 0.2$    &   1.2   &  $12.0 \pm 0.3$  \\
236  &      100.6078   &   9.7215   &     $59.5 \times 39.0$   &    90.2   &     11.8   &    $1.1 \pm 0.2$    &    $6.0 \pm 0.3$   &     0.16   &    $15.6 \pm 0.3$    &    $0.6 \pm 0.0$    &   3.2   &  $2.1 \pm 0.0$  \\
237  &      100.3463   &   9.6186   &     $41.9 \times 31.3$   &    134.1   &     10.1   &    $0.9 \pm 0.5$    &    $3.0 \pm 0.8$   &     0.11   &    $15.2 \pm 0.1$    &    $0.3 \pm 0.1$    &   4.8   &  $4.6 \pm 0.2$  \\
238  &      99.914   &   9.7416   &     $50.2 \times 30.3$   &    3.4   &     11.8   &    $0.8 \pm 0.3$    &    $4.0 \pm 0.6$   &     0.12   &    $14.6 \pm 0.1$    &    $0.4 \pm 0.1$    &   4.0   &  $3.1 \pm 0.1$  \\
239  &      100.4243   &   8.8565   &     $53.8 \times 41.1$   &    16.7   &     10.8   &    $1.3 \pm 0.3$    &    $7.0 \pm 0.6$   &     0.15   &    $13.6 \pm 0.1$    &    $0.8 \pm 0.1$    &   2.3   &  $3.5 \pm 0.1$  \\
240  &      100.1624   &   10.0982   &     $43.0 \times 31.3$   &    169.2   &     9.9   &    $0.9 \pm 0.2$    &    $3.0 \pm 0.3$   &     0.11   &    $14.4 \pm 0.2$    &    $0.3 \pm 0.0$    &   4.4   &  $2.5 \pm 0.1$  \\
241  &      99.9246   &   9.6363   &     $36.2 \times 28.7$   &    71.0   &     9.2   &    $0.9 \pm 0.3$    &    $2.0 \pm 0.4$   &     0.09   &    $14.5 \pm 0.1$    &    $0.3 \pm 0.1$    &   4.5   &  $3.7 \pm 0.0$  \\
242  &      100.3002   &   9.5817   &     $40.1 \times 33.1$   &    170.4   &     9.5   &    $16 \pm 8$    &    $49 \pm 10$   &     0.11   &    $12.7 \pm 0.3$    &    $5 \pm 1$    &   0.23   &  $33 \pm 3$  \\
243$^*$  &      100.0288   &   9.5179   &     $34.4 \times 23.7$   &    45.1   &     9.4   &    $0.9 \pm 0.4$    &    $2.0 \pm 0.5$   &     0.08   &    $14.1 \pm 0.1$    &    $0.2 \pm 0.1$    &   4.3   &  $3.6 \pm 0.1$  \\
244  &      100.2381   &   9.8821   &     $45.9 \times 40.7$   &    41.0   &     9.7   &    $1.3 \pm 0.8$    &    $6 \pm 1$   &     0.14   &    $16.9 \pm 0.4$    &    $0.6 \pm 0.1$    &   3.2   &  $5.1 \pm 0.2$  \\
245  &      100.1924   &   10.1409   &     $50.3 \times 31.8$   &    90.7   &     9.7   &    $1.1 \pm 0.2$    &    $4.0 \pm 0.3$   &     0.13   &    $14.6 \pm 0.2$    &    $0.4 \pm 0.0$    &   3.9   &  $2.7 \pm 0.1$  \\
246  &      100.5613   &   9.1736   &     $33.2 \times 31.1$   &    69.7   &     9.3   &    $1.7 \pm 0.9$    &    $5 \pm 1$   &     0.09   &    $13.5 \pm 0.1$    &    $0.5 \pm 0.1$    &   2.1   &  $5.2 \pm 0.3$  \\
247  &      100.6046   &   9.3906   &     $33.4 \times 25.8$   &    10.3   &     9.3   &    $1.0 \pm 0.5$    &    $2.0 \pm 0.6$   &     0.08   &    $13.9 \pm 0.1$    &    $0.2 \pm 0.1$    &   3.9   &  $4.7 \pm 0.0$  \\
248  &      100.3279   &   9.1959   &     $37.9 \times 28.7$   &    179.6   &     9.2   &    $0.6 \pm 0.1$    &    $2.0 \pm 0.2$   &     0.1   &    $14.4 \pm 0.1$    &    $0.2 \pm 0.0$    &   6.4   &  $2.8 \pm 0.0$  \\
249  &      100.2868   &   9.733   &     $60.7 \times 49.7$   &    170.2   &     11.1   &    $3 \pm 1$    &    $24 \pm 3$   &     0.18   &    $14.5 \pm 0.1$    &    $2.5 \pm 0.4$    &   0.87   &  $9.1 \pm 0.5$  \\
250  &      100.1913   &   9.8507   &     $37.9 \times 35.2$   &    160.4   &     9.2   &    $0.6 \pm 0.2$    &    $2.0 \pm 0.3$   &     0.11   &    $18.8 \pm 0.3$    &    $0.2 \pm 0.0$    &   8.2   &  $2.8 \pm 0.0$  \\
251  &      100.105   &   10.0504   &     $45.9 \times 34.5$   &    129.2   &     9.1   &    $0.7 \pm 0.2$    &    $3.0 \pm 0.3$   &     0.12   &    $15.1 \pm 0.1$    &    $0.3 \pm 0.0$    &   6.0   &  $2.1 \pm 0.0$  \\
252  &      100.0991   &   9.6237   &     $46.2 \times 36.4$   &    22.1   &     9.2   &    $1.1 \pm 0.6$    &    $5 \pm 1$   &     0.13   &    $16.0 \pm 0.1$    &    $0.5 \pm 0.1$    &   3.4   &  $5.6 \pm 0.1$  \\
253  &      100.1044   &   9.9756   &     $63.1 \times 44.1$   &    111.7   &     10.3   &    $1.0 \pm 0.4$    &    $6.0 \pm 0.8$   &     0.17   &    $15.0 \pm 0.2$    &    $0.6 \pm 0.1$    &   3.4   &  $2.4 \pm 0.2$  \\
254  &      100.2719   &   9.4106   &     $51.3 \times 42.5$   &    30.8   &     9.2   &    $3 \pm 2$    &    $14 \pm 3$   &     0.15   &    $14.2 \pm 0.1$    &    $1.5 \pm 0.3$    &   1.2   &  $7.8 \pm 0.4$  \\
255  &      100.0803   &   9.5105   &     $57.3 \times 48.6$   &    154.0   &     9.5   &    $2.3 \pm 0.6$    &    $15 \pm 1$   &     0.17   &    $14.0 \pm 0.1$    &    $1.6 \pm 0.1$    &   1.3   &  $4.0 \pm 0.2$  \\
256  &      99.8548   &   9.8456   &     $50.9 \times 37.3$   &    128.7   &     9.5   &    $0.7 \pm 0.1$    &    $3.0 \pm 0.3$   &     0.14   &    $14.5 \pm 0.1$    &    $0.3 \pm 0.0$    &   4.8   &  $2.1 \pm 0.0$  \\

\hline
\end{tabular} 
 } \\
  	\tablefootmark{a}{1 Av = $10^{21} \, \rm H_2 \, cm^{-2}$ } \\
  	\tablefootmark{b}{source deconvolved size at FWHM, $FWHM_{\rm dec}= (A\times B - 18.2^2)^{1/2}$, minimized at half the beam size} \\
  	$^*$ Clumps located outside the area where YSOs have been detected, excluded in the $nnd$ analysis. 
 \end{table*}

\end{appendix}

\end{document}